%% file: main.tex
\documentclass[manuscript, screen]{jair}

\setcopyright{none}

\JAIRAE{}
\JAIRTrack{}
\RequirePackage[
  datamodel=acmdatamodel,
  style=acmauthoryear,
  backend=biber,
  giveninits=true,
  uniquename=init
  ]{biblatex}

\usepackage{preamble}

\begin{document}

\title[Theoretical Foundations of \texorpdfstring{$\delta$}{delta}-margin Majority Voting]{Theoretical Foundations of \texorpdfstring{$\delta$}{delta}-margin Majority Voting for Quality Assurance in High-Stakes Machine Learning}

\author{Margarita Boyarskaya}
\orcid{0000-0002-8181-6030}
\email{mb6599@stern.nyu.edu}
\affiliation{%
  \institution{JP Morgan AI Research}
  \city{New York}
  \state{NY}
  \country{USA}
}

\author{Panos Ipeirotis}
\orcid{0000-0002-2966-7402}
\email{panos@stern.nyu.edu}
\affiliation{%
  \institution{New York University}
  \city{New York}
  \state{NY}
  \country{USA}
}

\renewcommand{\shortauthors}{Boyarskaya \& Ipeirotis}

\begin{abstract}
In high-stakes machine learning applications such as fraud detection, medical diagnostics, and content moderation, practitioners often rely on consensus-based approaches to control prediction quality. A particularly valuable technique---$\delta$-margin majority voting---collects votes sequentially until one candidate label exceeds alternatives by a predetermined threshold $\delta$, offering stronger confidence in uncertain cases than simple majority voting. Despite widespread adoption, this approach has lacked rigorous theoretical foundations, compelling practitioners to use heuristic estimates for critical performance metrics like expected accuracy and resource requirements.

This paper establishes a comprehensive theoretical framework for $\delta$-margin majority voting by explicitly formulating it as an absorbing Markov chain and leveraging classical results from Gambler's Ruin theory. Our contributions center on assembling a practical \emph{design calculus} for $\delta$-margin voting:
(1)~Closed-form expressions for consensus accuracy, expected voting duration (cost), variance, and the complete probability mass function of the stopping time, enabling model-based process design rather than heuristic trial-and-error.
(2)~A Bayesian extension that handles uncertainty in worker accuracy, allowing real-time monitoring of expected quality and cost as votes arrive, with both single-Beta and mixture-of-Betas priors.
(3)~Cost-calibration methods that determine how to achieve equivalent quality across worker pools with different accuracies, and how to set payment rates accordingly.

We validate our theoretical predictions using two real-world datasets, demonstrating close agreement between predicted and observed consensus quality and resource requirements. The framework provides practitioners with a rigorous, operational toolkit for designing $\delta$-margin voting processes---predicting expected accuracy, cost, and their distributions---thereby replacing ad-hoc experimentation with model-based process design in applications where quality control and cost transparency are essential.
\end{abstract}

\maketitle

\import{tex_files}{1-introduction}

\import{tex_files}{2-model}
\import{tex_files}{3-related}

\import{tex_files}{4-model-properties}

\import{tex_files}{5-unknown_p}
\import{tex_files}{6-payment}
\import{tex_files}{7-experiments}
\import{tex_files}{8-example}
\import{tex_files}{9-future}

\begin{acks}
We thank the anonymous reviewers and Associate Editor for their constructive feedback, which substantially improved the paper. We also thank our industry collaborator for providing the AML investigator dataset used in Section~\ref{sec:example}.
\end{acks}

\printbibliography

\appendix

\import{tex_files/appendices}{A-MC_simulations_table}
\import{tex_files/appendices}{A-payment-proof}

\end{document}

%% file: tex_files/1-introduction.tex
\section{Introduction}

Many decision-making pipelines require a binary judgment on individual items: is a flagged transaction fraudulent or benign? Are two database records duplicates? Should a piece of content be removed? When the cost of an error is high, a single worker's verdict is rarely sufficient, and the standard remedy is to \textbf{aggregate multiple independent votes to reach a consensus}. This pattern arises across a wide range of domains:

\begin{itemize}
    \item \emph{Financial Monitoring}: Transactions flagged as potential money laundering must be reviewed; workers decide whether to escalate each alert for further investigation.
    \item \emph{Security Systems}: User accounts exhibiting unusual activity are flagged as potentially compromised. Workers verify whether these accounts should be suspended.
    \item \emph{Database Management}: Algorithms identify potential duplicates in product catalogs or customer records. Workers confirm whether entries refer to the same entity before merging.
    \item \emph{Machine Learning Benchmarks}: Creating evaluation data and benchmarks requires minimizing label noise to ensure the integrity of the results.
    \item \emph{Citizen Science}: In domains where automated classification is unreliable---such as Zooniverse projects or disaster relief mapping---accurate human labeling remains essential.
\end{itemize}

\noindent The workers casting these votes may be human crowdsourkers, domain experts, or increasingly, Large Language Models (LLMs) and other automated classifiers. Regardless of the worker type, each individual vote is noisy, and ensuring high-quality outcomes requires aggregating multiple votes.\footnote{In early use cases, crowdsourcing with multiple workers was used for generating \emph{training data} for machine learning algorithms~\citep{russakovsky2015imagenet}. However, recent advances indicate that achieving high levels of machine learning performance is possible without relying on the repeated labeling of individual training data points~\citep{lin2016re}, even with noisy labels. Hence, in this paper we focus on repeated labeling for quality assurance in downstream tasks of ML pipelines (e.g., escalation of risky user behaviors or flagged content).} Although there is a wealth of literature proposing various vote aggregation schemes for quality control (e.g., \citeauthor{chilton2013cascade}, \citeyear{chilton2013cascade}; \citeauthor{dai2013pomdp}, \citeyear{dai2013pomdp}; \citeauthor{mortensen2013crowdsourcing}, \citeyear{mortensen2013crowdsourcing}), simple methods like majority voting or its variations are often preferred in practice due to their simplicity.


One variation of majority voting often encountered is the ``\textit{$\delta$-margin majority voting}.'' In this scheme, votes are collected until votes for one class outnumber votes for the other by a margin of $\delta$.\footnote{Note that the name \textit{$\delta$-margin voting} is our proposed terminology. This consensus scheme, widely discussed among the Mechanical Turk (MTurk) crowdsourcing community in its early days, was sometimes referred to as \textit{strong plurality} or \textit{adaptive} majority voting. The practitioners in the field criticized the scheme for its lack of theoretical guarantees, highlighting that reaching consensus with 2-vs-7 votes is not the same as reaching consensus with 1-vs-6 votes. Eventually, the consensus scheme was removed from the standard task design options on MTurk. In this work, we address the criticism of this consensus method and show how to leverage useful information about the exact combination of votes in order to gain insight about the workers accuracy.} Compared to standard majority voting, the $\delta$-margin approach demands stronger agreement among workers, imposing additional scrutiny which is harder to satisfy, especially for `difficult' items that lead to disagreement among workers. Despite its practical importance, the design of tasks using this scheme tends to be heuristic and ad hoc, lacking a solid theoretical foundation that describes the properties of the voting process.

\paragraph{Contributions:}

The main contribution of this paper is to assemble a comprehensive \emph{design calculus} for $\delta$-margin voting---a practical toolkit that enables decision-makers to answer key process-design questions \emph{ex ante} (i.e., before observing any results). The individual mathematical building blocks draw on well-established tools (absorbing Markov chains and the Gambler's Ruin model), but no prior work has combined them into a unified framework that covers quality prediction, cost estimation, distributional analysis, Bayesian treatment of unknown worker accuracy, and payment calibration across heterogeneous worker pools.\footnote{The alternative approach, which our work improves on, is to make an assumption about the voting process, and then to test the assumption's validity during (costly) initial experiments, and iterate this process to adjust the parameter values. Our approach allows for accurate process design with highly reliable estimates of key process properties, replacing heuristic design-level trial-and-error with model-based process design, given a prior on worker accuracy.} Specifically, this toolkit enables practitioners to answer:

\begin{itemize}
    \item \emph{How should a voting process be structured to achieve a desired level of accuracy?}
    \item \emph{What is the expected cost of running such a process?}
    \item \emph{Can a pool of less accurate workers achieve the same accuracy through aggregation?}
\end{itemize}

\noindent We demonstrate that, with minimal assumptions about the quality of workers or the difficulty of the items, we can estimate the process quality and duration before the labeling occurs. This contrasts with approaches that attempt to characterize individual worker accuracy, which can be challenging due to various factors (e.g., for human workers, factors like turnover and desire to maintain worker fungibility; and for LLM-based workers, the evolving nature of AI tools). Our method requires only a lightweight prior expectation on how well a \textbf{group} of workers is expected to perform on an item.

\paragraph{Broader Motivations:}

A critical gap in the design of vote-aggregation workflows is the ability to provide \emph{theoretical performance predictions} before any votes are collected. In practice, process designers must answer questions about quality, cost, and staffing without the luxury of extensive pilot studies. The $\delta$-margin voting framework addresses this gap by enabling accurate ex-ante estimation of key process properties---expected accuracy, cost distributions, and their sensitivity to design parameters---given a prior on worker accuracy. We highlight three areas where such predictions are especially valuable:\footnote{Throughout this paper, we use the term ``worker'' to refer to any entity that provides a single vote---whether a human labeler, an LLM, or another automated system. These votes, collected from multiple workers, are aggregated into a consensus label.}

\begin{itemize}
    \item \emph{Human-in-the-loop AI oversight.} As AI systems are deployed in high-stakes settings, humans play a crucial role in reviewing and correcting automated decisions. Vote aggregation is the standard mechanism for combining multiple reviewers' judgments, yet most deployed systems lack formal performance characterizations for this component. The $\delta$-margin framework provides closed-form expressions for the expected accuracy, cost, and duration of the review process, enabling principled workflow design.

    \item \emph{Multi-model AI systems.} When multiple LLMs or classifiers are applied to the same task---whether in parallel or in sequence---their outputs must be aggregated. The $\delta$-margin scheme offers a way to design such aggregation with theoretical performance predictions. While fixed-size majority voting provides exact binomial expressions for a predetermined number of votes, $\delta$-margin voting additionally enables \emph{adaptive stopping}, terminating as soon as sufficient agreement is reached and thereby avoiding unnecessary votes.

    \item \emph{Regulatory and compliance requirements.} In domains such as finance, healthcare, and content moderation, regulatory frameworks increasingly demand transparency and accountability in decision-making processes. Providing ex-ante performance predictions---including expected accuracy and its sensitivity to design parameters---supports compliance efforts by making the quality-control process auditable and quantifiable. We note, however, that the framework provides \emph{expected}-accuracy predictions conditional on the assumed prior, not worst-case or tail-risk bounds; for applications requiring such guarantees, the predictions should be supplemented with appropriate risk criteria (e.g., posterior credible intervals or escalation rules for low-confidence items).
\end{itemize}

\noindent As the adoption of hybrid human--AI systems grows, theoretical frameworks that characterize the expected performance of vote-aggregation components become increasingly valuable. Such frameworks should account for worker quality, the cost structure of the workflow, and the interplay between human and automated decision-making.

\paragraph{Paper Organization:}

The rest of the paper is organized as follows.
Section~\ref{sec:model} introduces the Markov chain formalism for the $\delta$-margin voting process.
Section~\ref{sec:litrev} discusses existing work on consensus aggregation design and quality assurance.
Section~\ref{sec:characteristics} presents theoretical equations for key characteristics of the $\delta$-margin voting process: quality of the results, expected number of votes required to reach consensus, variance of vote requirements, and other distribution moments.
Section~\ref{sec:unknown_p} examines how to operate when there's uncertainty about the quality of the underlying workers.
Section~\ref{sec:equiv} expands on these results to propose a quality-sensitive payment scheme that connects result quality with item-level worker accuracy.
Section~\ref{sec:theory_vs_exp} presents experimental validation with real data, comparing theoretical estimates with actual results.
Section~\ref{sec:example} applies the full framework to a real anti-money laundering operation with two investigator pools, answering concrete design questions---what margin threshold to use, and which pool is more cost-effective---directly from the theory, without pilot experiments.
Section~\ref{sec:future} explores immediate and future research developments that can build upon this work.

\paragraph{Relationship to prior work:} \emph{An earlier version of this work appeared as a conference paper~\citep{boyarskaya2024ci}. The present manuscript substantially extends that version with: a formal model definition with explicit assumptions (Section~\ref{sec:model}), an SPRT connection (Section~\ref{sec:relwork-sprt}), a Bayesian treatment of unknown~$p$ with mixture priors and deployment monitoring (Section~\ref{sec:unknown_p}), payment equivalence under unknown accuracy (Section~\ref{sec:pay_worker_unknown_acc}), expanded experiments with confidence intervals and stress tests (Section~\ref{sec:theory_vs_exp}), and a new real-world case study using an AML dataset (Section~\ref{sec:example}).}

%% file: tex_files/2-model.tex
\section{Markov chain model of \texorpdfstring{$\delta$}{delta}-margin majority voting}
\label{sec:model}

In the context of labeling items with a binary label using $\delta$-margin voting, we solicit votes regarding a specific item $i$ from a \emph{large worker pool}. We assume that each worker has an (unobserved) accuracy drawn once per worker-item pair from a distribution $\mathcal{A}_i$. We make no assumptions about the distribution $\mathcal{A}_i$ other than that it has an expected value $E[\mathcal{A}_i] = p_i$, meaning that the workers are expected to be correct for the item $i$ with expectation $p_i$ and incorrect with expectation $1-p_i$; \emph{we emphasize that this is the expectation across the pool, and not the accuracy of individual workers}. We continue to ask workers to assign labels to an item until the absolute difference between the number of votes for the two classes exceeds a predefined consensus threshold $\delta$. (In other words, until the number of votes for one class exceeds the number of votes for the other class by at least $\delta$.) We allow different items to have varying difficulty levels (that is, the distribution of worker accuracies $\mathcal{A}$ may differ between items); however, we assume that the \emph{mean} of the distribution $E[\mathcal{A}_i]$ remains stable across labels given for the same item $i$.

\textbf{Definition 1.} A \textit{$\delta$-margin voting} process for a binary-labeled item~$i$ operates as follows. Let $\delta \geq 1$ be a fixed positive integer threshold. Starting from state $S_0 = 0$, binary votes are collected sequentially; under Assumptions~A1--A3 (stated below), each vote is correct with probability~$p_i$ and incorrect with probability~$1-p_i$, where $p_i = E[\mathcal{A}_i]$ is the mean pool accuracy for item~$i$. A correct vote increments the state ($S_{t+1} = S_t + 1$) and an incorrect vote decrements it ($S_{t+1} = S_t - 1$). The process terminates at the stopping time $\tau = \min\{t \geq 1 : |S_t| = \delta\}$, and the consensus label is the class with the larger vote count. The indicator $\hat{Y}_i = \mathbf{1}[S_\tau = \delta]$ records whether the consensus is \emph{correct} (i.e., whether the correct class reached the~$+\delta$ boundary).

\begin{figure}[t]
\centering
\includegraphics[width=\columnwidth]{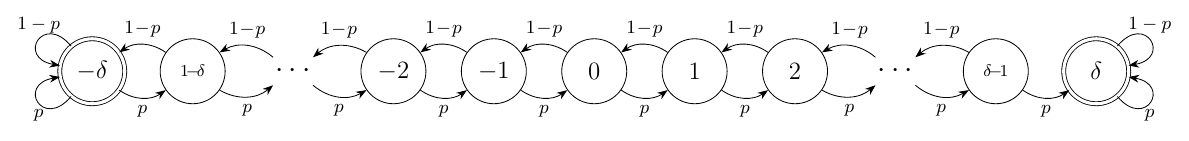}
\Description{Markov chain state diagram showing transitions between states from negative delta to positive delta, with absorbing states at both boundaries.}
\caption{A Markov chain diagram illustrating state transitions for $\delta$-margin majority voting on a single item, for which the average worker pool accuracy is $p$. Node labels indicate the difference between the numbers of correct ($n_1$) and incorrect ($n_0$) votes, $n_1-n_0$. Consensus is reached in absorbing states $\delta$ and $-\delta$, the former resulting in a correct consensus vote and the latter in an incorrect one.}
\label{fig:markov}
\end{figure}

\textbf{Example.} Set $\delta=2$. Then, a $\delta$-margin voting process might stop when the vote-count tuples $\langle n_{\mathit{correct}}, n_{\mathit{incorrect}}\rangle$ attain one of the following values: 
$\langle 2, 0\rangle$,
$\langle 0, 2\rangle$,
$\langle 3, 1\rangle$,
$\langle 1, 3\rangle$,
$\langle 4, 2\rangle$,
$\langle 2, 4\rangle$, ..., and so on. The consensus vote is correct in the states $\{
\langle 2, 0\rangle, \langle 3, 1\rangle,
\langle 4, 2\rangle, ...\}$, while the consensus vote is incorrect in the states $\{\langle 0, 2\rangle, \langle 1, 3\rangle, \langle 2, 4\rangle, ...\}$.

To further clarify our assumptions, consider a practical scenario. Imagine we have a large group of workers labeling images. Each worker has varying accuracy depending on their expertise or the specific content being labeled. We do not assume that all workers share the same accuracy---individual accuracies may vary substantially. Instead, we assume only that the average accuracy across the pool remains stable for each given image. For example, if an image is moderately difficult, workers' accuracies might vary significantly, with some workers highly accurate (e.g., 90\%) and others less so (e.g., 50\%). Despite this variability, as long as the overall average accuracy of workers labeling this particular image remains approximately constant (e.g., 70\%), and each vote is drawn independently from the pool (as formalized below), the theoretical results characterize the voting process accurately. Thus, the model explicitly allows individual worker accuracy to vary, relying on both the stability of the aggregate mean and the independence of successive draws.

A key implicit requirement is that successive votes are drawn independently. In practice, this holds when each vote is solicited from a fresh, randomly chosen worker---equivalent to sampling with replacement from the pool. The ``large worker pool'' assumption stated above ensures that repeated draws are approximately independent even without literal replacement, since the probability of re-sampling the same worker is negligible.

We now state the three assumptions that underpin the Markov chain model used throughout the paper.

\medskip
\noindent\fbox{\parbox{\dimexpr\columnwidth-2\fboxsep-2\fboxrule\relax}{%
\textbf{Assumptions.}
\begin{description}
\item[A1 (Item-level stationarity).] For each item~$i$, the mean pool accuracy $p_i = E[\mathcal{A}_i]$ remains constant throughout the adjudication of that item. Different items may have different values of~$p_i$.
\item[A2 (Conditional independence \& blind voting).] Votes on item~$i$ are conditionally independent given the item's true label~$Y_i$ and mean accuracy~$p_i$. Each worker casts their vote without observing the votes already cast by others, which prevents information cascades.
\item[A3 (Heterogeneity via mixing).] Individual worker accuracies may vary arbitrarily; the distribution~$\mathcal{A}_i$ may take any shape provided its mean is~$p_i$. The Markov chain transition probability depends only on this mean, not on the full shape of~$\mathcal{A}_i$.
\end{description}
}}
\medskip

\noindent \textbf{Why the mean accuracy suffices (under i.i.d.\ sampling).} Under A1--A3, the sequence of vote outcomes for item~$i$ forms an i.i.d.\ Bernoulli($p_i$) process, justifying the Markov chain model. To see this, note that each worker's accuracy~$a$ is drawn from~$\mathcal{A}_i$~(A3), so by the law of total expectation, $\Pr(\text{vote correct} \mid Y_i, p_i) = E_{a \sim \mathcal{A}_i}[a] = p_i$. By~A2 (conditional independence and blindness), successive votes are independent Bernoulli($p_i$) trials. By~A1, $p_i$ is constant during the adjudication of item~$i$, so the random walk $S_t$ is time-homogeneous: each step goes up with probability~$p_i$ and down with probability~$1 - p_i$, regardless of the current state or time step.

\emph{Scope of the reduction.} The i.i.d.\ Bernoulli($p_i$) reduction is exact when each vote is drawn independently from the pool---that is, when workers are sampled with replacement (or from a pool large enough that resampling is negligible). In operational settings with finite worker pools, repeated use of the same workers, or correlated error modes, a pool with the same mean~$p_i$ can exhibit different stopping-time distributions than the i.i.d.\ model predicts. Assumptions A1--A3 therefore accommodate arbitrary worker heterogeneity in accuracy \emph{levels}, but they do require that successive draws are independent. Investigating sensitivity to finite-pool and correlation effects is an important direction for future work (see Section~\ref{sec:future}).

The i.i.d.\ justification also breaks down if workers are selected adaptively based on interim vote tallies (violating~A2), workers observe and are influenced by earlier votes (violating~A2, causing information cascades), or the worker pool composition shifts mid-adjudication (violating~A1).

Since the only desideratum for the consensus decision is the difference $S_t$ between the numbers of correct and incorrect votes, the process can be modeled as a \emph{time-homogeneous Markov random walk with two absorbing states}: the process ends at $S_t = \delta$ (correct consensus) or $S_t = -\delta$ (incorrect consensus). In all other states, we procure additional votes, which transition the state from $k$ to $k+1$ (correct vote, probability~$p$) or from $k$ to $k-1$ (incorrect vote, probability~$1-p$).

\textbf{Parallelization.} Although the model describes votes arriving sequentially, the first $\delta$ votes can be collected in parallel because no early stopping is possible before $\delta$ votes have been cast. More generally, at any state~$k$, the minimum number of votes needed to reach either absorbing state is $l = \delta - |k|$; these $l$ votes can be solicited in parallel without affecting the stopping rule. This observation is useful in practice when latency---not just cost---is a concern. \textbf{Overshoot cost.} In batch or queue-based settings (e.g., parallel LLM calls), in-flight votes cannot always be cancelled once the boundary is crossed within a batch. If a batch of size~$l$ is dispatched and the boundary is reached before all $l$ responses arrive, the surplus votes represent wasted cost. The cost formulas in Sections~\ref{sec:characteristics}--\ref{sec:equiv} describe the \emph{sequential} stopping-time distribution; when votes are dispatched in batches, the realized cost will be at least as large, with the overshoot bounded by the batch size minus one. Practitioners operating in batch mode should account for this overhead when budgeting, for example by treating the batch overshoot as an additive correction to the expected cost.

This process has a state diagram illustrated in Figure~\ref{fig:markov} and is also known as the \emph{Gambler's Ruin} model.\footnote{Not to be confused with the \emph{Gambler's Fallacy} model, which describes sequential effects.} The model is a common introductory example for random walks and describes the probability of a gambler winning or losing a certain amount of money in a game of chance. (See~\cite[page~344]{FellerTextbook} for more details.) As we discuss in Section~\ref{sec:relwork-sprt}, the $\delta$-margin rule is closely related to the Sequential Probability Ratio Test~\citep{wald1945sequential}; indeed, when~$p$ is known, $\delta$-margin voting is equivalent to an SPRT with symmetric log-likelihood boundaries. However, to our knowledge, no prior work has assembled the complete design toolkit---quality, expected cost, variance, distributional analysis, payment equivalence, and Bayesian treatment of unknown~$p$---for the $\delta$-margin stopping rule in the context of label aggregation.

In Section~\ref{sec:characteristics}, we derive the theoretical properties of this process assuming that $p$ is known, and in Section~\ref{sec:unknown_p} we drop this assumption and show how to work with unknown~$p$, explicitly allowing~$p$ to vary across items.

\medskip
\noindent\fbox{\parbox{\dimexpr\columnwidth-2\fboxsep-2\fboxrule\relax}{%
\textbf{Notation Summary.}
Throughout the paper we use the following conventions.
\smallskip

\begin{tabular}{@{}l@{\quad}p{0.78\columnwidth}@{}}
$Y_i$ & True binary label for item~$i$. \\
$\mathcal{A}_i$ & Distribution of individual worker accuracies for item~$i$. \\
$p$ (or $p_i$) & Mean pool accuracy for item~$i$; $p_i = E[\mathcal{A}_i]$. \\
$\varphi$ & Odds of a correct vote: $\varphi = p/(1-p)$. \\
$\delta$ & Consensus threshold (positive integer). \\
$S_t$ & State (net correct$-$incorrect vote count) after $t$ votes; $S_0 = 0$. \\
$\tau$ & Stopping time: $\tau = \min\{t \geq 1 : |S_t| = \delta\}$. \\
$\hat{Y}_i$ & Consensus correctness indicator: $\hat{Y}_i = \mathbf{1}[S_\tau = \delta]$ (equals~1 iff the correct class wins). \\
$Q(\varphi,\delta)$ & Probability of a correct consensus vote (Theorem~\ref{th:Q_nonrand}). \\
$n_{\mathit{votes}}$ & Number of votes cast before consensus is reached. \\
$\mathbf{M},\mathbf{Q},\mathbf{R},\mathbf{N}$ & Canonical, transient, absorbing, and fundamental matrices of the Markov chain. \\
$\alpha,\beta$ & Parameters of the $\textit{Beta}(\alpha,\beta)$ prior on~$p$ (Section~\ref{sec:unknown_p}). \\
$\textit{pay}(\varphi)$ & Payment per vote for a worker pool with odds~$\varphi$ (Section~\ref{sec:pay_equiv}). \\
\end{tabular}
}}
\medskip

Before providing the analytical results for several key characteristics of $\delta$-margin voting, we overview the related literature in the following Section~\ref{sec:litrev}.

%% file: tex_files/3-related.tex
\section{Literature review}
\label{sec:litrev}

In this section, we offer a concise summary of relevant research in label aggregation in crowdsourcing (Section~\ref{sec:relwork-quality}), followed by references to related work on $\delta$-margin voting (Section~\ref{sec:relwork-delta-margin}), and a discussion of the connection to sequential hypothesis testing (Section~\ref{sec:relwork-sprt}). To our knowledge, no prior work provides the combined theoretical toolkit---quality, cost, variance, distributional, and payment analysis---for the $\delta$-margin stopping rule, particularly in the practically important case where worker accuracy is unknown.

\subsection{Literature on quality control in label aggregation}
\label{sec:relwork-quality}

A substantial amount of literature on crowdsourcing proposes various quality controls for aggregating worker expertise. However, most of this work is experimental~\citep{kazai2011crowdsourcing, hansen2013quality, yin2014monetary}, and the proposed quality control mechanisms are often empirical, lacking theoretical guarantees ~\citep{kucherbaev2016relauncher, de2017efficiently, dai2013pomdp}. Moreover, many studies rely on accurate priors of essential process or workforce parameters~\citep{abassi2017adaptive, tao2018domain, heer2010crowdsourcing, dalvi2013aggregating, laureti2006information, jung2011improving, rutchick2020does}, which are costly to obtain~\citep{de2017crowd, bonald2016minimax}, and which furthermore can be vulnerable to adversarial attacks by a colluding group of workers~\citep{checco2020adversarial}. Prominent examples of crowdsourced datasets (\cite{russakovsky2015imagenet, zhou2019hype, krishna2017visual}) tacitly support the thesis that majority voting is the de facto method of choice used to aggregate labels and ensure quality. Much of this work also acknowledges that many iterations of experimenting with crowdsourced process design are required to perfect the data collection and choose (without guarantees) best-performing parameters for the voting procedure. The absence of a good prior for what accuracy one is to expect given a voting method may be one of the reasons that compels most practitioners to opt for simple majority voting, often in conjunction with rapid judgment techniques~\citep{krishna2016embracing}. In constructing ImageNet~\citep{russakovsky2015imagenet}, the authors acknowledged the challenge inherent in the fact that different item categories require different levels of consensus among users. They address this by dynamically determining the number of agreeing votes needed for a given category of images using an initial sample of items and then requiring the chosen confidence of agreement for the remaining items. Using $\delta-$margin voting eliminates the need for this expensive preliminary discovery step.

Another strand of literature examines non-computational approaches to data quality assurance in crowdsourcing. A prominent direction here is the study of incentives (usually in form of monetary reward) for workers \cite{mason2009financial, daniel2018quality, hossfeld2014crowdsourcing, heer2010crowdsourcing, singer2013pricing}. Here too, our work offers a novel contribution: the formulas for expected accuracy, parametrized by a chosen decision threshold and the accuracy of responses, give rise to a formula describing \textit{a ratio of payments for two worker pools}, such that the results have the same expected quality. This offers a principled way to set the incentives for various tasks in alignment with the resulting label quality, as we describe in Section~\ref{sec:equiv}. 

Some notable work in the study of compensation incentives \citep{mason2009financial} suggests that increased financial
reward increases the quantity, but not the quality, of crowdsourced work. However, other research \citep{kazai2011search} presents evidence to the contrary, showing that higher pay encourages better work, especially among qualified workers \citep{kazai2013analysis}. Findings such as in \cite{mason2009financial} do not threaten the results we present in Section~\ref{sec:equiv}, where we propose a way to relate the payment rates for two queues of workers depending on the expected accuracy of the outcome. Here we do not assume that by paying certain workers more we will incentivize them to perform better -- in fact, we assume that the performance will remain stable for the pool of workers labeling a given item. What we propose is a way to use the expected level of performance and this given payment rate as an `anchor' to set a different level of payment for another voting process on a different item (given that we might require more votes on that second item to attain the same accuracy of results).

In the broader literature on the quality and costs of crowdsourcing, a subset of work provides theoretical guarantees. For instance,~\cite{khetan2016achieving} formulates a theoretical trade-off between budget and accuracy in voting processes with adaptive task assignment, where tasks are assigned based on data collected up to the point of assignment~\citep{barowy2012automan}. The authors present an adaptive assignment scheme that achieves the fundamental limit.
Meanwhile,~\cite{berend2014consistency} establishes consistency bounds for the weighted majority voting scheme. Note that in addition to focusing on \textit{consistency} (i.e., the guarantees and the rate of error decay to zero), this work applies to a majority voting setup in which a fixed committee of workers each vote on an item. Our work, in contrast, focuses on flexible workflows where the voting process may terminate with various number of votes solicited, instead of a fixed committee size. 

Another example of theoretical work in this area is provided by~\cite{manino2018efficiency}, who explore the adaptive assignment of the next workers to an item. They formulate an accuracy gap between the uniform allocation of workers~\citep{karger2014budget}, adaptive allocation, and an assignment that maximizes information gain~\citep{simpson2015bayesian}. Their work derives tight but not exact bounds on the accuracy of this assignment policy. 

Additionally,~\cite{livshits2014saving} is a theoretical study that examines the costs of completing voting tasks. The authors use power analysis to obtain ex-ante estimates for the number of votes needed to resolve each item with a specific level of statistical significance. Also focusing on the costs of running a voting process, \cite{liu2022doubly} use a supervised learner to imitate workers, building a doubly robust estimator for the score function that captures all workers' response accuracy (i.e., an estimator that is unbiased and has minimal achievable variance). Their work is complementary to ours: while they focus on \emph{estimating} worker accuracy from observed labels, we take the (estimated or prior) accuracy as input and derive the downstream properties of the voting process. The work of \cite{venanzi2016time} is similar to ours in that they too examine a method to estimate the final labels quality and the tasks duration simultaneously using Bayesian inference. While the authors do not use $\delta$-margin voting and instead study a method called BCCT, the more important distinction is that Venazi et al. focus on estimating the reliability of individual workers, whereas we work with estimating quality and time for an item given few votes from multiple workers (even if a given worker only contributed a single vote in the process).  

While worker wages are not often addressed in the literature, a notable exception is the paper from~\cite{singer2013pricing}. The authors present mechanisms compatible with incentives, maximizing the number of tasks within a budget constraint and minimizing worker payments given a fixed number of tasks.

Two strands of related theoretical work deserve explicit comparison. \emph{Bayesian Classifier Combination} methods, exemplified by BCCT~\citep{venanzi2016time}, maintain and dynamically update posterior estimates of individual worker reliabilities, using these to derive posterior confidence in each item's label. In contrast, our framework requires no per-worker reliability model: it operates at the pool level, taking the mean accuracy~$p$ (or a prior over~$p$) as input and producing closed-form, ex-ante guarantees on quality, cost, and variance. This distinction is practically important in settings where workers are numerous, transient, or deliberately fungible---common in crowdsourcing platforms and LLM ensembles---and where per-worker modeling is infeasible or unnecessary.

\emph{Budget-optimal allocation} methods~\citep{karger2014budget, khetan2016achieving} take a complementary approach: given a global budget and a set of items, they optimize the assignment of workers to items to maximize dataset-wide accuracy. These methods excel when the practitioner controls the global allocation, but they require solving an optimization problem over the entire item set and typically provide asymptotic rather than exact guarantees. The $\delta$-margin framework instead characterizes each item independently, providing exact closed-form expressions for quality and cost as functions of~$(p, \delta)$. The two approaches are not mutually exclusive: a budget-optimal allocator could use $\delta$-margin voting as the item-level stopping rule, combining global allocation efficiency with per-item quality guarantees.

In summary, our work focuses on the $\delta$-margin stopping rule and provides exact expressions for the probability of error, expected time until consensus, and its variance. This allows for a more detailed consideration of the requester's utility, including the workers' wages.

\subsection{Literature on \texorpdfstring{$\delta$}{delta}-margin voting}
\label{sec:relwork-delta-margin}

The literature on group decision-making and voting mechanisms~\citep{laruelle2011majorities} distinguishes between two types of majority consensus vote: \emph{simple} majority and \emph{absolute} majority. A simple majority is achieved when the number of votes for an option $A$ exceeds the number of votes for an alternative option $B$ ($s_A>s_B$), while an absolute majority requires that the number of votes for an option $A$ is greater than half of all votes ($s_A>\frac{n}{2}$).\footnote{Alternatively, the strength of the majority condition can be defined as a multiplier $k$ on the inequality requirements, which are modified to $s_A>k \cdot s_B$ for $k>1$ and $s_A>k \cdot n$ for $\frac{1}{2} \leq k \leq 1$.} In the case of binary (or \emph{dichotomic}) voting, the distinction between simple and absolute majorities only exists when some workers abstain or cast neutral votes.

In~\cite{dietrich2007judgment}, majority voting is extended to a broader category of rules known as ``quota rules'' (also known as ``$k$-unanimity'' or ``$k$-majority'' in cybernetics and discrete mathematics~\citep{alon2006dominating, scheidler2015k}). According to quota rules, an item is assigned a particular label if the number of workers who vote for that label exceeds a threshold value of $k$. In situations where all workers vote simultaneously within a fixed workforce, the quota rule is similar to $\delta$-margin voting, with two notable differences: (a)~the margin may not be satisfied at the end of the voting process, and (b)~the process may accumulate more votes than needed, which can be a disadvantage when each vote incurs a cost.

The connection between $\delta$-margin voting and supermajority rules can be made precise. In a binary vote with $n$ ballots cast, the quota rule requiring $s_A > k$ for some threshold $k$ is equivalent to requiring that the margin $s_A - s_B > 2k - n$; setting $\delta = 2k - n + 1$ gives the same acceptance region as $\delta$-margin voting, but without early stopping.\footnote{Conversely, given a $\delta$-margin rule and a fixed committee of $n$ voters, the equivalent quota is $k = \lceil(n + \delta)/2\rceil$.} This places $\delta$-margin voting within the broader family of supermajority mechanisms studied in social choice theory~\citep{brandt2016handbook}. In particular, \citet{fey2003condorcet} extends the Condorcet Jury Theorem to supermajority rules, showing that large-jury correctness holds when average voter competence exceeds the supermajority threshold. Our Theorem~\ref{th:Q_nonrand} provides the finite-sample counterpart for $\delta$-margin voting: $Q(\varphi,\delta) = \varphi^\delta/(1+\varphi^\delta)$, which gives the exact quality for any~$\delta$ and any accuracy level~$p > 0.5$. The key additional feature of the $\delta$-margin rule relative to fixed-committee supermajority is \emph{adaptive stopping}: the process terminates as soon as the margin is reached, avoiding unnecessary votes.

The $\delta$-margin voting aggregation method is common in industry practice but remains under-theorized: it is typically applied in empirical or experimental settings, without rigorous performance guarantees. This indicates a missed opportunity for principled workflow design based on theoretical estimation of the scheme's benefits. Notably, while the rule is common among practitioners, it has received little attention in the theoretical literature---outside of crowdsourcing, $\delta$-margin majority voting is called ``the forgotten decision rule''~\citep{llamazares2006forgotten, garcia2001majority}, and the earliest acknowledgments of the $\delta$-margin voting scheme can be found in~\citep{fishburn2015theory} and~\citep{saari1990consistency}, though these are brief. According to~\cite{de2017crowd}, in experimental comparisons of typical crowdsourcing tasks, the $\delta$-margin method, called ``Beat-By-$K$'' by~\cite{goschin2014stochastic}, provides very accurate results for relatively high values of $\delta$, but it is expensive to run in settings that prioritize utility without any budget constraints. Other consensus aggregation methods in the literature capture similar ideas to $\delta$-margin voting but require stronger, more confident agreement among workers. For example, in the "Automan" scheme~\citep{barowy2012automan}, the requester keeps sampling votes until the voting process reaches a given statistical confidence value.

\subsection{Connection to sequential hypothesis testing}
\label{sec:relwork-sprt}

The $\delta$-margin voting process has a natural interpretation in the framework of sequential hypothesis testing. Given a binary item with unknown true label~$Y_i$, deciding~$Y_i$ from sequential votes is equivalent to testing two simple hypotheses about the per-vote Bernoulli parameter: $H_c$: $\Pr(\text{correct vote}) = p$ (the true label is the one that the majority of votes will tend toward) versus $H_i$: $\Pr(\text{correct vote}) = 1-p$ (the true label is the minority class). Under~$H_c$, the majority is likely correct; under~$H_i$, it is likely incorrect.

\paragraph{Mapping to SPRT.}
The classical Sequential Probability Ratio Test (SPRT) of~\citet{wald1945sequential} accumulates the log-likelihood ratio $\Lambda_t = \sum_{k=1}^{t} \ell_k$ after $t$ observations, where each increment is $\ell_k = +\!\log\varphi$ for a correct vote and $\ell_k = -\!\log\varphi$ for an incorrect vote, with $\varphi = p/(1-p)$. The SPRT terminates the first time $\Lambda_t$ crosses an upper boundary~$\log A$ (accept~$H_c$) or a lower boundary~$-\!\log B$ (accept~$H_i$). Since $\Lambda_t = S_t \cdot \log\varphi$, where $S_t$ is the net vote count from Definition~1, the $\delta$-margin stopping rule $|S_t| = \delta$ is equivalent to an SPRT with \emph{symmetric} boundaries $A = B = \varphi^\delta$.

Moreover, because the random walk~$S_t$ moves in unit steps $\pm 1$, the boundaries are hit exactly---there is no overshoot. This yields the exact error probability $1/(1+\varphi^\delta) = 1 - Q(\varphi,\delta)$ from Theorem~\ref{th:Q_nonrand}, slightly tighter than Wald's classical approximation of~$\varphi^{-\delta}$~\citep{wald1945sequential}. The optimality of SPRT among all sequential tests with the same error constraints was established by~\citet{waldwolfowitz1948}.

\paragraph{Key differences.}
Despite this equivalence when~$p$ is known, the two frameworks differ in important ways:
\begin{itemize}
\item \textbf{Design direction.} SPRT fixes the desired error rates~$\alpha$ and~$\beta$ and derives boundaries that depend on~$p$. In contrast, $\delta$-margin voting fixes the integer threshold~$\delta$---a single, interpretable design parameter---and accepts the resulting error rate as a function of~$p$. This parameterization is more natural for a practitioner who controls the process design but may not know~$p$ precisely.

\item \textbf{Unknown~$p$.} The classical SPRT requires specifying the parameter values under both hypotheses. When the worker accuracy~$p$ is unknown, the SPRT boundaries cannot be computed. The framework developed in Section~\ref{sec:unknown_p} addresses this by placing a prior on~$p$ and computing posterior expected quality and cost, yielding practical decision tables without requiring~$p$ in advance.

\item \textbf{Distributional and cost analysis.} Beyond the expected sample size---the primary SPRT performance metric---our analysis provides closed-form expressions for the variance (Theorem~\ref{th:var_nonrand}) and the full probability mass function (Theorem~4) of time to consensus, as well as a payment-equivalence framework across worker pools of different quality (Section~\ref{sec:equiv}). These quantities are essential for operational planning and budgeting in crowdsourcing and human-in-the-loop workflows, and, while analogous results exist in the broader sequential analysis literature, are not part of the standard SPRT design toolkit as typically applied in practice.
\end{itemize}

\paragraph{Relation to modern sequential inference.}
More recently, the framework of \emph{anytime-valid inference}~\citep{ramdas2023gametheoretic} generalizes sequential testing by constructing test supermartingales (e-processes) that allow optional stopping at any data-dependent time while maintaining Type-I error control. This provides a flexible alternative to SPRT's fixed-boundary design. Our work is complementary: rather than developing a new sequential test, we provide a complete operational toolkit for a specific, widely deployed stopping rule that is already used in practice. The contribution lies in the \emph{design calculus}---translating the theoretical properties of this rule into actionable guidance for choosing~$\delta$, estimating costs, and setting payments across heterogeneous worker pools, including under uncertainty about~$p$.

%% file: tex_files/4-model-properties.tex
\section{Theoretical characteristics of \texorpdfstring{$\delta$}{delta}-margin voting processes}
\label{sec:characteristics}

In this section, we summarize \emph{existing} results from the Gambler's Ruin model to characterize the following aspects of the $\delta$-margin voting processes:
\begin{itemize}
    \item The quality of the consensus label obtained from the $\delta$-margin voting process.
    \item The expected number of votes necessary to reach a consensus, i.e., the expected time until consensus.
    \item The variance of the votes needed to reach a consensus.
    \item The probability mass function of the votes required to reach a consensus.
\end{itemize}

\noindent We begin by assuming that the mean $p$ of the distribution of worker accuracies for a fixed item is known. Readers already familiar with the Gambler's Ruin model may prefer to skip and move to Section~\ref{sec:unknown_p}, where we relax this assumption by allowing $p$ to be a random variable, starting with some prior beliefs about the quality of the worker pool, which are updated during the voting process. 

As discussed in Section~\ref{sec:relwork-sprt}, the individual results in this section are well known from the Gambler's Ruin model and, equivalently, from Wald's Sequential Probability Ratio Test~\citep{wald1945sequential}. Our contribution is not the formulas themselves but the \emph{design calculus} that connects them: taken together, they provide a practitioner with exact expressions for quality, cost, variance, and timing as functions of two interpretable parameters ($p$ and $\delta$), enabling principled workflow design. In Sections~\ref{sec:unknown_p} and~\ref{sec:equiv}, we extend this toolkit to the practically important cases where~$p$ is unknown and where multiple worker pools must be compared.

We state the results using notation from the theory of Markov chains with absorbing states~\citep{grinstead1997ch}. Since this section is background in nature, we do not provide proofs; the interested reader can find them in introductory textbooks, for example, in~\citep[page~344]{FellerTextbook}.

\subsection{Background on Markov Chains with absorbing states}
\label{sec:markovchains}

We demonstrated that a $\delta$-margin voting process can be modeled as a Markov chain with two absorbing states and $2(\delta-1)+1$ transient states. This section presents the necessary background~\citep{grinstead1997ch} for working with such Markov chains to obtain relevant results. To begin, we define the \emph{canonical form $\mathbf{M}$} of the transition matrix for this process:
\begin{displaymath}
\mathbf{M} = \left(\begin{smallmatrix}
\mathbf{Q} & \mathbf{R} \\
\textbf{0} & I_2 
\end{smallmatrix}\right)
\end{displaymath}

\noindent The $\mathbf{Q}$ is a matrix of size $(2\delta-1)\times(2\delta-1)$ that contains the probabilities of transitioning from a transient state $i$ to a transient state $j$, and $\mathbf{R}$ is a $(2\delta-1)\times2$ matrix of transition probabilities from transient into absorbing states. Formally:
\begin{displaymath}
\mathbf{Q} = \left(
\begin{smallmatrix}
0 & p & 0 & 0 & \cdots & 0\\
1-p & 0 & p & 0 & \cdots & 0\\
0 & 1-p & 0 & p & \cdots & 0\\
0 & 0 & 1-p & 0 & \cdots & 0\\
\vphantom{\int\limits^x}\smash{\vdots} & \vdots & \vdots &\vdots & \ddots & \vdots\\
0 & 0 & 0 & 0 & 1-p & 0
\end{smallmatrix}\right)
\end{displaymath}
\begin{displaymath}
\mathbf{R} = \left(
\begin{smallmatrix}
1-p & 0 & 0 & \cdots & 0 & 0\\
0 & 0 & 0 & \cdots & 0 & p
\end{smallmatrix}\right)^\top
\end{displaymath}

\noindent Finally, $\mathbf{I}_2$ is a $2\times 2$ identity matrix, and \textbf{0} is a $2\times(2\delta-1)$ null matrix. The definition of the \emph{fundamental matrix $\mathbf{N}$} of this absorbing matrix chain is:
\begin{displaymath}
\mathbf{N}:=\sum_{k=0}^{\infty}\left(\mathbf{Q}\right)^k ,
\end{displaymath}

\noindent which can be written~\citep{grinstead1997ch} as: 
\begin{displaymath}
\mathbf{N}=\left(\mathbf{I}_{2\delta-1}-\mathbf{Q} \right)^{-1}
\end{displaymath}

\noindent The elements $\mathbf{N}_{ij}$ correspond to the expected number of visits to the state $j$, given that the initial state was $i$. We can now define the essential properties of the voting process by utilizing the fundamental matrix.

\subsection{Quality of the consensus vote}
\label{sec:quality}

The first property of interest is the \textit{quality of the final consensus vote}. To calculate this value, we want to estimate the probabilities of reaching the absorbing states $\delta$ and $-\delta$. Following the notation introduced in the previous section, we define the matrix $\mathbf{B}$:
\begin{displaymath}
\mathbf{B}:= \mathbf{N} \cdot \mathbf{R}
\end{displaymath}

\noindent The matrix $\mathbf{B}$ is a row-stochastic matrix, with $2\delta-1$ rows and two columns. The matrix's $(i,j)$-entry contains the probability of eventually reaching absorbing state $j$ when starting from transient state $i$. The process typically starts at state $0$ (no votes), so we are interested in the contents of the row corresponding to the state 0.\footnote{In particular cases, we may want to start at a different state. For example, we may require $\delta=3$ to flag an account for money laundering while dismissing such an alert with $\delta=1$. In such a case, we may start the process from state -1, setting $\delta=2$. Deriving the results for this case is reasonably simple, but for brevity, we do not provide the details in this paper.} Through standard algebraic manipulations, we get that the probabilities of reaching states $\delta$ and $-\delta$, when starting from state 0 are, respectively:  
\begin{align*}
\mathbf{B}_{0,\delta} & =  \frac{\varphi^{\delta}}{1+\varphi^{\delta}} \\
\mathbf{B}_{0,-\delta} & =  \frac{1}{1+\varphi^{\delta}}
\end{align*}

\noindent where $\varphi=\frac{p}{1-p}$ are the \emph{odds} of the average worker's vote on the item being correct.\footnote{The proof is also readily available in many introductory texts on Markov Chains, e.g., in~\cite[page~344]{FellerTextbook}, without using the approach that relies on the canonical transition matrix $\mathbf{M}$ and the fundamental matrix $\mathbf{N}$.} Using these quantities, we can easily estimate the expected labeling quality of an item:

\begin{theo}

\label{th:Q_nonrand}
For voting on a single item using a $\delta$-margin voting process with mean accuracy $p$ of the worker responses and a consensus threshold $\delta$, the probability $Q(\varphi,\delta)$ of obtaining a correct consensus vote is: 

\begin{equation}
Q(\varphi,\delta) := \frac{\varphi^{\delta}}{1+\varphi^{\delta}}
\label{eq:Q_nonrand}
\end{equation}

\noindent where $\varphi=\frac{p}{1-p}$ are the \emph{odds} of an average worker giving a correct vote on the item. 
\end{theo}

\begin{corollary}[Weak jury theorem for $\delta$-margin voting]
\label{cor:weak_jury}
The consensus threshold $\delta$ is a direct design knob for quality:
\begin{itemize}
    \item If $p > \frac{1}{2}$ (i.e., $\varphi > 1$), then $Q(\varphi,\delta) \to 1$ as $\delta \to \infty$, with the error probability $1 - Q = (1+\varphi^{\delta})^{-1}$ vanishing at exponential rate $\varphi^{-\delta}$.
    \item If $p < \frac{1}{2}$ (i.e., $\varphi < 1$), then $Q(\varphi,\delta) \to 0$ as $\delta \to \infty$.
    \item If $p = \frac{1}{2}$ (i.e., $\varphi = 1$), then $Q = \frac{1}{2}$ for every $\delta$.
\end{itemize}
\noindent Thus, for any pool of workers who are better than random, any desired quality level can be guaranteed simply by choosing $\delta$ large enough.
\end{corollary}

\paragraph{Items with $p < 0.5$.}
In real datasets, some items may be ``confusing'' enough that the average worker is more likely to be wrong than right ($p < 0.5$, i.e., $\varphi < 1$). For such items, $Q(\varphi, \delta) < 0.5$: the consensus label is more likely incorrect, and increasing~$\delta$ makes matters worse ($Q \to 0$). This is itself informative for the practitioner: a low estimated quality signals an item where the consensus should not be trusted and may warrant \emph{label inversion} (accepting the minority vote), abstention, or escalation to domain experts. When using the Bayesian framework of Section~\ref{sec:unknown_p} with a prior that allows mass on $[0, 0.5]$, the design calculus should account for the possibility that some items fall in this regime; the per-hypothesis posterior (Section~\ref{sec:pure-prior}) naturally distinguishes these cases.

\begin{figure}[t]
\centering
\includegraphics[width=0.8\columnwidth]{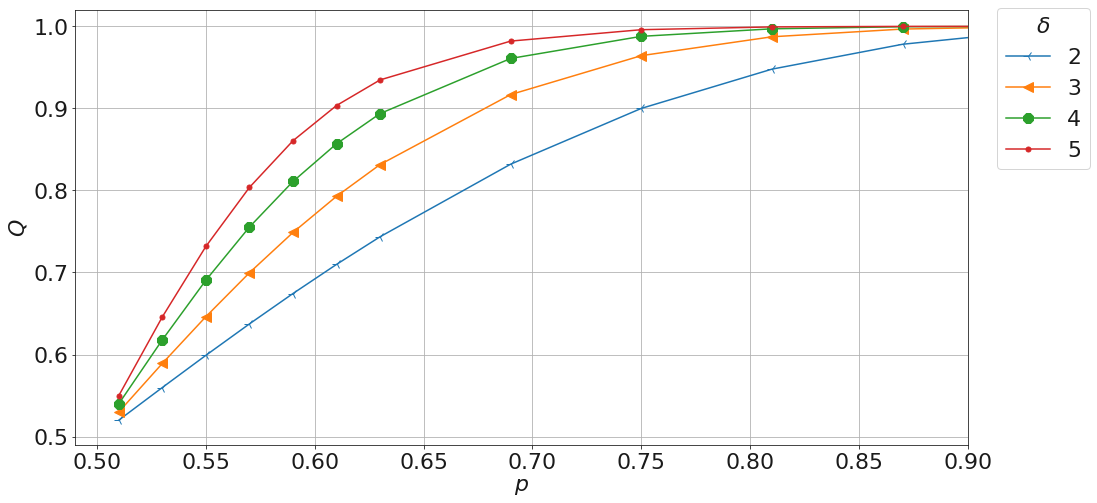}
\Description{Line plot of consensus quality Q as a function of worker accuracy p for delta values 2 through 5, showing S-shaped curves that steepen with increasing delta.}
\caption{Theoretical values of quality $Q$ of consensus vote (Theorem~\ref{th:Q_nonrand}) as a function of the probability of a correct answer $p$, for a fixed consensus threshold $\delta$.}
\label{fig:Q_of_p}
\end{figure}

\paragraph{Discussion.}
Figure~\ref{fig:Q_of_p} plots the dependence of consensus vote quality $Q$ on item difficulty $p$ for $\delta \in \{2, 3, 4, 5\}$.
Note that the parameter $\delta$ plays a significant role in the quality of the consensus vote: based on Equation~\ref{eq:Q_nonrand}, the \emph{odds} of the \emph{consensus} vote being correct are~$\varphi^{\delta}$. Therefore by increasing $\delta$, we exponentially increase the odds that the consensus vote is correct. 
For example, consider a pool of workers with expected response accuracy $p=0.75$ (i.e., $\varphi=\frac{0.75}{1-0.75}=3$) for a given item.
If we set $\delta=2$, the expected quality of the overall classification will be $Q(3,2)=0.9$ (i.e., odds 9 to 1 being correct). If we increase $\delta$ to $\delta=3$, then $Q(3,3)=0.964$ (i.e., odds 27 to 1), and if we increase to $\delta=4$, then $Q(3,4)=0.9878$ (i.e., odds 81 to 1). 

Next, we discuss how the cost of the process changes when we change $\delta$, and we show that we achieve exponential improvements in quality with a mostly linear increase in cost.

\subsection{Time until consensus} 

The next set of properties we want to explore relate to the cost of running the process -- namely, the number of votes one would have to procure in order to reach a consensus. We start by examining the expected number of votes until the process terminates. While the expected number of votes to completion is useful for characterizing the voting process, it describes just the average across runs. In addition to the expectation, we also want to know the robustness of a process and how reliably it will finish within the expected time frame. For this reason, it is important to know the \emph{variance} of the number of votes required to complete the process and, more generally, the distribution's overall \emph{pdf}.

\subsubsection{Expected Time until consensus}
\label{sec:votes}

What is the \emph{expected number of steps} until consensus is reached? Following our usual notation, we define the vector $\mathbf{t}$, which contains in position $i$ the expected number of steps until consensus when starting in transient state $i$. 

\begin{displaymath}
\mathbf{t} = \mathbf{N} \cdot \mathbf{1}
\end{displaymath}

\noindent where $\mathbf{1}$ is a vector of ones of length $2\delta-1$. Since the process starts from state $S_0 = 0$ (Definition~1), the closed-form expression below corresponds to the component of~$\mathbf{t}$ for the initial state~$0$. Through algebraic simplification, we get the following:

\begin{theo}
\label{th:ET}
The expected number of votes $n_{votes}$ it takes to reach a consensus (correct or incorrect) when classifying an item using a $\delta$ margin voting scheme with item-level expected worker accuracy $p\ne\frac{1}{2}$ and consensus threshold $\delta$ is:

\begin{equation}
\label{eq:exp_votes}
\mathbb{E}[n_{votes}|\varphi,\delta]=\delta\cdot\frac{\varphi+1}{\varphi-1} \cdot \frac{\varphi^{\delta}-1}{\varphi^{\delta}+1}
\end{equation}

\noindent where $\varphi=\frac{p}{1-p}$ are the \emph{odds} of an average worker giving a correct vote on the item. 
When $p=0.5$, we have $\mathbb{E}[n_{votes}|\varphi,\delta]=\delta^2$.
\end{theo}

\begin{figure}[t]
\centering
\includegraphics[width=0.8\columnwidth]{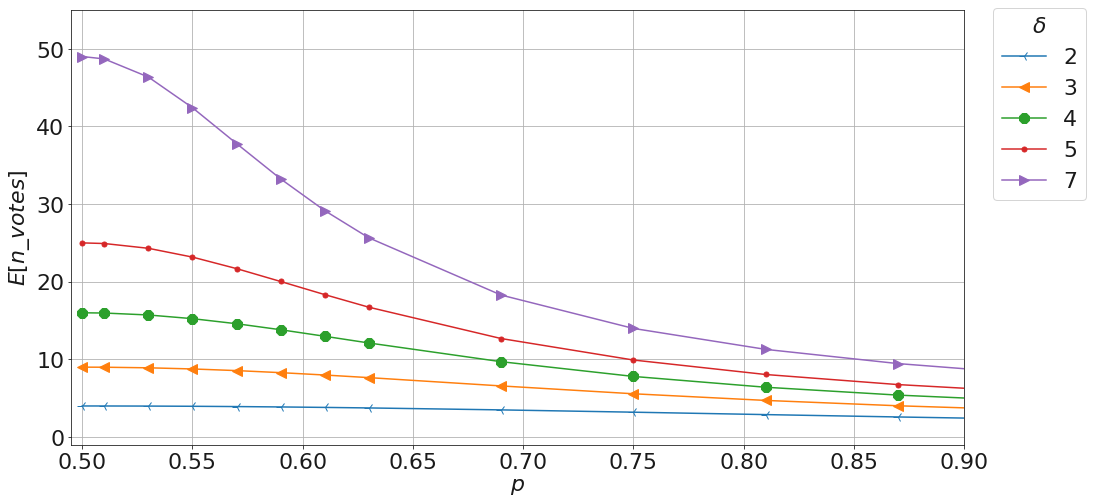}
\Description{Line plot of expected number of votes to consensus as a function of worker accuracy p for delta values 2 through 5, showing peaks near p equals 0.5.}
\caption{Expected time to reach consensus as a function of the probability of correct answer $p$, for a fixed consensus threshold $\delta$.}
\label{fig:ET(d)}
\end{figure}

\paragraph{Discussion.} Note that the expected time to termination increases \emph{mostly} linearly with $\delta$. When $\varphi$ gets close to 1 (that is, mostly random worker responses), the expected time to consensus peaks at $\delta^2$. The fact that the maximum expected time to completion is non-infinite also shows that the process is guaranteed to finish; Section~\ref{sec:pdf} shows how to derive the full \emph{pmf} of the distribution of times to consensus. 
Figure~\ref{fig:ET(d)} illustrates the time until consensus $\mathbb{E}(n_{votes})$ as a function of the mean of worker accuracies $p$, for a given consensus threshold $\delta$.

\subsubsection{Variance of time until consensus}
\label{sec:var}

The variance in the number of steps required to reach consensus when starting in a state $i$ is the $i$-th entry of vector $\bar{v}$, defined as:

\begin{displaymath}
\bar{v} = (2\mathbf{N}-\mathbf{I}_{t})\mathbf {t} -\mathbf {t} _{\operatorname {sq} }
\end{displaymath}

\noindent where $\mathbf {t}_{\operatorname{sq} }$ is the vector of squared elements of $\mathbf {t}$. Applying this  result to our model for $\delta$-margin voting, and starting with the state with no votes, we have the following expression for the variance of time until consensus:

\begin{theo}
For $p\ne\frac{1}{2}$, the variance of the number of votes it takes to reach consensus using the $\delta$-margin voting process is:

\begin{equation}
\label{fla:var_nonrand}
Var[n_{votes}|\varphi,\delta] = 
4\delta\varphi\left(\frac{\varphi+1}{\varphi^\delta+1}\right)^2
\cdot\left[ \left \lfloor{\frac{\delta^2}{4}}\right \rfloor \cdot\varphi^{\delta-2}
+ \sum_{i=1}^{\delta-2} \Big( h(\delta-i)(\varphi^{\delta+i-2}+\varphi^{\delta-i-2}) \Big)\right] \nonumber 
\end{equation}
\label{th:var_nonrand}

\noindent When $p=\frac{1}{2}$ (i.e., $\varphi = 1$), taking the limit gives:
\begin{displaymath}
Var[n_{votes}|\varphi{=}1,\delta] \;=\; \frac{2\,\delta^2(\delta^2-1)}{3}
\end{displaymath}

\end{theo}

The sequence of coefficients defined by $h(z):=\left \lfloor{\frac{z^2}{4}}\right \rfloor$ is also known as the \emph{quarter squares} sequence and can be defined as the interleaving of square numbers and pronic numbers (the latter defined as the product of two consecutive integers) \citep{losanitsch1897isomerie, sloane2019line}.

Equation~\ref{fla:var_nonrand} simplifies the formula proposed by \cite{andvel2012variance} while agreeing with it numerically. Moreover, the polynomial in square brackets exhibits a symmetric ``pyramidal'' structure governed by the quarter-squares sequence: reading the coefficients from the highest to the lowest power of~$\varphi$, they follow the pattern $h(2), h(3), \ldots, h(\delta-1), h(\delta), h(\delta-1), \ldots, h(3), h(2)$. That is, the coefficients ascend through the quarter-squares values from $h(2)=1$ up to a peak of $h(\delta)$ at the center, then descend symmetrically. For example, for $\delta=7$ the coefficient sequence is $1, 2, 4, 6, 9, 12, 9, 6, 4, 2, 1$, corresponding to $h(2)$ through $h(7)$ and back down. This palindromic structure arises because the variance formula sums contributions from symmetric pairs of transient states equidistant from the origin. To provide the reader with further intuition, we list explicit formulas for $Var(n_{votes})$ for the first few values of $\delta$ in Table~\ref{tab:var_examples}.

\begin{table}[t]
  \begin{center}
  \bgroup
    \def\arraystretch{2.2}
    \begin{tabular}{c|c} 
      $\bm{\delta}$ & $\mathbf{Var(n_{votes})}$ \\\hline\hline
      2 & $8\varphi\left(\frac{\varphi+1}{\varphi^2+1}\right)^2                 $  \\
      \hline
      3 & $12\varphi\left(\frac{\varphi+1}{\varphi^3+1}\right)^2 \left( \varphi^2 + 2\varphi + 1 \right) $ \\\hline
      4 & $16\varphi\left(\frac{\varphi+1}{\varphi^4+1}\right)^2 \left( \varphi^4 + 2\varphi^3 + 4\varphi^2 + 2\varphi+ 1 \right) $ \\\hline
      5 & $20\varphi\left(\frac{\varphi+1}{\varphi^5+1}\right)^2 \left( \varphi^6 + 2\varphi^5 + 4\varphi^4 + 6\varphi^3 + 4\varphi^2 + 2\varphi+ 1 \right) $ \\\hline
      6 & $24\varphi\left(\frac{\varphi+1}{\varphi^6+1}\right)^2 \left( \varphi^8 + 2\varphi^7 + 4\varphi^6 + 6\varphi^5 + 9\varphi^4 + 6\varphi^3 + 4\varphi^2 + 2\varphi+ 1 \right) $ 
      \\\hline
      7 & $28\varphi\left(\frac{\varphi+1}{\varphi^7+1}\right)^2 \left( \varphi^{10} + 2\varphi^9 + 4\varphi^8 + 6\varphi^7 + 9\varphi^6 + 12\varphi^5 + 9\varphi^4 + 6\varphi^3 + 4\varphi^2 + 2\varphi+ 1 \right) $ 
      \\\hline
    \end{tabular}
    \caption{Particular values of variance on the number of votes until consensus}
    \label{tab:var_examples}
    \egroup
  \end{center}
\end{table}

\begin{figure}[t]
\centering
\includegraphics[height=0.33\columnwidth]{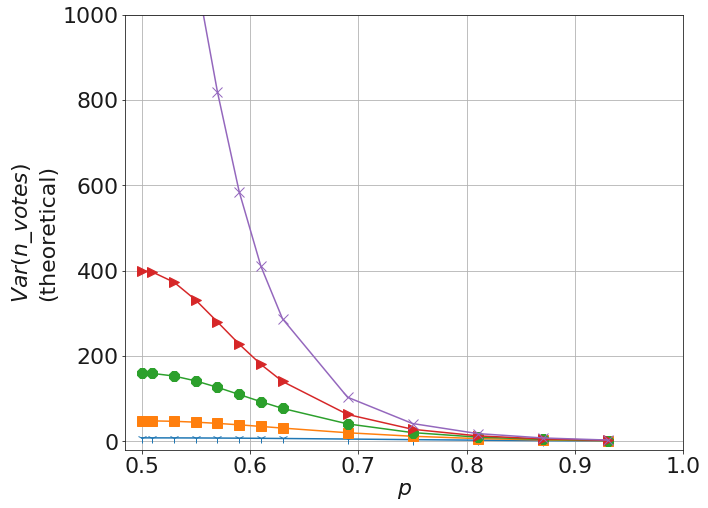}
\includegraphics[height=0.33\columnwidth]{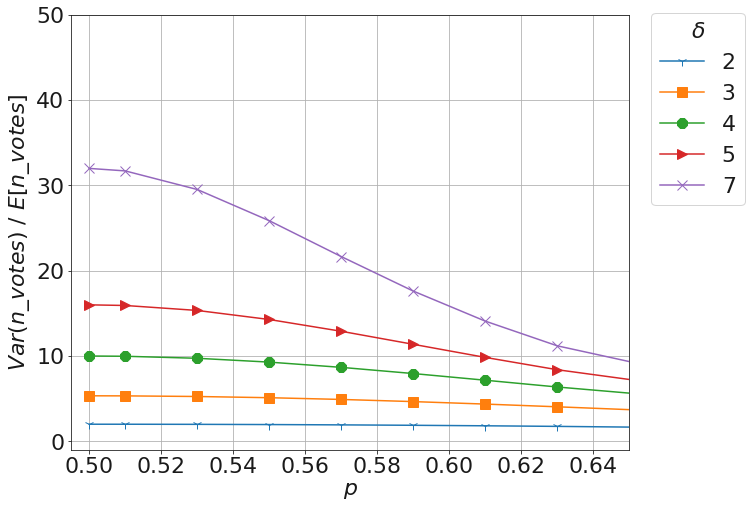}
\Description{Two panels: left shows variance of votes to consensus versus worker accuracy p for several delta values; right shows the ratio of variance to expected votes.}
\caption{$Var(n_{votes})$ (left panel) and $Var(n_{votes})/\mathbb{E}(n_{votes})$ (right panel) as a function of probability of correct answer $p$, for a fixed consensus threshold $\delta$. (Note the differing scales.)}
\label{fig:Var_by_d}
\end{figure}

\begin{figure}[b]
\centering
\includegraphics[width=0.32\columnwidth]{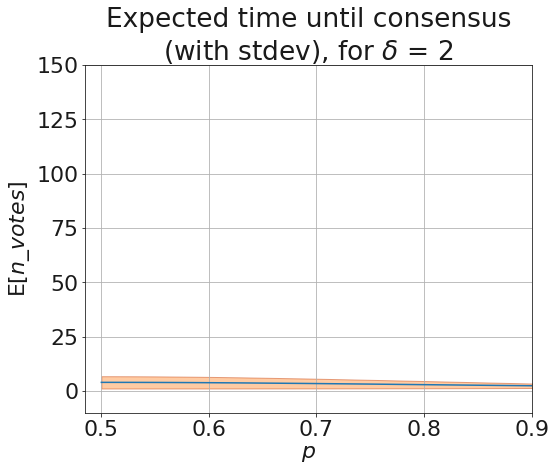}
\includegraphics[width=0.32\columnwidth]{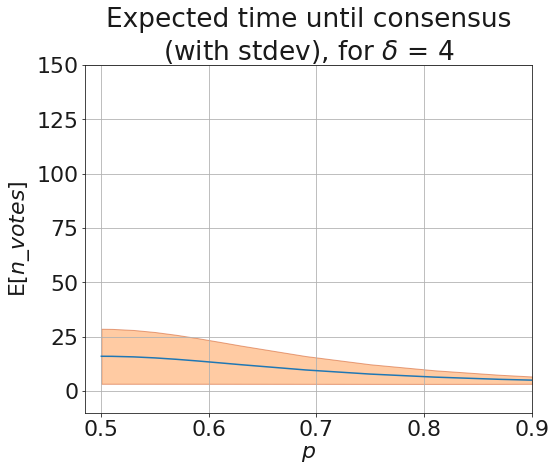}
\includegraphics[width=0.32\columnwidth]{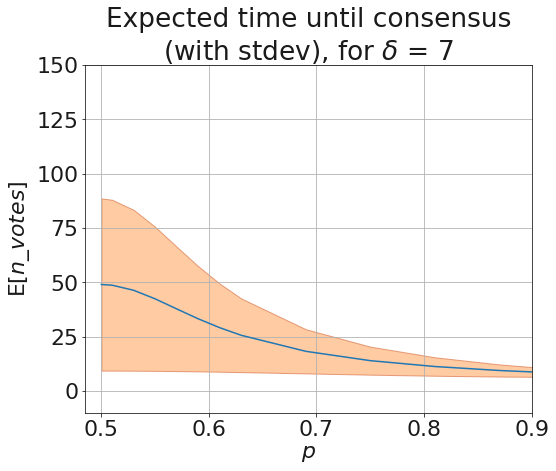}
\Description{Three panels showing expected time to consensus with standard deviation bands for delta equals 2, 4, and 7 as a function of worker accuracy p.}
\caption{Expected time until reaching consensus, with standard deviation bounds, for selected values of $\delta$.}
\label{fig:ET_with_std}
\end{figure}

The formulation in Equation~\ref{fla:var_nonrand} allows us to elaborate upon the plots in Figure~\ref{fig:ET(d)} by adding bands of the size $2\sqrt{Var(n_{votes})}$ to each trajectory of $\mathbb{E}[n_{votes}]$. Detailed plots for selected values of $\delta$ are shown in Figure~\ref{fig:ET_with_std}.

\subsubsection{Distribution of time to consensus:}
\label{sec:pdf}

Using the notation used in the previous subsections and known results about discrete phase-type distributions from queuing theory~\citep{neuts1994matrix, latouche1999introduction}, we can describe the probability mass function of time until consensus.\footnote{Since the stopping time $m$ is a discrete (integer-valued) random variable, the distribution is properly called a probability mass function (pmf). We use the notation $\textit{pmf}(m)$ throughout.} The absorption probabilities at time~$m$ are captured by the following vector:
\begin{displaymath}
T_m = \mathbf{z} \cdot \mathbf{Q}^{m-1} \cdot \mathbf{R}
\end{displaymath}

\noindent where $\mathbf{z}$ is a vector of length $2\delta-1$ that encodes the initial state of the process. The expression $T_m$ is a $1\times2$ vector, specifying the probabilities of the voting process terminating in exactly $m$ steps in each of the two absorbing states---i.e., reaching a consensus with a correct label and an incorrect one. To obtain the total probability of termination after exactly $m$ steps, regardless of which absorbing state is reached, we sum the two entries of $T_m$:

\begin{theo}
\label{th:pdf}
The probability $\textit{pmf}(m)$ that a $\delta$-margin majority voting process will terminate after exactly $m$ votes when starting from zero votes is:
\begin{equation}
\textit{pmf}(m) = \boldsymbol{z} \cdot \boldsymbol{Q}^{m-1} \cdot \boldsymbol{R} \cdot \boldsymbol{1}
\label{eq:pdf_m}
\end{equation}

\noindent where $\boldsymbol{z}$ is a vector of length $2\delta-1$ with
$\boldsymbol{z} := \langle z_{-\delta+1}, \ldots , z_{\delta-1} \rangle$; $z_{0} = 1$ and $z_i = 0$ for $i\ne 0$; and $\boldsymbol{1}$ is a column vector of ones of length $2$. Since the random walk starts at the origin and moves $\pm 1$ at each step, absorption requires at least $\delta$ steps and the parity of the step count must match that of~$\delta$; hence $\textit{pmf}(m)=0$ for $m < \delta$ and for $m \not\equiv \delta \pmod{2}$, i.e., the support is $m \in \{\delta,\, \delta{+}2,\, \delta{+}4,\, \ldots\}$.

\end{theo}

\begin{figure}[t]
\centering
\includegraphics[width=0.7\columnwidth]{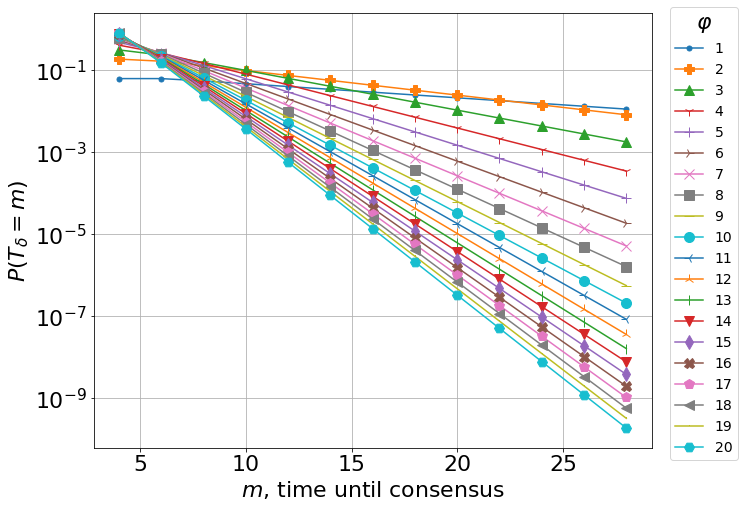}
\Description{Probability mass function of reaching consensus in exactly m steps for delta equals 4, plotted for varying worker accuracy odds phi.}
\caption{Probability mass function of reaching consensus in exactly $m$ steps (i.e., any consensus, correct or incorrect) $\textit{pmf}(m)$ for $\delta=4$, as a function of the number of votes $m$, for varying item-level odds of worker expected accuracy $\varphi$.}
\label{fig:pdf_Tk}
\end{figure}

Figure~\ref{fig:pdf_Tk} visualizes the results of this computation for $\delta = 4$ and for various values of $\varphi$.

%% file: tex_files/5-unknown_p.tex
\section{Modeling uncertainty in \texorpdfstring{$p$}{p}}
\label{sec:unknown_p}

In the prior section, we presented results that assumed that the mean labeling quality $p$ is exogenously known. This section removes this assumption by modeling the mean worker pool accuracy $p$ as a random variable.
In our case, the quantity $p$ corresponds to the average accuracy when labeling an item: we assume that the average accuracy is constant when labeling a single item, but we do not know its exact value. Instead, we have some prior expectations about $p$, and based on the votes that come in, we update our beliefs. Note that we only need the assumption of ``constant $p$'' within the context of a single item. We explicitly allow $p$ to be different across different items.

We split our discussion into two parts. First, Section~\ref{sec:pure-prior} presents the results when the prior belief of $p$ is Beta distribution. Then, in Section~\ref{sec:mixture-priors}, we show how we can incorporate priors of arbitrary complexity using a \emph{mixture of Betas} approach for the priors.

\subsection{Working with a simple prior for \texorpdfstring{$p$}{p}}
\label{sec:pure-prior}

Without loss of generality, we can assume that we have a time-homogeneous Markov process, which means that the mean of the workers' accuracy distribution remains stable while labeling the item. (In other words, the next worker labeling a given item is as good, on average, as the previous one.) 

The discussion in the previous section relied on the assumption that we know the value of $p$.
Now, we are making a more realistic assumption that $p$ is a quantity we estimate during voting.

\paragraph{Why the Beta distribution.}
Since $p \in (0,1)$ is the success probability of a Bernoulli trial, the natural conjugate prior is the \emph{Beta} distribution: if $p \sim \textit{Beta}(\alpha,\beta)$, then the posterior after observing votes is also Beta, enabling tractable closed-form updates throughout the voting process. The two parameters $\alpha$ and $\beta$ can be interpreted as ``pseudo-counts'' of prior correct and incorrect votes, respectively, encoding the practitioner's belief about the workforce before any votes are observed.

\paragraph{Why symmetric priors ($\alpha = \beta$).}
During voting, we observe labels but do not know which class is the ground-truth correct one. In our generative model, $p = P(\text{vote matches true label})$ is the accuracy relative to the (unknown) ground truth. If we were to flip which class is treated as the true label, accuracy $p$ would become $1-p$. Since we do not know a priori which class is correct, a symmetric prior $\textit{Beta}(\alpha, \alpha)$---which is invariant under $p \mapsto 1-p$---is the appropriate default. (Section~\ref{sec:mixture-priors} shows how mixture priors can relax this constraint when asymmetric beliefs are warranted.)

\paragraph{Choosing $\alpha$.} A few practical guidelines:
\begin{itemize}
    \item $\alpha = 1$ (uniform): maximum ignorance---all values of $p$ are equally likely \emph{a priori}. A reasonable default when no historical data is available.
    \item $\alpha > 1$ (e.g., $\alpha=2$ or $\alpha=5$): concentrates mass around $p=0.5$, encoding a belief that the task is difficult and workers perform near chance. Larger $\alpha$ means stronger prior conviction.
    \item $\alpha < 1$ (e.g., $\alpha=0.5$): bimodal, placing mass near $p=0$ and $p=1$---appropriate when items are expected to be either very easy or very hard, with few intermediate cases.
\end{itemize}
\noindent When historical accuracy data from similar tasks is available, one can set $\alpha$ by matching moments: if the expected accuracy is $\bar{p}$ with variance $\sigma^2$, then $\alpha = \beta = \tfrac{1}{2}\bigl(\bar{p}(1-\bar{p})/\sigma^2 - 1\bigr)$ (for the symmetric case where $\bar{p} \approx 0.5$, the formula simplifies to $\alpha \approx 1/(8\sigma^2) - 1/2$). When prior knowledge indicates that workers are above chance, an \emph{asymmetric} prior such as $\textit{Beta}(2,1)$ ($\mathbb{E}[p] = 2/3$) or $\textit{Beta}(3,1)$ ($\mathbb{E}[p] = 3/4$) can be used instead; Appendix~\ref{app:informative_prior_tables} provides complete decision tables for these informative priors.

With this motivation, we model $p$ using the $\textit{Beta}(\alpha, \beta)$ distribution as a conjugate prior. The posterior distribution of $p$ after collecting $n_1$ correct and $n_2$ incorrect votes is:
\begin{displaymath}
P(p | \alpha, \beta, n_1, n_2) \sim \textit{Beta}(\alpha + n_1, \beta + n_2)
\end{displaymath}

\noindent Note that, since we do not know which of the two classes is correct during voting, our priors should be symmetric, i.e., we always start with $\alpha=\beta$. (This limits the shape of priors we can use, and we alleviate this concern in Section~\ref{sec:mixture-priors}.) Nevertheless, even operating with a simple prior $\textit{Beta}(\alpha, \alpha)$ can be remarkably effective, as we illustrate in the experimental evaluation in Section~\ref{sec:theory_vs_exp}.

We can now leverage the results from Section~\ref{sec:characteristics} together with the posterior for $p$ to obtain Bayesian estimates of quality, cost, and variance at any point during the voting process.

\begin{propo}[Bayesian estimates of quality and cost]
\label{prop:bayesian_estimates}
Let the prior for $p$ be $\textit{Beta}(\alpha, \beta)$, and suppose that $n_1$ correct and $n_2$ incorrect votes have been observed in a $\delta$-margin voting process. Then the posterior expected quality of the consensus vote is:
\begin{align}
\label{equ:quality_unknown_p}
Q(\delta, \alpha, \beta, n_1, n_2)
&= \int_0^1 Q\bigl(\varphi(p),\delta\bigr) \cdot f_{\textit{Beta}}\bigl(p;\, \alpha{+}n_1,\, \beta{+}n_2\bigr)\, dp \nonumber \\
&= \int_0^1 \frac{\varphi(p)^{\delta}}{1+\varphi(p)^{\delta}} \cdot \frac{p^{\alpha-1+n_1} \cdot (1-p)^{\beta-1+n_2}}{B(\alpha + n_1, \beta + n_2)}\, dp
\end{align}

\noindent where $Q(\varphi,\delta)$ is the known-$p$ quality from Theorem~\ref{th:Q_nonrand} and $f_{\textit{Beta}}$ is the Beta density. Similarly, the posterior expected number of remaining votes to consensus is:
\begin{align}
\label{equ:nvotes_unknown_p}
\mathbb{E}[n_{\mathit{votes}} \mid \delta, \alpha, \beta, n_1, n_2]
&= \int_0^1 \mathbb{E}[n_{\mathit{votes}} \mid \varphi(p),\delta] \cdot f_{\textit{Beta}}\bigl(p;\, \alpha{+}n_1,\, \beta{+}n_2\bigr)\, dp
\end{align}

\noindent where $\mathbb{E}[n_{\mathit{votes}} \mid \varphi,\delta]$ is given by Theorem~\ref{th:ET}. The same approach extends to the variance (Theorem~\ref{th:var_nonrand}) and the probability mass function (Theorem~\ref{th:pdf}).
\end{propo}

\noindent\textit{Proof sketch.} Each formula in Theorems~\ref{th:Q_nonrand}--\ref{th:pdf} is a deterministic function $g(p)$ of the mean accuracy. Since $p$ is uncertain with posterior $\textit{Beta}(\alpha{+}n_1, \beta{+}n_2)$, the law of total expectation gives $\mathbb{E}[g(p) \mid \text{data}] = \int_0^1 g(p)\, f_{\textit{Beta}}(p;\,\alpha{+}n_1,\,\beta{+}n_2)\, dp$. The integrals generally lack closed-form solutions but can be evaluated numerically or via Monte Carlo integration.\hfill$\square$

\noindent Proposition~\ref{prop:bayesian_estimates} formalizes a key practical advantage: the practitioner can monitor expected quality and remaining cost \emph{during} the voting process, updating estimates as each vote arrives. The integrals can be pre-computed for the modest range of $(n_1, n_2)$ states that arise in practice.\footnote{In many practical settings discussed in the introduction, the threshold for agreement is likely to be as small as $\delta=2$ or $\delta=3$, keeping the number of states small. Alternatively, we can assume that $p>0.5$ and integrate over $[0.5, 1]$, but in that case we must multiply the integrand by $1/\bigl[1 - F_{\text{Beta}}(0.5;\, \alpha+n_1,\, \beta+n_2)\bigr]$, where $F_{\text{Beta}}$ is the Beta CDF.}

\paragraph{From observed votes to quality estimates.}
During voting, the ground-truth label~$Y_i$ is unknown, so ``correct'' and ``incorrect'' are not directly observable---we observe only the running vote counts for each class, which we denote $n_{\max} = \max(n_1, n_2)$ and $n_{\min} = \min(n_1, n_2)$. To apply Proposition~\ref{prop:bayesian_estimates}, we must relate these observable counts to the latent correct/incorrect counts. This requires reasoning about two complementary hypotheses:
\begin{itemize}
    \item Under $H_c$ (the majority class is correct): ``correct'' votes = $n_{\max}$, ``incorrect'' = $n_{\min}$, yielding a posterior $\textit{Beta}(\alpha + n_{\max}, \beta + n_{\min})$.
    \item Under $H_i$ (the majority class is incorrect): the roles are reversed, yielding $\textit{Beta}(\alpha + n_{\min}, \beta + n_{\max})$.
\end{itemize}
\noindent The next proposition formalizes how to combine these per-hypothesis estimates into a single deployment quantity~$\hat{Q}$---the probability that the consensus label is correct---without access to ground truth.

\begin{propo}[Model-averaged quality from observed votes]
\label{prop:deployment_quality}
Let the prior for $p$ be $\textit{Beta}(\alpha, \beta)$, and suppose we observe $n_{\max}$ votes for the majority class and $n_{\min}$ votes for the minority class, with equal prior probability on the two class labels. Then:

\begin{enumerate}
    \item \textbf{Posterior hypothesis probability.} The posterior probability that the majority class is correct is:
    \begin{equation}
    \label{equ:Hc_posterior}
    P(H_c \mid n_{\max}, n_{\min}) = \frac{B(\alpha + n_{\max},\, \beta + n_{\min})}{B(\alpha + n_{\max},\, \beta + n_{\min}) + B(\alpha + n_{\min},\, \beta + n_{\max})}
    \end{equation}
    where $B(\cdot,\cdot)$ is the Beta function.

    \item \textbf{Model-averaged quality.} The probability that the consensus label is correct is:
    \begin{equation}
    \label{equ:Q_hat}
    \hat{Q} = P(H_c \mid \text{data}) \cdot \mathbb{E}[Q \mid H_c] + P(H_i \mid \text{data}) \cdot \mathbb{E}[Q \mid H_i]
    \end{equation}
    where $\mathbb{E}[Q \mid H_c] = Q(\delta,\, \alpha + n_{\max},\, \beta + n_{\min})$ and $\mathbb{E}[Q \mid H_i] = Q(\delta,\, \alpha + n_{\min},\, \beta + n_{\max})$ are computed via Proposition~\ref{prop:bayesian_estimates}.
\end{enumerate}
\end{propo}

\noindent\textit{Proof.}
Under $H_c$, the observed vote counts arise from $n_{\max}$ correct and $n_{\min}$ incorrect votes. The marginal likelihood under $H_c$ is:
\begin{displaymath}
P(\text{data} \mid H_c) = \textstyle\binom{n}{n_{\max}} \int_0^1 p^{n_{\max}} (1{-}p)^{n_{\min}} \, f_{\textit{Beta}}(p;\, \alpha,\beta)\, dp
= \textstyle\binom{n}{n_{\max}} \frac{B(\alpha + n_{\max},\, \beta + n_{\min})}{B(\alpha,\beta)}
\end{displaymath}
\noindent Under $H_i$, the correct votes are $n_{\min}$ and the incorrect are $n_{\max}$, giving $P(\text{data} \mid H_i) = \binom{n}{n_{\max}} {B(\alpha + n_{\min},\, \beta + n_{\max})}/{B(\alpha,\beta)}$. Since we assume equal class priors, Bayes' rule gives Equation~\ref{equ:Hc_posterior} after canceling the common $\binom{n}{n_{\max}}$ and $B(\alpha,\beta)$ terms. Equation~\ref{equ:Q_hat} then follows from the law of total expectation over the hypothesis.\hfill$\square$

\noindent\textbf{Symmetric priors.}  When $\alpha = \beta$ (as recommended in Section~\ref{sec:pure-prior}), $B(\alpha{+}n_{\max},\, \alpha{+}n_{\min}) = B(\alpha{+}n_{\min},\, \alpha{+}n_{\max})$ because the Beta function is symmetric in its arguments. Therefore $P(H_c \mid \text{data}) = 1/2$ regardless of the vote counts, and $\hat{Q} = 1/2$. Intuitively, a symmetric prior assigns equal mass to ``workers are accurate'' ($p > 0.5$) and ``workers are systematically wrong'' ($p < 0.5$), so the observed vote split cannot distinguish the two hypotheses.

This has two practical consequences:
\begin{enumerate}
    \item \textbf{Design-time planning} with a symmetric prior should use $\mathbb{E}[Q \mid H_c]$, which conditions on the majority being correct and integrates over $p$ with the posterior $\textit{Beta}(\alpha{+}n_{\max},\, \alpha{+}n_{\min})$. The tables in Appendix~\ref{app:MC_sim_table} report this conditional quantity. Equivalently, one can restrict the prior to $p > 0.5$ (using a truncated Beta), which enforces the assumption that workers are better than random.
    \item \textbf{Deployment monitoring} where model-averaged $\hat{Q}$ is needed requires an \emph{asymmetric} prior ($\alpha \neq \beta$, e.g., $\textit{Beta}(2,1)$ or $\textit{Beta}(3,1)$), which breaks the symmetry by encoding a prior belief that workers are above chance. With an asymmetric prior, $P(H_c \mid \text{data}) > 1/2$ whenever the majority has more votes, and $\hat{Q}$ provides a meaningful, conservative estimate of consensus quality.
\end{enumerate}

\paragraph{Worked example (asymmetric prior).}
Consider a $\textit{Beta}(2,1)$ prior (encoding a prior belief that $\mathbb{E}[p] = 2/3$), $\delta = 2$, and observed votes $(n_{\max}, n_{\min}) = (3, 1)$.
\begin{enumerate}
    \item \emph{Posterior hypothesis probability.}
    $B(5, 2) = 1/30$, \; $B(3, 4) = 1/60$, so
    $P(H_c \mid \text{data}) = \frac{1/30}{1/30 + 1/60} = 2/3$.
    \item \emph{Per-hypothesis quality.}
    Numerical integration of Equation~\ref{equ:quality_unknown_p} gives
    $\mathbb{E}[Q \mid H_c] \approx 0.805$ (posterior $\textit{Beta}(5,2)$) and
    $\mathbb{E}[Q \mid H_i] \approx 0.390$ (posterior $\textit{Beta}(3,4)$).
    \item \emph{Model-averaged quality.}
    $\hat{Q} = \tfrac{2}{3} \times 0.805 + \tfrac{1}{3} \times 0.390 \approx 0.667$.
\end{enumerate}
\noindent The model-averaged estimate $\hat{Q} = 0.667$ is more conservative than the $H_c$-conditional estimate of $0.805$, reflecting the one-third posterior probability that the majority class is in fact incorrect.

For instance, Table~\ref{tab:example_q_nvotes_bayesian} shows the expected quality of the final consensus and the number of steps until completion, when we start with a $\textit{Beta}(1, 1)$ prior, and for different combinations of the collected votes.\footnote{For the complete set of numerically integrated expectations of quality, number of remaining votes to consensus, and the variance thereof for various values of $\delta$ and a $\textit{Beta}(1,1)$ prior for $p$, please refer to Appendix~\ref{app:MC_sim_table}. Appendix~\ref{app:informative_prior_tables} provides analogous tables for the informative priors $\textit{Beta}(2,1)$ and $\textit{Beta}(3,1)$, demonstrating how prior knowledge of worker quality affects the expected quality and cost of the voting process.} Notice that as we update our beliefs about $p$ while running the voting process, the probability of labeling an item correctly will be different for different combinations of outcomes. For example, for $\delta=2$, a voting process that terminates with a combination $(2,0)$ of two correct and zero incorrect votes will have a different estimate than a process that terminates with a combination $(3,1)$ of three correct votes and one incorrect vote -- and different form $(5,7)$. This intuitively makes more sense than the results we would have obtained assuming a $p$ that is fixed.

\begin{table}[t]
    \centering
    \def\arraystretch{1.7}
    \caption{Expected quality of the consensus vote and expected remaining number of steps to completion, for various values of $\delta$ and collected votes $n_1$ and $n_2$, when starting with a prior distribution $\textit{Beta}(1, 1)$ for the variable $p$, and assuming $p>0.5$.
    \label{tab:example_q_nvotes_bayesian}}
        \begin{tabular}{cccc}
            \includegraphics[width=0.23\columnwidth]{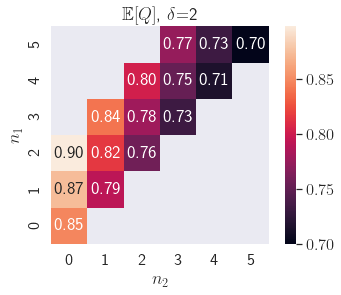}
            & \includegraphics[width=0.23\columnwidth]{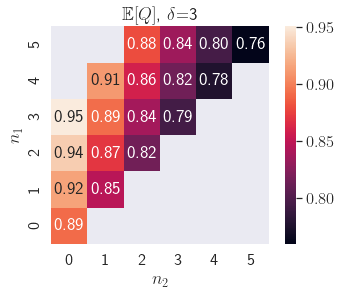}
            & \includegraphics[width=0.23\columnwidth]{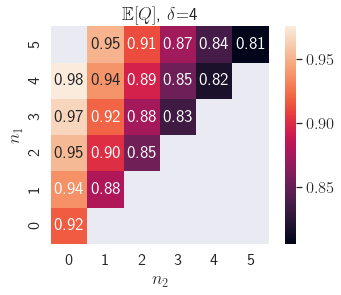}
            & \includegraphics[width=0.23\columnwidth]{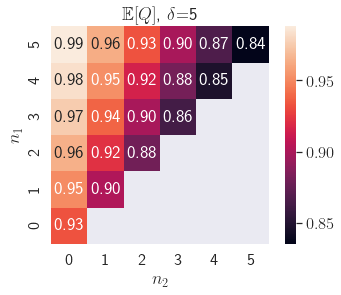}\\[-4pt]
            \includegraphics[width=0.23\columnwidth]{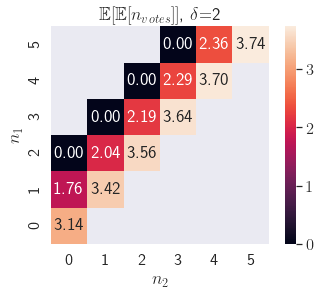}
            & \includegraphics[width=0.23\columnwidth]{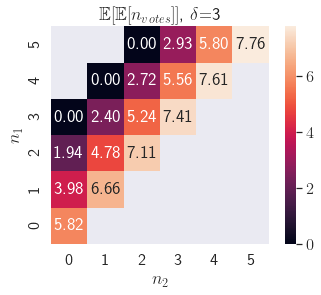}
            & \includegraphics[width=0.23\columnwidth]{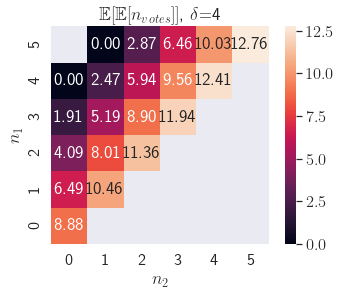}
            & \includegraphics[width=0.23\columnwidth]{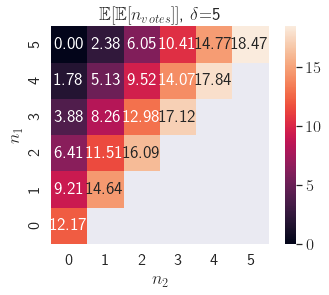}\\[-4pt]
            \Description{Eight heatmap panels showing expected quality and expected remaining votes for delta values 2 through 5 as a function of vote counts under a Beta(1,1) prior.}
        \end{tabular}%
\end{table}

\subsection{Working with mixture priors for \texorpdfstring{$p$}{p}}
\label{sec:mixture-priors}

\begin{figure}[t]
    \centering 
    \begin{minipage}[c]{0.45\columnwidth}
    \includegraphics[width=\columnwidth]{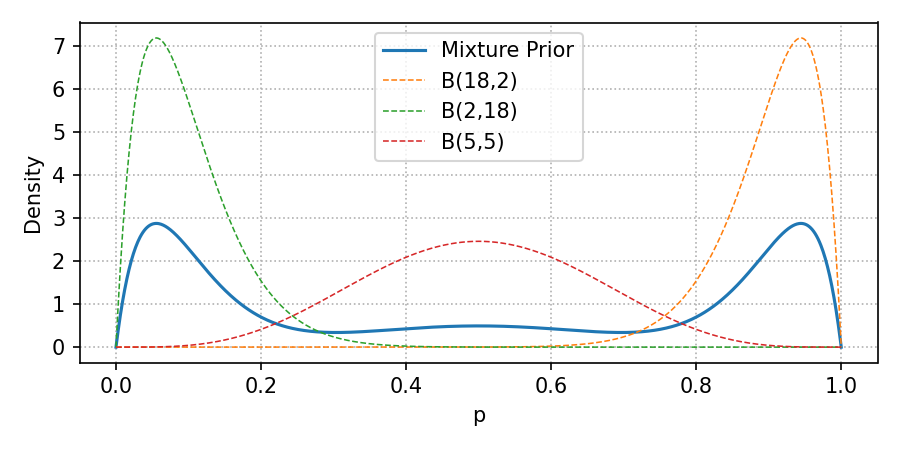}
    \Description{Line plot of a mixture-of-Betas prior distribution combining three components with peaks near different accuracy levels.}
    \caption{An example of a mixture prior distribution, mixing three Beta distributions: $0.4 \cdot B(18,2) + 0.4 \cdot B(2,18) + 0.2 \cdot B(5,5)$.}
    \label{fig:mixture_of_betas}
    \end{minipage}
    \hfill
    \begin{minipage}[c]{0.45\columnwidth}
    \includegraphics[width=\columnwidth]{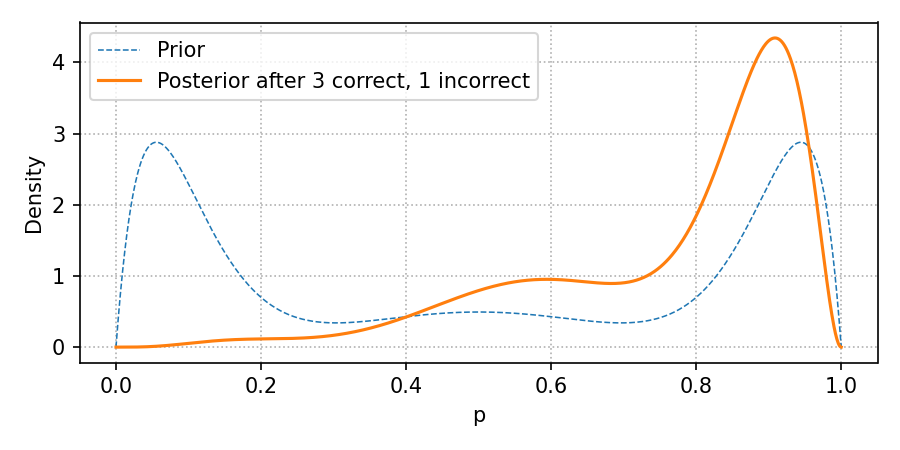}
    \Description{Line plot of the posterior mixture-of-Betas distribution after observing votes, showing updated component weights.}
    \caption{The posterior distribution, starting with prior from Figure~\ref{fig:mixture_of_betas} and updating after receiving three correct out of four votes.}
    \label{fig:mixture_of_betas_posterior}
    \end{minipage}

\end{figure}

Previous section demonstrated the basic principle of estimating $p$ given a prior estimate of the worker pool quality and the information from the votes for the item being adjudicated. The approach works for any prior of the form $\textit{Beta}(\alpha, \beta)$, but this is often insufficient to express our prior knowledge of worker quality. For example, considering a prior distribution with two peaks at the low and high ends of the accuracy may be more realistic ---a mixture of two symmetrically skewed $Beta$ distributions. For example, a more realistic prior may have peaks at 0.1 and 0.9, as in Figure~\ref{fig:mixture_of_betas}), with a mixture of random votes, reflecting a prior belief that the workers are often 95\% accurate, but there are still items that are confusing for the workers.

When the prior distribution of $p$ comprises a blend of Beta distributions, then we are working with a blend of \emph{conjugate} priors. In this case, updating the parameters of the individual Beta distributions as well as the mixture weights can be done concurrently, simplifying the process. Specifically, if we assume that the prior distribution $P(\theta)$ is a mixture of $K$ Beta distributions with mixing proportions $w = \{w_1, w_2, ..., w_k\}$ and corresponding parameters $\{(\alpha_1, \beta_1), (\alpha_2, \beta_2), ..., (\alpha_k, \beta_k)\}$, then the prior distribution can be written as:
\begin{displaymath}
P(\theta) = \sum_{i=1}^k w_i \cdot Beta(\theta; \alpha_i, \beta_i) 
\end{displaymath}

\noindent where $\theta$ is the probability of success in a Bernoulli trial. To jointly update both the parameters of the individual Beta distributions and the mixture weights, we can use the following procedure:

\begin{enumerate}

    \item \textbf{Update the parameters of each Beta distribution}: For the $i$-th component of the mixture, we have: $\alpha^{post}_i = \alpha_i + n_1$ and $\beta^{post}_i = \beta_i + n_2$, where $n_1$ is the number of correct votes in the observed data, $n_1+n_2$ is the total number of votes, and $(\alpha_i, \beta_i)$ are the prior parameters for the $i$-th component of the mixture.

    \item \textbf{Update the mixture weights}: The posterior weight of each component is proportional to its prior weight times its marginal likelihood $L_i$. For the $i$-th component:

        \begin{displaymath}
        w^{post}_i = \frac{w_i \cdot L_i}{\sum_{j=1}^k  w_j \cdot L_j }
        \quad\text{where}\quad
        L_i = \frac{B(\alpha_i + n_1, \beta_i + n_2)}{B(\alpha_i, \beta_i)}
        \end{displaymath}

        \noindent Here, $B(\cdot,\cdot)$ is the Beta function. The denominator $B(\alpha_i, \beta_i)$ is the normalizing constant of the $i$-th prior component; it is essential when the mixture components have different prior parameters, as it ensures that the marginal likelihoods are correctly scaled.

    \item \textbf{Specify the posterior distribution}: Once the updated mixture weights and Beta parameters have been obtained, the posterior distribution can be calculated as:

        \begin{displaymath}
        P(\theta | n_1,n_2) = \sum_{i=1}^k w^{post}_i \cdot Beta(\theta; \alpha^{post}_i, \beta^{post}_i)
        \end{displaymath}

\end{enumerate}

\paragraph{Worked example.}
Consider a bimodal prior expressing the belief that items are either easy (workers highly accurate) or hard (workers near chance), with a small probability of intermediate difficulty:
\begin{displaymath}
P(p) = 0.4 \cdot \textit{Beta}(p;\, 18, 2) + 0.4 \cdot \textit{Beta}(p;\, 2, 18) + 0.2 \cdot \textit{Beta}(p;\, 5, 5)
\end{displaymath}

\noindent The three components have means $0.9$, $0.1$, and $0.5$, respectively. Suppose we observe $n_1 = 3$ correct and $n_2 = 1$ incorrect votes. The update proceeds as follows:

\begin{enumerate}
    \item \emph{Update Beta parameters.} Each component's parameters are shifted:
    $(18,2) \to (21,3)$, \; $(2,18) \to (5,19)$, \; $(5,5) \to (8,6)$.

    \item \emph{Compute marginal likelihoods.}  For each component~$i$, $L_i = B(\alpha_i + 3,\, \beta_i + 1)\, /\, B(\alpha_i, \beta_i)$:
    \begin{align*}
    L_1 &= B(21, 3) / B(18, 2) = 6.44 \times 10^{-2}, \\
    L_2 &= B(5, 19) / B(2, 18) = 2.03 \times 10^{-3}, \\
    L_3 &= B(8, 6) / B(5, 5) = 6.12 \times 10^{-2}.
    \end{align*}

    \item \emph{Update mixture weights.}  The posterior weights are proportional to $w_i \cdot L_i$:
    \begin{displaymath}
    w_1^{\textit{post}} = 0.664, \quad w_2^{\textit{post}} = 0.021, \quad w_3^{\textit{post}} = 0.315
    \end{displaymath}
    (after normalizing so that the weights sum to~1).
    The data favor the ``easy item'' component, shifting mass from the low-accuracy component ($0.4 \to 0.021$) to the high-accuracy component ($0.4 \to 0.664$).
\end{enumerate}

\noindent This example illustrates the key advantage of mixture priors over a single Beta: they allow the posterior to distinguish between qualitatively different regimes (easy vs.\ hard items) after just a few votes, rather than averaging them into a single unimodal posterior.

\paragraph{Numerical stability.}
When the number of observed votes is large, the Beta function values $B(\alpha_i + n_1, \beta_i + n_2)$ can underflow in floating-point arithmetic. A standard remedy is to work in log-space: compute $\log L_i = \log B(\alpha_i + n_1, \beta_i + n_2) - \log B(\alpha_i, \beta_i)$ using the log-Beta function (available in all standard numerical libraries as \texttt{lbeta} or \texttt{betaln}), and then use the log-sum-exp trick to normalize the posterior weights: $\log w_i^{\textit{post}} = \log w_i + \log L_i - \log\!\bigl(\sum_j \exp(\log w_j + \log L_j)\bigr)$. This avoids numerical overflow and underflow for any practical vote count.

\noindent We state this formally:

\begin{propo}[Conjugacy of mixture-of-Betas priors]
\label{prop:mixture_conjugacy}
Let the prior for~$p$ be a mixture of $K$ Beta distributions, $P(p) = \sum_{i=1}^{K} w_i \cdot \textit{Beta}(p;\, \alpha_i, \beta_i)$. After observing $n_1$ correct and $n_2$ incorrect votes, the posterior is also a mixture of $K$ Beta distributions:
\begin{displaymath}
P(p \mid n_1, n_2) = \sum_{i=1}^{K} w_i^{\textit{post}} \cdot \textit{Beta}\bigl(p;\, \alpha_i + n_1,\, \beta_i + n_2\bigr)
\end{displaymath}
\noindent where the posterior weights are $w_i^{\textit{post}} = {w_i \cdot L_i}\big/{\sum_{j=1}^{K} w_j \cdot L_j}$ and $L_i = B(\alpha_i + n_1, \beta_i + n_2) / B(\alpha_i, \beta_i)$ is the marginal likelihood of the $i$-th component. All Bayesian estimates from Proposition~\ref{prop:bayesian_estimates} extend to the mixture case via:
\begin{displaymath}
\mathbb{E}[g(p) \mid n_1, n_2] = \sum_{i=1}^{K} w_i^{\textit{post}} \int_0^1 g(p) \cdot \textit{Beta}\bigl(p;\, \alpha_i{+}n_1,\, \beta_i{+}n_2\bigr)\, dp
\end{displaymath}
\end{propo}

\noindent\textit{Proof.} Each Beta component is a conjugate prior for Bernoulli observations, so the posterior within each component is $\textit{Beta}(\alpha_i{+}n_1,\, \beta_i{+}n_2)$. By Bayes' rule applied to the discrete mixture index, the posterior weight of component~$i$ is proportional to $w_i$ times the marginal likelihood $L_i = \int_0^1 p^{n_1}(1{-}p)^{n_2}\,\textit{Beta}(p;\,\alpha_i,\beta_i)\,dp = B(\alpha_i{+}n_1,\,\beta_i{+}n_2)/B(\alpha_i,\beta_i)$. The extension to arbitrary $g(p)$ follows from the law of total expectation over the mixture components.\hfill$\square$

\noindent The procedure above is analogous to updating a single Beta prior but includes an additional step of updating the mixture weights. The updated mixture weights reflect the extent to which each individual Beta distribution contributes to the posterior distribution based on the observed data.

\paragraph{Model-averaging with mixture priors.}
The deployment quality estimate from Proposition~\ref{prop:deployment_quality} extends directly to mixture priors. The posterior hypothesis probability (Equation~\ref{equ:Hc_posterior}) generalizes by replacing the single-component marginal likelihoods with the mixture marginal likelihoods: $P(\text{data} \mid H_c) = \sum_{i=1}^{K} w_i \cdot B(\alpha_i{+}n_{\max},\, \beta_i{+}n_{\min}) / B(\alpha_i, \beta_i)$ and analogously for $P(\text{data} \mid H_i)$. Notably, asymmetric mixture priors---those not invariant under $p \mapsto 1-p$---naturally break the symmetry that causes $P(H_c \mid \text{data})$ to be degenerate under symmetric priors. (The bimodal prior in the worked example above is itself symmetric under $p \mapsto 1-p$, so it does \emph{not} break the degeneracy; an asymmetric mixture such as $0.6 \cdot \textit{Beta}(18,2) + 0.2 \cdot \textit{Beta}(2,18) + 0.2 \cdot \textit{Beta}(5,5)$ would.)


%% file: tex_files/6-payment.tex
\section{Equivalence conditions for two \texorpdfstring{$\delta$}{delta}-margin voting processes}
\label{sec:equiv}

In this section, we build on the results presented so far to address a common design problem: a requester has access to multiple worker pools with different accuracies and wants to achieve a target quality at minimum cost. Specifically, we explore:

\begin{itemize}
    \item How should we set the consensus threshold $\delta$ for each pool to match quality across pools with different known accuracies $p_1$ and $p_2$ (Section~\ref{sec:quality_equivalence})?

    \item Given a quality target, what per-vote payment for each pool equalizes the requester's total cost per item (Section~\ref{sec:pay_equiv})?

    \item How can we apply these results when worker accuracies are unknown and estimated via priors (Section~\ref{sec:pay_worker_unknown_acc})?
\end{itemize}

\noindent Following the presentation pattern from earlier sections, we first assume that the parameters $p_1$ and $p_2$ for the two worker pools are exogenously given and known. Then, we show how to relax this assumption.

\subsection{Quality equivalence for processes with different values of \texorpdfstring{$\delta$}{delta}}
\label{sec:quality_equivalence}

Suppose we have two sets of workers, one with accuracy $p_1$ and another with accuracy $p_2$. The first set of workers operates a $\delta$-margin majority voting scheme with threshold parameter $\delta_1$. Assuming for now that the requester only cares about the quality of the resulting work, what value of $\delta_2$ should she choose so that the set of workers with accuracy $p_2$ generate the same quality of the results as the first set of workers?

\begin{theo}
\label{theo:delta}
If workers classify an item with accuracy $p_1\ne\frac{1}{2}$ and threshold $\delta_1$, we can achieve the same quality of results by a set of workers with accuracy $p_2\ne\frac{1}{2}$ if we set the threshold $\delta_2$ to be:
\begin{displaymath}
\delta_2 = \delta_1 \cdot \frac{\ln{\varphi_1}}{\ln{\varphi_2}}
\end{displaymath}

\noindent where $\varphi_1=\frac{p_1}{1-p_1}$ and $\varphi_2=\frac{p_2}{1-p_2}$ are the \emph{odds} of a single worker classifying the item correctly.
\end{theo}

\begin{proof}[Proof sketch]
Setting $Q(\varphi_1,\delta_1) = Q(\varphi_2,\delta_2)$ via Theorem~\ref{th:Q_nonrand} gives $\varphi_1^{\delta_1}/(1+\varphi_1^{\delta_1}) = \varphi_2^{\delta_2}/(1+\varphi_2^{\delta_2})$, which requires $\varphi_1^{\delta_1} = \varphi_2^{\delta_2}$. Taking logarithms of both sides yields $\delta_1 \ln\varphi_1 = \delta_2 \ln\varphi_2$, and solving for $\delta_2$ gives the result.
\end{proof}

\noindent\textbf{Remark (integer constraint).} Since $\delta$ must be a positive integer, when the formula yields a non-integer $\delta_2$, rounding up to $\lceil \delta_2 \rceil$ guarantees quality at least as high as the target $Q(\varphi_1, \delta_1)$, while rounding down to $\lfloor \delta_2 \rfloor$ yields slightly lower quality. In practice, rounding up is the conservative choice when a minimum quality threshold must be met.

\subsection{Cost-equivalent payment across worker pools}
\label{sec:pay_equiv}

In practice, the requester cares not only about quality but also about \emph{cost}. We now formalize the design problem that arises when a requester has access to multiple worker pools.

\paragraph{Design problem.} A requester labels items using a $\delta$-margin voting process. She has access to two worker pools with accuracies $p_1$ and $p_2$ ($p_1 \ne p_2$, both $\ne \frac{1}{2}$), and pays workers in pool~$i$ a non-negative per-vote rate $\textit{pay}(\varphi_i) \geq 0$. The \emph{requester's total expected expenditure} to label one item using pool~$i$ is:
\begin{displaymath}
\mathbb{E}[\textit{Cost} \mid \varphi_i,\delta_i] = \textit{pay}(\varphi_i) \cdot \mathbb{E}[n_{\mathit{votes}} \mid \varphi_i,\delta_i]
\end{displaymath}

\noindent Using Theorem~\ref{th:ET}, this becomes:
\begin{displaymath}
\mathbb{E}[\textit{Cost}|\varphi,\delta] = \textit{pay}(\varphi) \cdot \delta\cdot\frac{\varphi+1}{\varphi-1} \cdot \frac{\varphi^{\delta}-1}{\varphi^{\delta}+1}
\end{displaymath}

\noindent The requester's goal is to set the thresholds $\delta_i$ and per-vote payments $\textit{pay}(\varphi_i)$ so that both pools produce results of equal quality at equal expected cost:
\begin{enumerate}
    \item \textbf{Quality constraint:} $Q(\varphi_1,\delta_1) = Q(\varphi_2,\delta_2) \geq Q^*$ \quad (solved by Theorem~\ref{theo:delta});
    \item \textbf{Cost constraint:} $\mathbb{E}[\textit{Cost} \mid \varphi_1,\delta_1] = \mathbb{E}[\textit{Cost} \mid \varphi_2,\delta_2]$ \quad (solved below).
\end{enumerate}

\noindent We assume that the requester is \emph{risk-neutral} and cares only about the expected quality of the result and the expected cost.\footnote{It is straightforward to extend the results to the case of risk-averse requesters by incorporating the variance of time until consensus (Theorem~\ref{th:var_nonrand}) as a proxy for cost uncertainty.} We also assume that setting payments differently for a group of workers has minimal effects on the level of effort, and hence accuracy, of worker performance (see Section~\ref{sec:litrev}).

From Section~\ref{sec:quality_equivalence}, we can set $\delta_2$ to adjust for the different worker accuracy $p_2$, thereby assuring that the quality of results is the same: $Q(\varphi_1,\delta_1)=Q(\varphi_2,\delta_2)$. Of course, as shown in Section~\ref{sec:votes}, a different consensus threshold $\delta_2$ also changes the expected number of votes required to reach consensus. When $\delta_2 = \delta_1 \cdot \frac{\ln{\varphi_1}}{\ln{\varphi_2}}$, we have results of equal quality; we can ensure equal costs by setting: 

\begin{displaymath}
\mathbb{E}[\textit{Cost}|\varphi_1,\delta_1]  = \mathbb{E}\left[\textit{Cost}|\varphi_2,\delta_1 \cdot \frac{\ln{\varphi_1}}{\ln{\varphi_2}}\right]
\end{displaymath}

\noindent With a few simple algebraic manipulations, and knowing that $\varphi_2^{\delta_1 \cdot \frac{\ln{\varphi_1}}{\ln{\varphi_2}}} = \varphi_1^{\delta_1}$, we get:

\begin{theo}
\label{theo:pay_eq}
If workers with response accuracy $\varphi_1$ are paid $\textit{pay}(\varphi_1)$ per vote, then the same quality of result can be generated by workers with accuracy $\varphi_2$ at the same total cost, if the ratio of the payments is:
\begin{displaymath}
\frac{\textit{pay}(\varphi_1)}{\textit{pay}(\varphi_2)}=\frac{\ln\varphi_1}{\ln\varphi_2}\cdot\frac{\varphi_2+1}{\varphi_1+1}\cdot\frac{\varphi_1-1}{\varphi_2-1}
\end{displaymath}

\noindent where $\varphi_i=\frac{p_i}{1-p_i}$ are the odds that a worker in pool $i$ classifies an item correctly with $p_i\ne\frac{1}{2}$. Based on the above, we can infer that 

\begin{displaymath}
\textit{pay}(\varphi) \propto  \ln\varphi \cdot \frac{\varphi-1}{\varphi+1}
\end{displaymath}

\noindent and if expressed as a function of $p$:

\begin{displaymath}
\textit{pay}(p) \propto  \left( \log{(p)} - \log{(1-p)} \right) \cdot \left( p - (1 - p) \right)
\end{displaymath}

\noindent which can also be interpreted as a measure of the information gain provided by the workers while adjudicating the task.
\end{theo}

\noindent\textbf{Remark (integer constraint).} Theorem~\ref{theo:pay_eq} derives the payment ratio by substituting $\delta_2 = \delta_1 \cdot \ln\varphi_1 / \ln\varphi_2$ from Theorem~\ref{theo:delta}. Because $\delta$ must be a positive integer, exact quality matching across pools is generally impossible (see the rounding remark after Theorem~\ref{theo:delta}); the payment ratio is therefore a continuous approximation. In practice, one rounds $\delta_2$ to the nearest integer and recomputes the exact payment ratio from the resulting (slightly mismatched) qualities, or treats the formula as a first-order calibration guide.

\subsection{The case of unknown \texorpdfstring{$p$}{p}}

In this subsection, we relax the assumption of knowing a priori the accuracy values of the worker pool. We follow the same pattern as in Section~\ref{sec:unknown_p}: we start with a prior belief about the quality of the worker pool, and we update the belief as we receive votes. First, we examine the case where the worker pool has a stable but unknown accuracy. Then, we examine the case where the accuracies of the worker pools are dependent and vary across items.

\paragraph{Important: relative vs.\ absolute payments.} Throughout this subsection, all payment values are defined only up to a proportionality constant. The framework determines \emph{payment ratios} between pools or between vote outcomes---it does not prescribe absolute dollar amounts. To convert a ratio into actual payments, the practitioner must anchor one pool's per-vote rate to a known market rate and scale the other pool's rate accordingly.

\subsubsection{Stable but unknown accuracy for the worker pool} \label{sec:pay_worker_unknown_acc}
Suppose we want to estimate the payment for a worker pool with an accuracy value $p$. In that case, we can use the definition of $\textit{pay}(p) $ from Theorem~\ref{theo:pay_eq} and assume that $p$ is a random variable following a $\textit{Beta}(\alpha,\beta)$ distribution. In that scenario, the value $\frac{p}{1-p}$ follows a Beta Prime distribution, and the log of a Beta Prime ($\log\frac{p}{1-p}$) is known to follow the logistic-beta distribution. In this scenario, if we treat each worker pool independently and integrate over $\textit{pay}(p) $ with $p \sim \textit{Beta}(\alpha,\beta)$~\citep{archer14a}, then the expected payment is:

\begin{equation}
\textit{pay}(\alpha,\beta) \propto \frac{\alpha-\beta}{\alpha+\beta} \cdot \big(\psi(\alpha) - \psi(\beta) \big) + \frac{2}{\alpha+\beta}
\label{equ:pay_p_beta}
\end{equation}

\noindent where $\psi(\cdot)$ is the digamma function. A complete derivation is given in Appendix~\ref{app:payment_proof}. The additive $\frac{2}{\alpha+\beta}$ term is a covariance correction that arises because $(2p-1)$ and $\ln\varphi(p)$ are positively correlated under the Beta distribution; it depends on $(\alpha,\beta)$ and cannot be absorbed by the proportionality constant.

\begin{figure}[t]
\centering
\includegraphics[width=0.4\columnwidth]{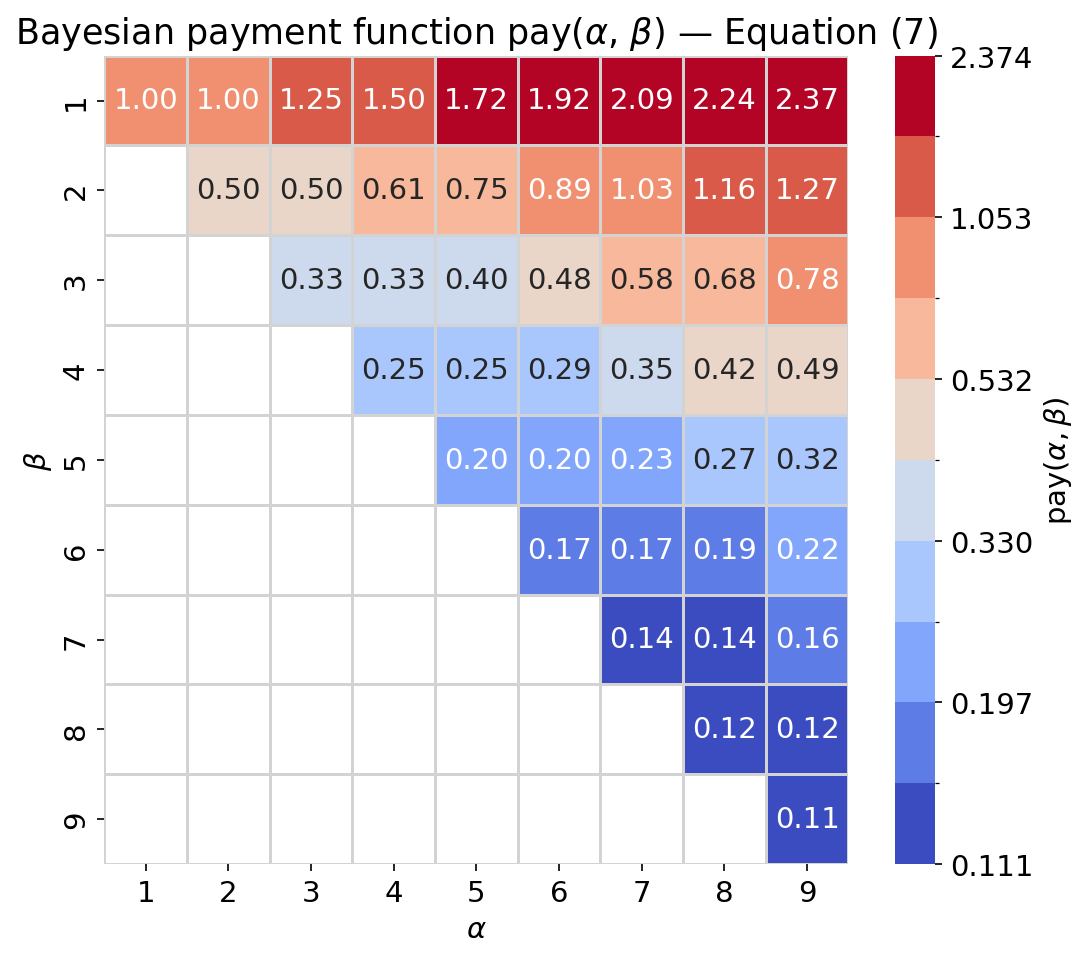}
\Description{Heatmap of payment function values for different posterior Beta parameters alpha and beta, with higher values along the diagonal where alpha exceeds beta.}
\caption{The values of the payment function $\textit{pay}(p|\alpha,\beta)$ from Equation~\ref{equ:pay_p_beta} for different values of $\alpha$ and $\beta$.}
\label{fig:pay_ab}
\end{figure}

Figure~\ref{fig:pay_ab} illustrates $\textit{pay}(\alpha,\beta)$ for various posterior parameter values. Since the formula defines payments only up to a proportionality constant, the numerical values in the figure are meaningful only as \emph{ratios}: the ratio of any two cells gives the relative per-vote payment that equalizes the requester's expected cost across two voting outcomes of equal quality. For instance, consider a pool that adjudicates an item with $\delta=3$, a voting result of 3-0, and a prior of $\textit{Beta}(1,1)$. The posterior parameters are $(\alpha+n_1, \beta+n_2) = (4,1)$, giving $\textit{pay}(4,1) = 1.50$. If another pool adjudicates the same item with $\delta=4$ and a voting result of 5-1, the posterior parameters are $(6,2)$, giving $\textit{pay}(6,2) = 0.89$. Despite the different voting results and payment levels, the expected quality of both outcomes is the same: $\mathbb{E}[Q \mid 3,0] = \mathbb{E}[Q \mid 5,1] = 0.951$, as per Equation~\ref{equ:quality_unknown_p}. (Detailed values for $\mathbb{E}[Q]$ can be found in Appendix~\ref{app:MC_sim_table}.) The payment ratio $1.50 / 0.89 = 1.69$ tells the requester that per-vote compensation for the first pool should be about 69\% higher than for the second, given these observed vote patterns. In contrast, an adjudication with a voting score of 5-2 results in $\textit{pay}(6,3) = 0.48$ and a lower expected quality of $\mathbb{E}[Q \mid 5,2] = 0.88$.

\subsubsection{Payment ratios under a joint distribution of worker accuracies} Suppose we aim to determine the optimal balance of payments across two distinct worker pools with unknown accuracies, such as a pool of "novice" and "experienced" workers. To achieve this, we must comprehend how each pool would adjudicate the same items. The process we follow is outlined below:

\begin{enumerate}

    \item We start with two worker pools, each with our prior beliefs about their accuracy, as discussed in Section~\ref{sec:unknown_p}. Let's denote the priors for the two pools as $\textit{Beta}(\alpha_1,\beta_1)$ and $\textit{Beta}(\alpha_2,\beta_2)$, respectively.
    
    \item For each item $i$, we proceed as follows:
    
        \begin{enumerate}
        
        \item Item $i$ is independently adjudicated by both worker pools.
        
        \item From the first pool, we receive $y_{i1}$ correct votes and $n_{i1} - y_{i1}$ incorrect votes (here ``correct'' and ``incorrect'' are determined relative to the ground truth, as in Section~\ref{sec:model}).

        \item Similarly, for the second pool, we receive $y_{i2}$ correct votes and $n_{i2} - y_{i2}$ incorrect votes.
        
        \item We then use Equation~\ref{equ:pay_p_beta} to estimate the payments for each pool.
        
        \item The payment for the first pool, $\textit{pay}_{i1}$, is calculated as $\textit{pay}(\alpha_1 +  y_{i1},\beta_1 + n_{i1} - y_{i1})$.

        \item The payment for the second pool, $\textit{pay}_{i2}$, is calculated as $\textit{pay}(\alpha_2 +  y_{i2},\beta_2 + n_{i2} - y_{i2})$.
        
        \end{enumerate}

\end{enumerate}

The payment ratio between two pools can be calculated using either micro- or macro-aggregation across items. Macro-aggregation first calculates the overall payment for each pool and then computes the ratio: $\frac{\sum_i \textit{pay}_{i1}}{\sum_i \textit{pay}_{i2}}$. On the other hand, micro-aggregation first calculates the ratios $\frac{\textit{pay}_{i1}}{\textit{pay}_{i2}}$ for each item and then aggregates these ratios. In the case of micro-aggregation, the geometric mean of the ratios is typically more stable than the arithmetic mean.

At the end of this process, we obtain a payment ratio that ensures a "fair" balance of payments between the two worker pools. When the pools produce similar quality outcomes, they receive the same payment per item.

\paragraph{Scope of the payment framework.}
The payment equivalence result (Theorem~\ref{theo:pay_eq}) and its Bayesian extension (Equation~\ref{equ:pay_p_beta}) are intended for \emph{ex-ante pool-level calibration}: setting the per-vote rate for each worker pool so that, in expectation across items, the requester's cost per correct label is equalized. The framework is not designed for \emph{ex-post item-level compensation} in its raw form, because $\textit{pay}(\varphi) \to 0$ as $p \to 0.5$: applying this at the item level would assign negligible pay for ambiguous items, which could incentivize workers to cherry-pick easy tasks. In practice, item-level payment schemes should incorporate a base pay (participation fee) plus a bonus component derived from the payment ratio, ensuring that all workers receive adequate compensation regardless of item difficulty.

%% file: tex_files/7-experiments.tex
\section{Experimental evaluation}
\label{sec:theory_vs_exp}

While theoretically sound and asymptotically accurate, the above results invite the question: How effectively do they mirror the outcomes of a genuine $\delta$-margin voting process? Our models rely on simplifying assumptions (A1--A3) formalized in Section~2 and operationalized by the closed-form quantities in Section~\ref{sec:characteristics} --- in particular, the i.i.d.\ Bernoulli trial model. The purpose of this section is to \emph{check that the closed-form formulas produce accurate predictions when applied to realistic per-item accuracy values drawn from actual crowdsourcing data}. Specifically, we draw items whose empirical accuracies~$p_i$ span a wide range (including items where $p_i < 0.5$), simulate the $\delta$-margin voting process via sampling with replacement (thereby restoring i.i.d.\ draws), and compare the theoretical predictions with the simulation outcomes. This design validates the formulas as an operational tool for process design; it does not, by itself, test robustness to finite-pool or dependent-error effects, which remains a direction for future work (see Section~\ref{sec:future}).

\subsection{Data set description}
\label{sec:bluebirds}

\begin{figure}[t]
\centering
\includegraphics[width=0.45\columnwidth]{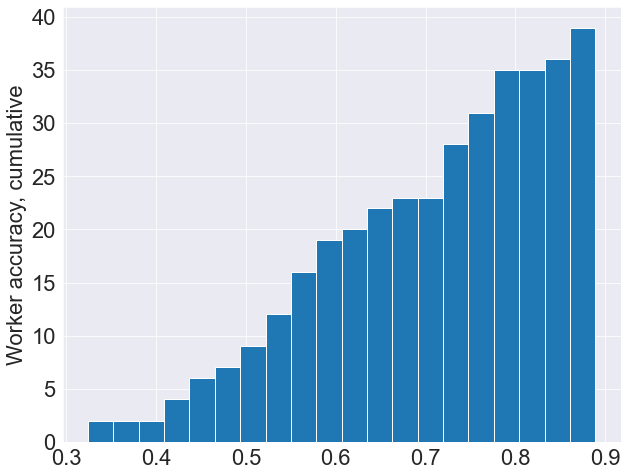}
\includegraphics[width=0.46\columnwidth]{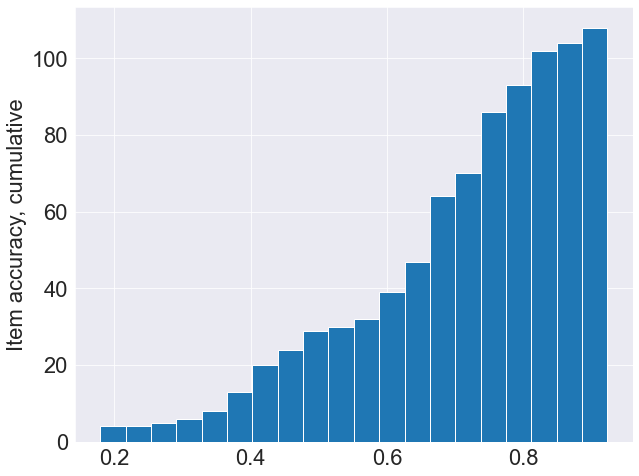}
\Description{Two cumulative histograms showing distribution of worker accuracy and item difficulty in the Bluebirds dataset.}
\caption{Cumulative histograms describing the \emph{Bluebirds} dataset. Left: worker accuracy (calculated as proportion of correct answers among the votes on all items for a given worker). Right: item difficulty (average accuracy of all worker responses for a given item).}
\label{fig:worker_acc_hist}
\end{figure}

We use a publicly available dataset of real MTurk worker votes for our evaluation -- the \emph{Bluebirds} dataset~\citep{bluebirds}. This dataset contains 39 binary labels from different workers for each of the 108 unique images. It also includes the ground truth labels for each item (with a 44:56 size ratio of the \emph{True} and \emph{False} classes).\footnote{For more information on the dataset, please see~\cite{bluebirds}. The data set is available at \url{https://github.com/welinder/cubam}.} The histogram in Figure~\ref{fig:worker_acc_hist} shows the empirical distribution of individual accuracy levels among workers, computed as the frequency of labeling an item correctly. The mean worker accuracy stands at 0.636, with significant heterogeneity in the labeling quality for each item. The distribution of computed per-item rates of correct response suggests that a third of items are ``confusing'' for the average worker (i.e., the frequency of correct responses across all workers for an item is below 0.5). 

\begin{figure}[t]
\centering
\includegraphics[width=0.9\columnwidth]{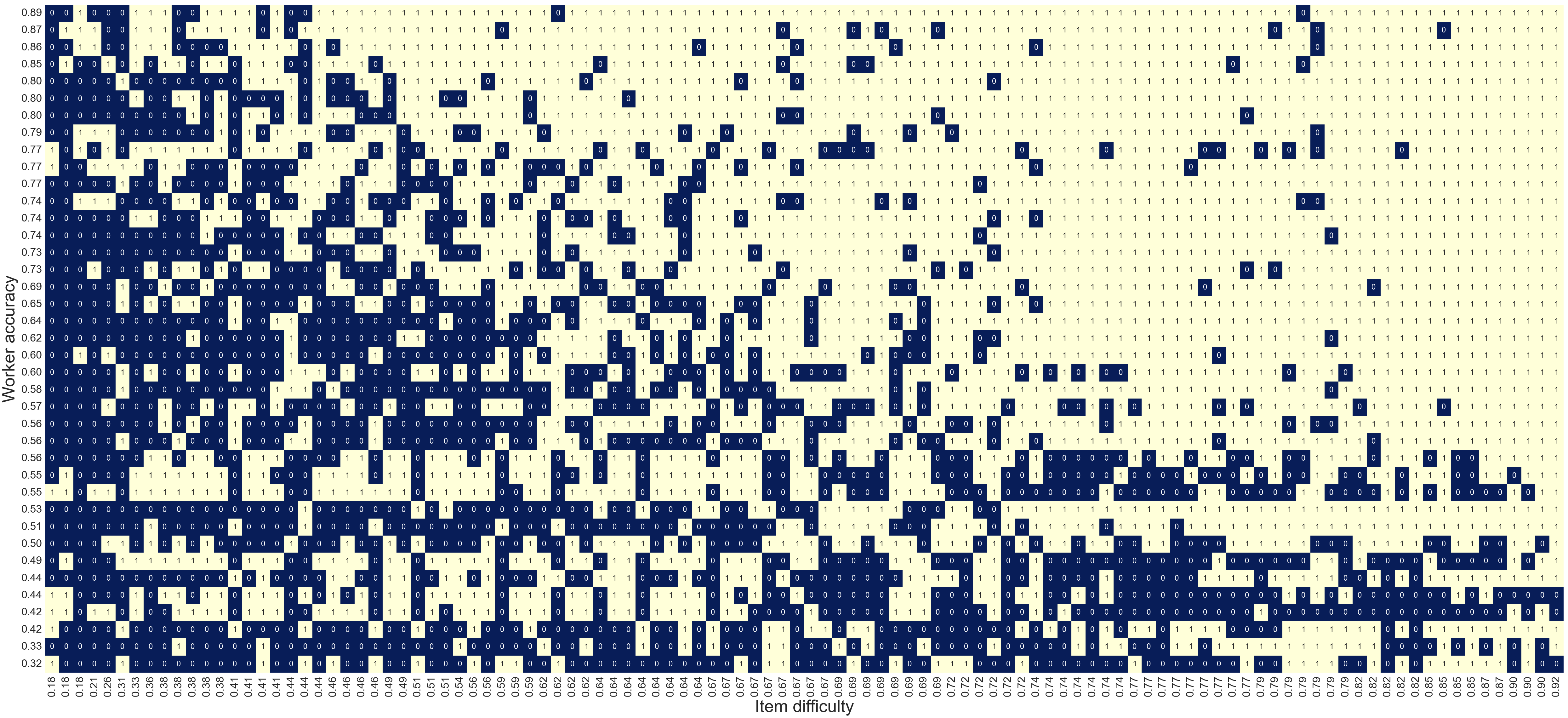}
\Description{Heatmap of all votes in the Bluebirds dataset with items ranked by difficulty and workers ranked by accuracy.}
\caption{All votes in the \texttt{Bluebirds} dataset, relative to ground truth; light color means correct answer; dark color means incorrect answer. The $x$-axis shows each item in the dataset; the items are ranked based on response correctness across workers in ascending order, with the easiest items being on the right. The $y$ axis shows the individual workers; the workers are ranked based on response correctness across items in ascending order, with the most accurate workers at the top.}
\label{fig:all_votes}
\end{figure}

One of the advantages of the \emph{Bluebirds} dataset is a relatively large number of labels per item. While no realistic process will ever collect 39 labels for an item, by having such a large number of labels per item, we can perform multiple simulations drawing a small(er) number of votes and examining the outcomes of the voting process. This allows multiple runs of the $\delta$-margin voting process to be conducted, allowing us to understand whether the model predictions are accurate. Figure~\ref{fig:all_votes} shows the total distribution of all votes in the \texttt{Bluebirds} dataset, relative to ground truth (also given in the dataset).

\subsection{Simulation setup}
\label{sec:simulation_setting}

We first want to know how well the quality estimate $Q(\varphi, \delta)$ in Equation~\ref{eq:Q_nonrand} and the time estimate $\mathbb{E}[n_{votes}|\varphi,\delta]$ in Equation~\ref{eq:exp_votes} match the actual results of a crowdsourcing process when people vote in a labeling task.  We compare the theoretical estimates with the results of our simulations that use real data. We simulate multiple voting runs using random draws from our dataset. We then compare the results of the simulated runs with our quality estimate to examine the accuracy of our prediction. 

Specifically, for a given $\delta \in [1, 2, .., 11]$, the process we used for simulating $\delta$-margin voting is as follows:

\begin{itemize}

    \item We run $r = 1{,}000$ simulations of the $\delta$-margin voting process for each item.
    
    \item For a given item, we iteratively draw votes (with replacement) from the \emph{Bluebirds} dataset labels (which are real-life observational votes given by real workers on this item) until we attain the agreement threshold $\delta$. 
    
    \item We perform sampling \emph{with replacement}. This choice intentionally restores the i.i.d.\ Bernoulli trial assumption (A2) that underpins the Gambler's Ruin formulas: each draw is independent with success probability equal to the item's empirical accuracy~$p_i$. It also prevents exhausting the 39 available votes per item for high values of $\delta$. Consequently, the tight agreement between theory and experiment reported below is primarily a \emph{consistency check}---confirming that the closed-form derivations are correctly implemented on realistic~$p_i$ values---rather than a test of robustness to finite-pool or dependent-error effects.

    \textbf{Quantitative justification.} For moderate $\delta$ and a pool of $N = 39$ workers, the difference between sampling with and without replacement is small. The finite-population correction factor for the variance of the sample mean after $n$ draws is $(N-n)/(N-1)$; at the median number of votes ($n \approx 3{-}4$ for $\delta = 2$), this factor exceeds $0.92$, implying less than an $8\%$ reduction in variance relative to the with-replacement case. Empirically, we verified this by comparing $1{,}000$ simulations per item under both sampling regimes: for $\delta \leq 3$, the absolute difference in consensus quality between with-replacement and without-replacement is less than $2.3$ percentage points, and the difference in expected votes is less than $0.1$. For $\delta = 5$, the gap widens to approximately $2.9$ percentage points in quality and $1.2$ votes, and pool exhaustion occurs in ${\sim}2\%$ of runs. In typical deployments with larger worker pools (hundreds or thousands of available workers), these finite-pool effects are negligible for all practical~$\delta$.
    
    \item We compute the consensus vote quality (i.e., is it correct or not) and the number of votes to completion for the $1{,}000$ simulations for each item.
\end{itemize}

Once we have the results of the simulation, we can compare them with our theoretical predictions and see how accurate our model is. To quantify the precision of the empirical estimates, we report 95\% bootstrap confidence intervals (CIs): for each item and~$\delta$, we resample the $1{,}000$ simulation outcomes with replacement $B = 2{,}000$ times, compute the mean of each resample, and take the 2.5th and 97.5th percentiles as the CI bounds.

\subsection{Theoretical vs. experimental \emph{quality of the consensus label}: the case of known \texorpdfstring{$p$}{p}}
\label{sec:q_vs_exp}

First, we want to evaluate how well the estimate $Q(\varphi, \delta)$ (Equation~\ref{eq:Q_nonrand}) predicts the actual outcome of a voting process. For us to calculate $Q(\varphi, \delta)$, we first assume that $p$ is known for the item: this is the \emph{expected} worker quality when labeling the item. We take the item difficulty $p$ to be equal to the average accuracy of all 39 workers' responses for this item in the \emph{Bluebirds} dataset (regardless of whether a vote was solicited from all 39 workers during our simulated voting). Using these values of $p$ and $\delta$, we compute the \textbf{theoretical value of consensus vote quality} from Equation~\ref{eq:Q_nonrand}. 

For example, for the first item in the \emph{Bluebird} data, the oracle (ground-truth) item-level correctness of crowdsourced responses across all 39 workers is $0.692$, and $\varphi \approx \frac{0.692}{1-0.692} \approx 2.247$. Hence for a voting process with $\delta=3$, we get $Q_{th}(\varphi, \delta) \approx \frac{2.247^3}{1+2.247^3} \approx 0.918$. We then compare this value against the average correctness of the consensus vote obtained across $1{,}000$ simulated runs.

\begin{figure}[t]
\centering
\includegraphics[width=0.32\columnwidth]{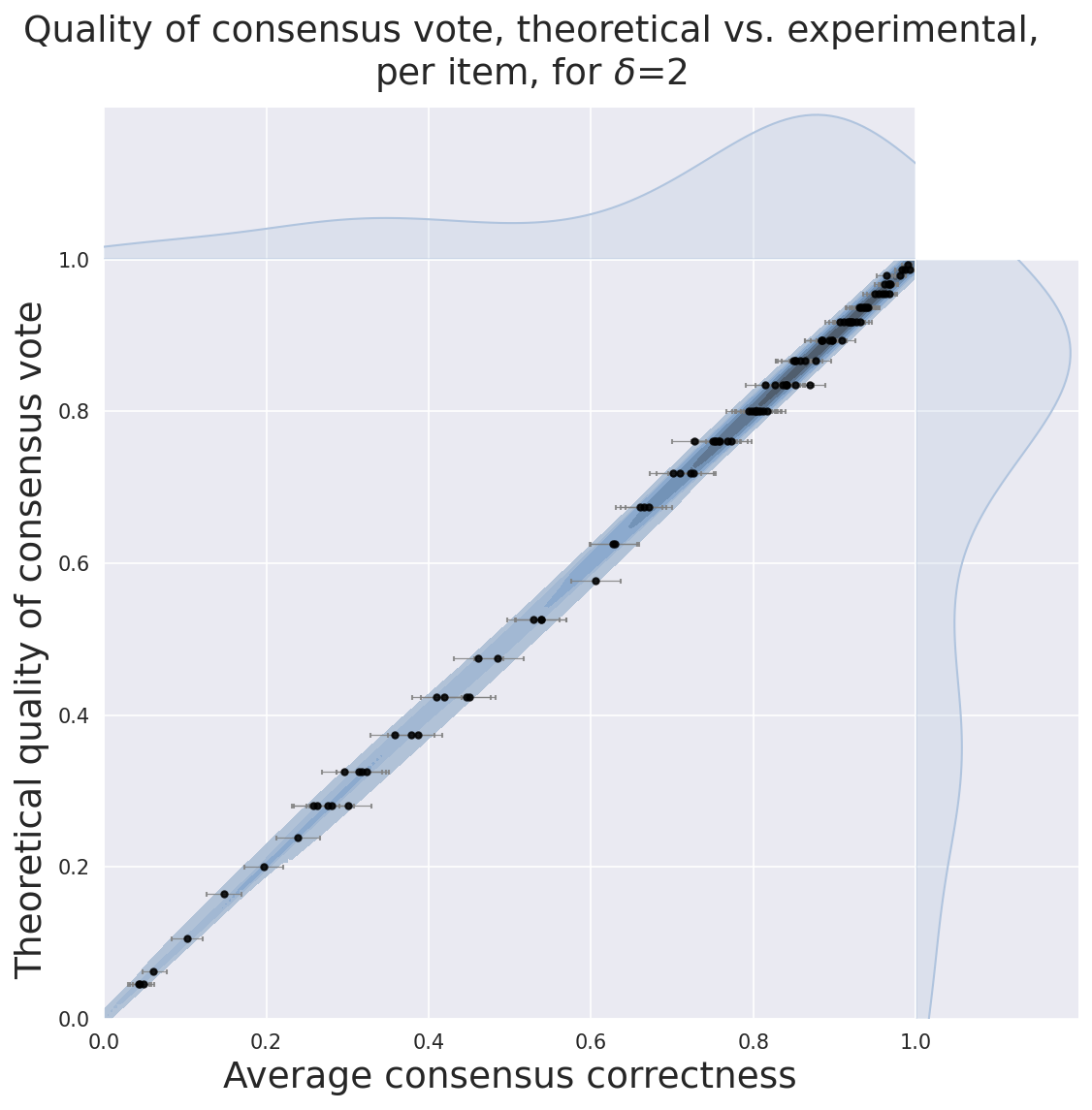}
\includegraphics[width=0.32\columnwidth]{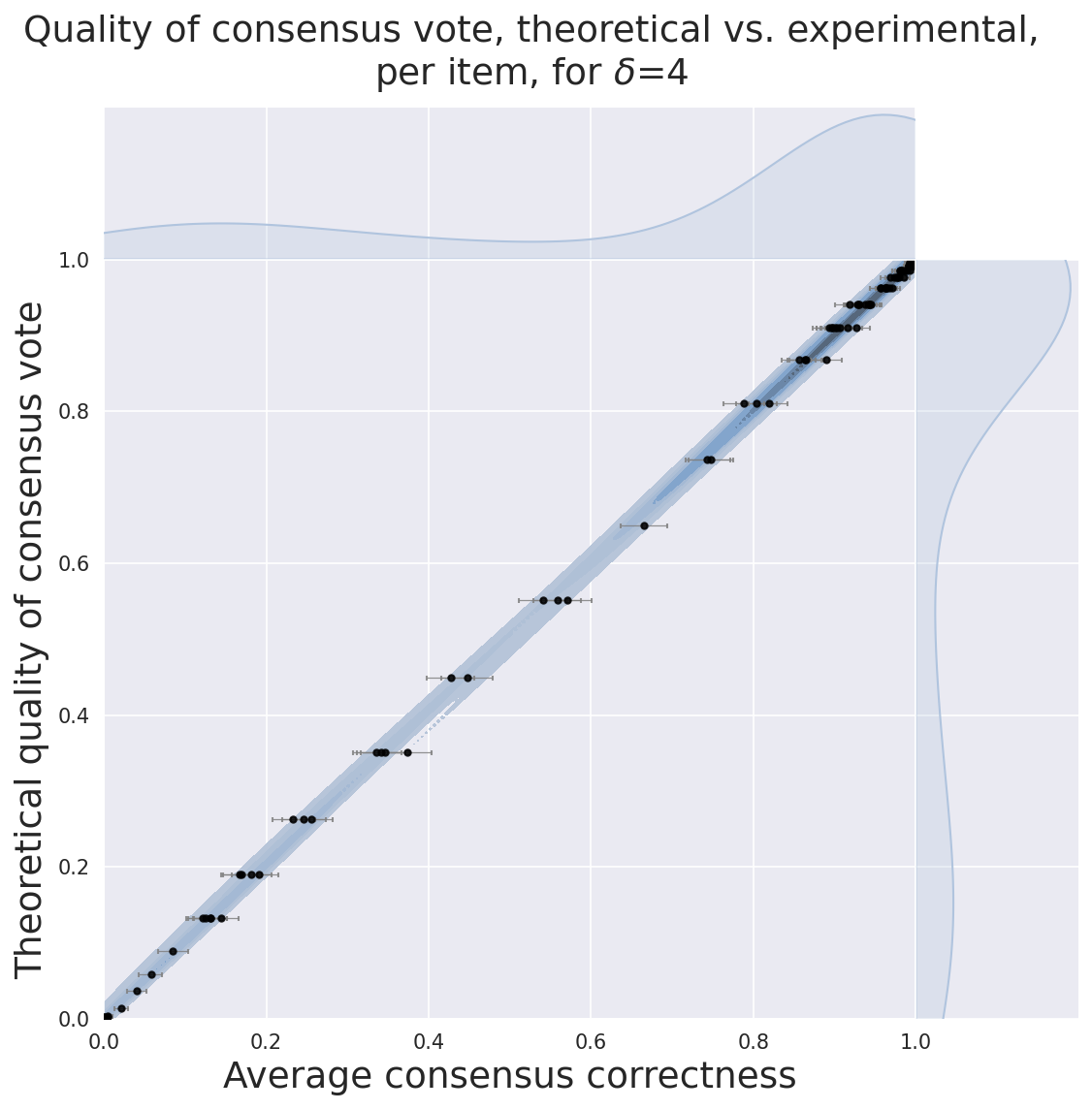}
\includegraphics[width=0.32\columnwidth]{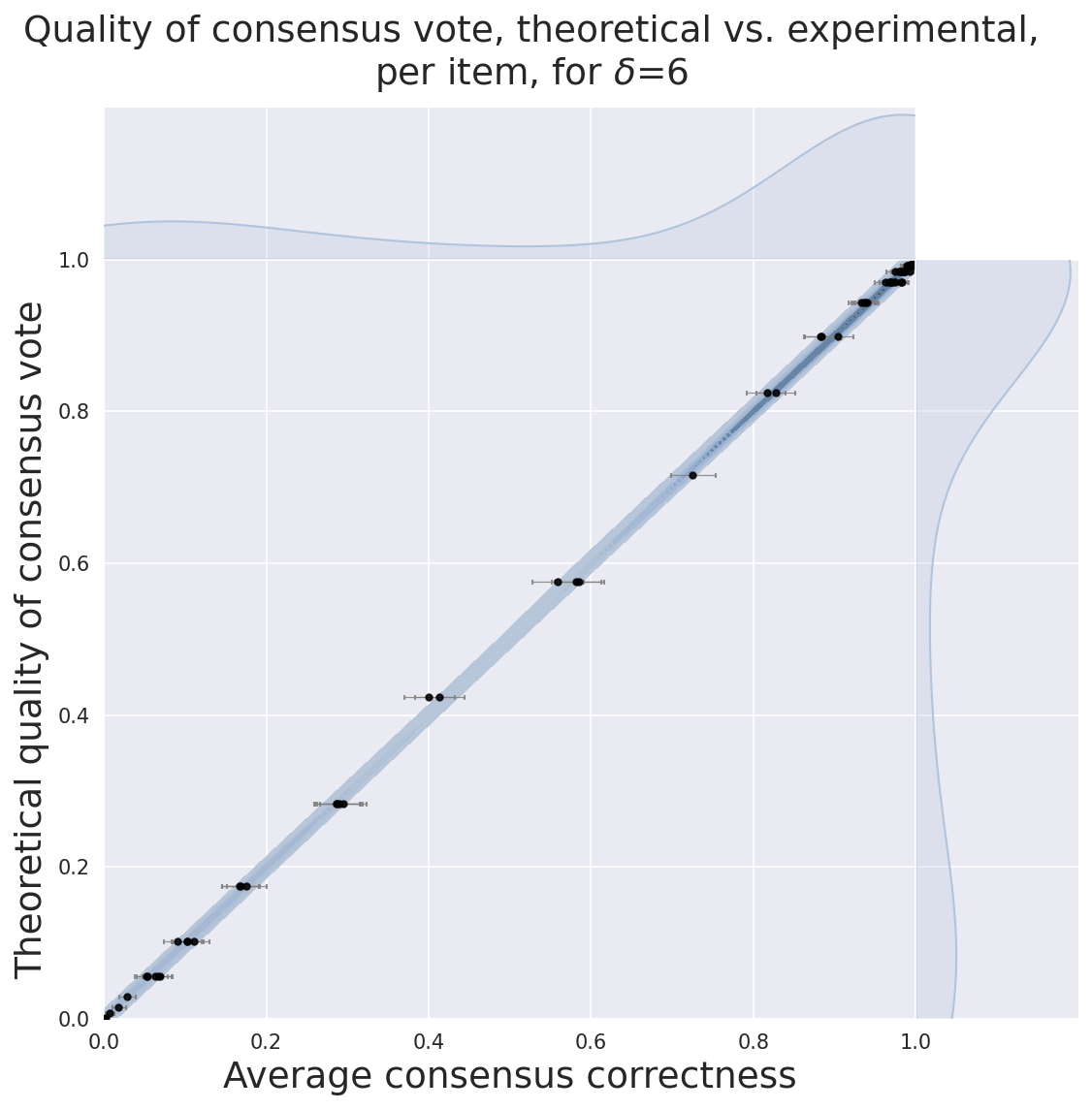}
\Description{Three scatter plots comparing theoretical versus empirical consensus quality for delta equals 2, 4, and 6.}
\caption{Comparison between theoretical and empirical values of the overall resulting label accuracy for various values of consensus threshold $\delta$. Each scatter plot point represents an item, where the $x$-coordinate is the calculated average correctness for $1{,}000$ experiments on the \emph{Bluebirds} data; horizontal error bars show 95\% bootstrap confidence intervals. The color indicates the estimated density of the joint distribution of empirical vs.\ theoretical results.}
\label{fig:q_th_vs_q_exp}
\end{figure}

Figure~\ref{fig:q_th_vs_q_exp} demonstrates the results of these experiments; horizontal error bars show 95\% bootstrap CIs for each item's empirical estimate. Empirical consensus label quality deviates minimally from the theoretical values, with the variance of the difference becoming smaller for stronger consensus requirements (greater $\delta$). The gap is also smaller for more `certain' items, which are frequently classified either correctly or incorrectly (the discrepancy between theoretical and empirical values appears to be concentrated around `divisive' items close to the $(0.5,0.5)$ region of the plot). For $\delta=2$, the theoretical prediction $Q(\varphi,\delta)$ falls within the 95\% bootstrap CI for 94\% of items; for $\delta=3$ the coverage is 95\%.

Table~\ref{tab:th_vs_exp_table} contains a sample of the detailed results, showing the theoretical and empirical values of the consensus label quality $Q(\varphi,\delta)$ along with $\pm$ half-widths of the 95\% bootstrap CIs for the empirical estimates. The deviation from the theoretical values is small for every value of $\delta$.

%
\begin{table}[t]
\def\arraystretch{1.4}
\centering
\resizebox{\columnwidth}{4cm}{
\begin{tabular}{|c|c|c|*{5}{c|}*{4}{c|}*{1}{c||}*{5}{c|}*{5}{c|}}
\hline
\multicolumn{3}{|c|}{} & 
\multicolumn{5}{c|}{\multirow{2}{*}{\begin{tabular}[c]{@{}c@{}}$Q$\\ Empirical\end{tabular}}} &
\multicolumn{5}{c||}{\multirow{2}{*}{\begin{tabular}[c]{@{}c@{}}$Q$\\ Theoretical\end{tabular}}} &
\multicolumn{5}{c|}{\multirow{2}{*}{\begin{tabular}[c]{@{}c@{}}$n_{votes}$\\ Empirical\end{tabular}}} &
\multicolumn{5}{c|}{\multirow{2}{*}{\begin{tabular}[c]{@{}c@{}}$\mathbb{E}[n_{votes}]$\\ Theoretical\end{tabular}}}\\
\multicolumn{3}{|c|}{} & \multicolumn{5}{c|}{} & \multicolumn{5}{c||}{} & \multicolumn{5}{c|}{} & \multicolumn{5}{c|}{}\\
\cline{3-23}
\multicolumn{2}{|c|}{} & $\delta$ & 
2 & 3 & 4 & 5 & 6 & 
2 & 3 & 4 & 5 & 6 & 
2 & 3 & 4 & 5 & 6 & 
2 & 3 & 4 & 5 & 6  \\
\cline{1-23}
 Item ID & $p$ & \multicolumn{21}{c|}{} \\
\cline{1-2}\cline{4-23}
\input{tex_files/utils/bigtable.tex}\\
\hline
\end{tabular}}
\caption{Results of empirical vs.\ theoretical computation of consensus label accuracy $Q(\varphi,\delta)$ and time (number of votes used) until voting completion $n_{votes}$. The empirical results are averaged across $1{,}000$ experiments per item of the \emph{Bluebirds} dataset; $\pm$ values are 95\% bootstrap CI half-widths. A sample of 12 items is shown.}
\label{tab:th_vs_exp_table}
\end{table}

\subsection{Theoretical vs. experimental \emph{time until completion}: the case of known \texorpdfstring{$p$}{p}}

This section compares the theoretical and empirical values of the votes required to reach a consensus. To estimate the theoretical value $\mathbb{E}[n_{votes}|\varphi,\delta]$ (Equation~\ref{eq:exp_votes}) for each item, we again estimate $p$ for each item using the average accuracy of the workers that labeled it. We then compare the theoretical estimate with the actual number of votes it took to reach consensus, averaged over the $1{,}000$ runs.

\begin{figure}[t]
\centering
\includegraphics[width=0.32\columnwidth]{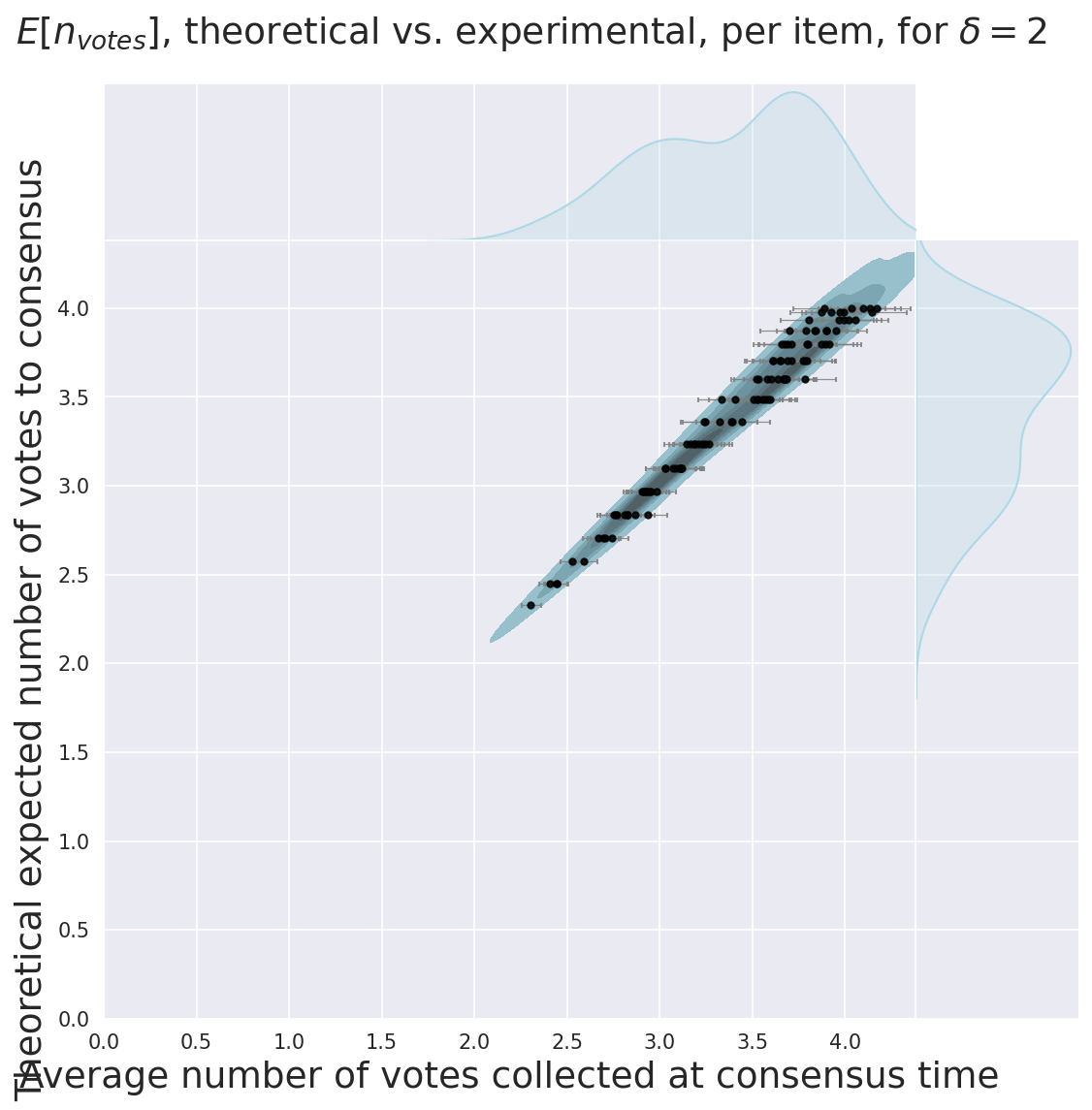}
\includegraphics[width=0.32\columnwidth]{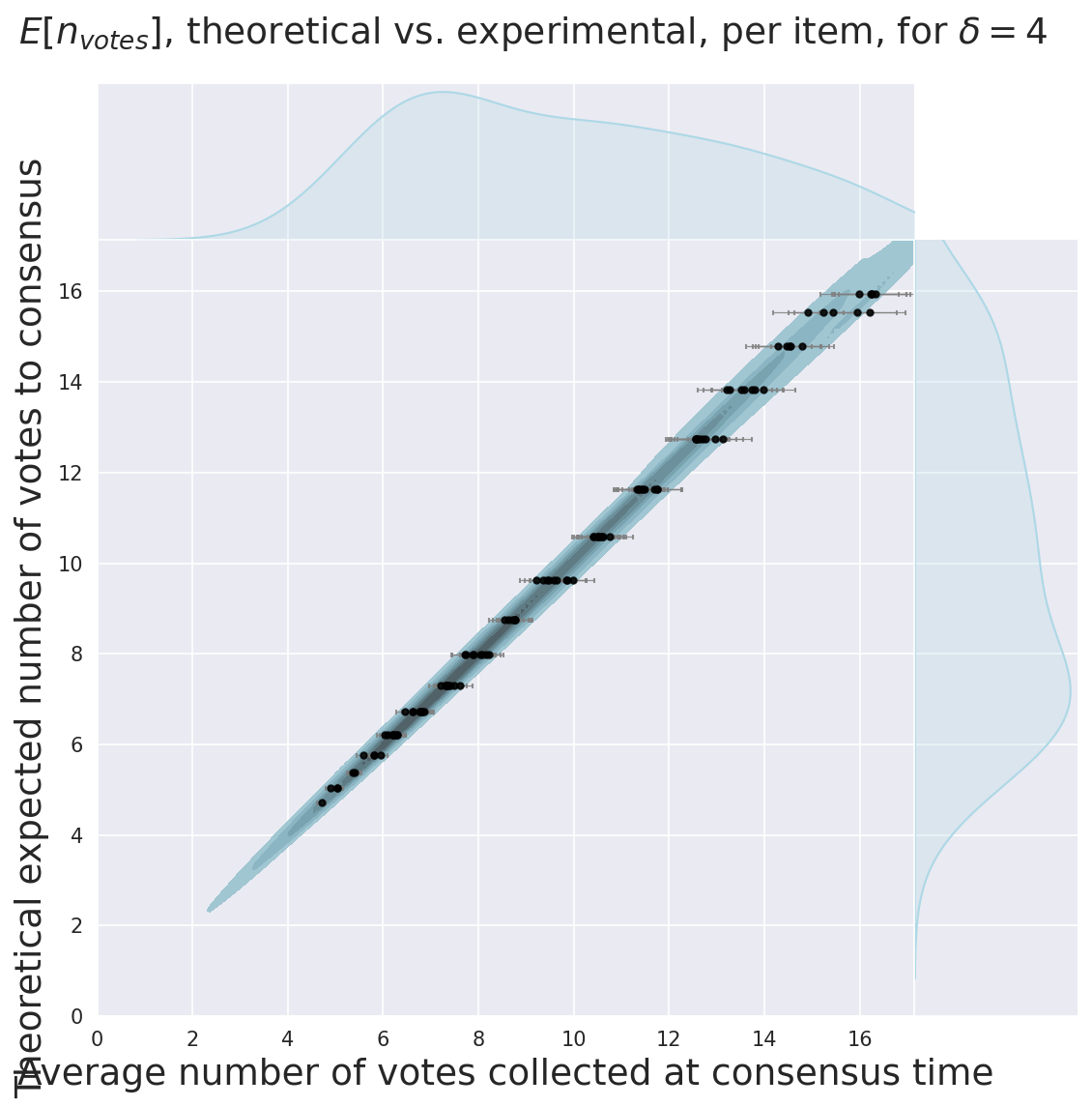}
\includegraphics[width=0.32\columnwidth]{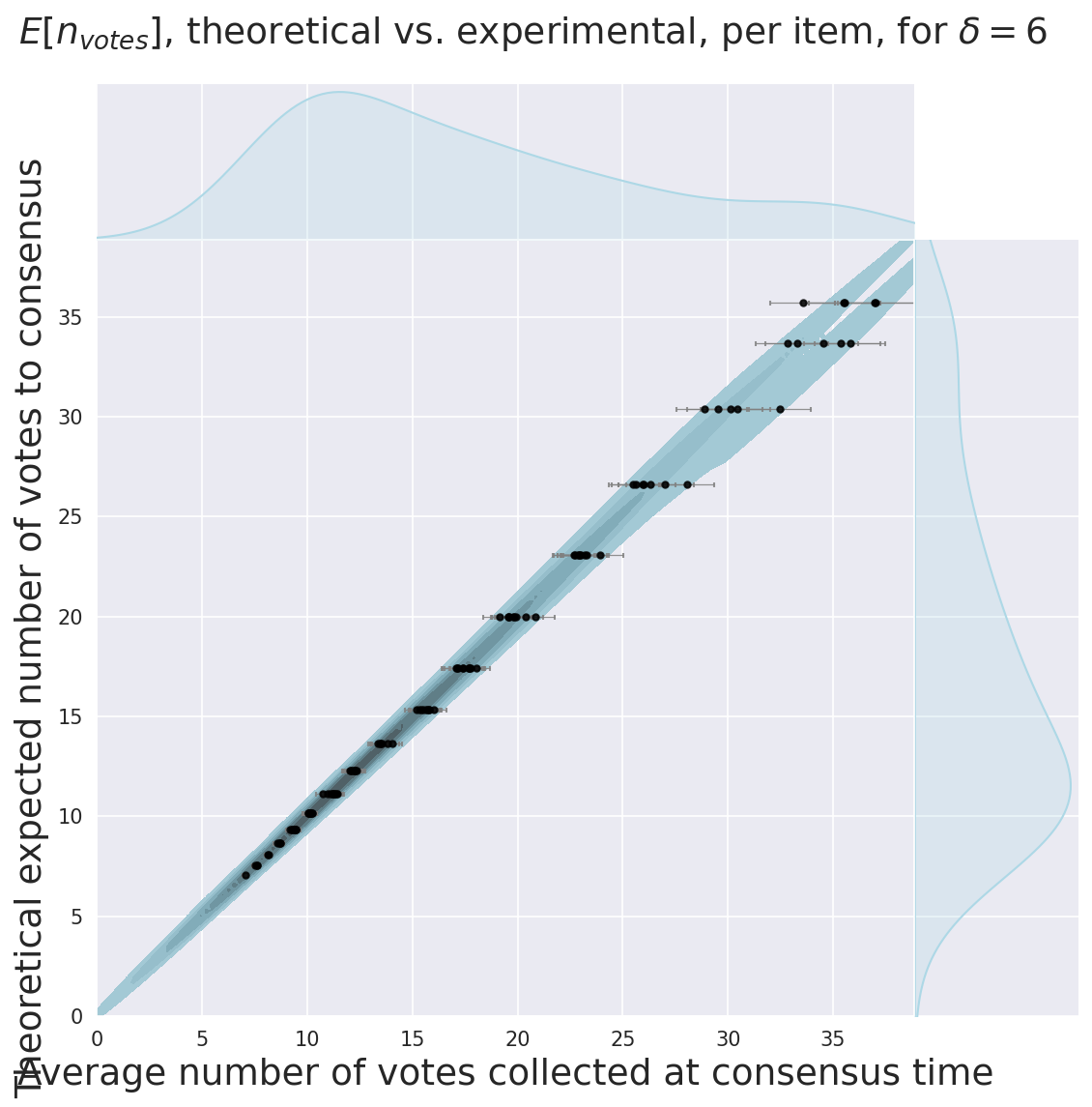}
\Description{Three scatter plots comparing theoretical versus empirical expected votes to consensus for delta equals 2, 4, and 6.}
\caption{Comparison between theoretical and empirical values of time until completion (i.e., number of votes it takes to reach consensus), for various values of consensus threshold $\delta$. Each scatter plot point represents an item, where the $x$-coordinate is calculated as the average number of votes taken to reach consensus across $1{,}000$ experiments using the \emph{Bluebirds} data; horizontal error bars show 95\% bootstrap confidence intervals. The color indicates density. (Note the differing scales.)}
\label{fig:nvotes_th_vs_exp}
\end{figure}

Figure~\ref{fig:nvotes_th_vs_exp} illustrates the results; horizontal error bars again show 95\% bootstrap CIs. For the `easy' items, any sample of workers is likely to agree. Hence, a series of $\delta$ votes is likely enough to reach a $\delta$ majority -- these are the cases in which the discrepancy between the theoretically expected number of votes and the experimental results is the smallest. Considering the change in scale between the left and right panels in Figure~\ref{fig:nvotes_th_vs_exp}, one can conclude that for higher threshold values $\delta$, a few items may occasionally take a greater number of votes to reach consensus. Still, even such ``outlier'' cases only drive up the discrepancy between the mean value of $N_{votes}$ and the theoretically expected value by at most $15\%$. We should also note that the variance (see Equation~\ref{fla:var_nonrand}) also increases with $\delta$, so observing such ``outliers'' is expected. Across all values of~$\delta$, the theoretical $\mathbb{E}[n_{votes}]$ falls within the 95\% bootstrap CI for over 92\% of items.

Table~\ref{tab:th_vs_exp_table} also includes the results of our experiments simulating the $\delta$-margin voting on the \emph{Bluebirds} observational votes, side-by-side with the theoretical quantities for the corresponding item. The discrepancy between the theoretical and empirical values of time (i.e., number of votes used) until voting completion $n_{votes}$ is small for every value of $\delta$.

\subsection{Monte-Carlo estimates using simulated sequential voting process with \emph{Bluebirds} data. (The case of unknown \texorpdfstring{$p$}{p})}
\label{sec:unknown_p_experiments_MC_vs_Bluebirds}

We now revisit the relaxation of modeling assumptions discussed in Section~\ref{sec:unknown_p}. Recall that we no longer assume that $p$ is known; instead, we treat it as a random variable estimated using Bayesian updating for a given prior. Below we describe the experiments we ran using Monte Carlo estimation for a $Beta(1,1)$ prior on $p$ and a Bernoulli likelihood, where the sequential voting process is simulated using real-life observational data on the votes from the \emph{Bluebirds} dataset.

The experimental process is as follows: For a given value of $\delta$, ($2\le \delta \le 5$) and each of the 108  items in the dataset, run $1{,}000$ experiments. In each experiment, simulate the sequential voting process by sampling votes randomly with replacement from the worker responses (until reaching a consensus). With each new vote collected, we compute the expectation of the quality $Q(\delta, \alpha, \beta, y, n-y)$ using Monte-Carlo estimation (see Equation~\ref{equ:quality_unknown_p}), given the current state $(y, n-y)$. Similarly, we estimate the expected remaining number of votes until consensus.\footnote{Note that the results in Section~\ref{sec:characteristics} are formulated for estimating the quality, time until consensus, and variance when starting from the initial voting state---zero votes. However, using the matrix formulation in Section~\ref{sec:votes}, it is straightforward to obtain the formulas for the number of steps remaining when starting at any current combination of votes $(y, n-y)$, which corresponds to the transient state $2y-n$. The expected remaining votes are computed via Monte-Carlo integration over the posterior distribution of~$p$.}

\begin{table}[t]
    \centering
    \def\arraystretch{1.3}
\begin{tabular}{c|c|c|c|c}
Terminal vote counts  & Runs  & Ground truth $Q$, averaged  &  95\% CI  & $Q(\delta, \alpha, \beta, y, n-y)$, Monte-Carlo     \\
\hline
2-0	& 50{,}289 & 0.843	& $\pm$0.003  & 0.848 \\
3-1	& 18{,}842 & 0.829	& $\pm$0.005  & 0.840 \\
4-2	& 7{,}576  & 0.796	& $\pm$0.009  & 0.802 \\
5-3	& 3{,}050  & 0.809	& $\pm$0.014  & 0.774 \\
\end{tabular}
\caption{Summary of results of experiments for estimating quality when $p$ is unknown, for $\delta=2$. Terminal vote counts are shown as (majority votes)$-$(minority votes), i.e., winner/loser counts irrespective of which label is correct. ``Runs'' indicates the number of simulation runs (across all items) that terminated in each state. Ground truth $Q$ is the empirical fraction of consensus labels that match the true label, averaged across all runs terminating with a given vote pattern; the 95\% CI column shows the half-width of bootstrap confidence intervals ($B = 2{,}000$ resamples). The Monte-Carlo column shows the corresponding Bayesian estimate from Equation~\ref{equ:quality_unknown_p} with a $Beta(1,1)$ prior, integrating over the full posterior on $[0,1]$ (not restricted to $p>0.5$). For terminal states close to $\delta$ (e.g., 2-0), the posterior retains non-negligible mass below $p=0.5$, causing these values to be slightly lower than the truncated-prior values in Appendix~\ref{app:MC_sim_table}.}
\label{tab:Q_true_vs_EQ_MC}
\end{table}

\begin{figure}[t]
\centering
\includegraphics[width=0.45\columnwidth]{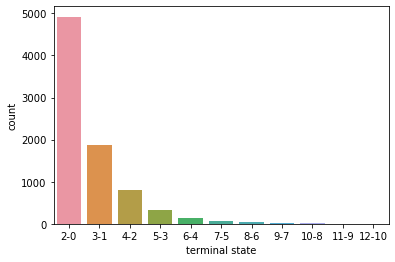}
\includegraphics[width=0.45\columnwidth]{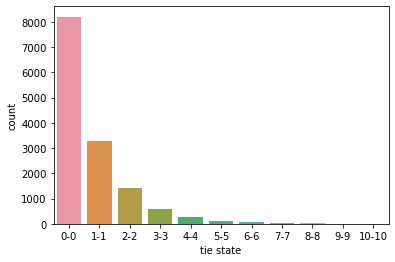}
\Description{Two bar charts showing frequency of terminal vote pairs and tied intermediate states for delta equals 2.}
\caption{Left: Frequency of occurrence for all possible terminal vote pairs in $\delta$-margin voting with $\delta=2$, across all $108 \times 1{,}000 = 108{,}000$ simulated runs. Each bar counts the number of \emph{runs} (not total visits) that terminated in the given state. Right: Frequency of occurrence for the (initial or intermediate) tied-state vote pairs, counted as the number of runs whose random walk visited each state at least once.}
\label{fig:terminal_initial_states_d2}
\end{figure}

Table~\ref{tab:Q_true_vs_EQ_MC} compares the Monte-Carlo estimated expectation of the quality of the consensus vote, using a $Beta(1,1)$ prior, versus the ground truth quality of the consensus vote. The results are very close, even with a relatively uninformed quality prior. The estimation accuracy can be further improved when the prior distribution for the quality is closer to the actual quality distribution of the labeling process. We should note that the apparent uptick in ground truth $Q$ at the 5-3 terminal state (0.809 vs.\ 0.796 at 4-2) is within statistical noise: the 95\% bootstrap CI half-width for the 5-3 state is $\pm 0.014$ (based on approximately 3{,}050 runs), compared to $\pm 0.009$ for the 4-2 state (7{,}576 runs). The smaller sample size at rarer terminal states naturally leads to wider confidence intervals (see Figure~\ref{fig:terminal_initial_states_d2} for the frequency distribution of terminal states).

Table~\ref{tab:nvotes_true_vs_EEnvotes_MC} compares the Monte-Carlo theoretical estimates and the empirical number of votes to completion. In contrast to the results on quality demonstrated above, we show the results for various ``intermediate states'' for the time to completion, showing that our formulations can provide a good estimate of how much longer the voting process is expected to take. The results are similar for other intermediate states and values of $\delta$.

\begin{table}[t]
    \centering
    \def\arraystretch{1.2}
\begin{tabular}{c||c|c}
   	\begin{tabular}{@{}c@{}}(Quasi-) initial \\ vote counts\end{tabular}
    &  
        \begin{tabular}{@{}c@{}} Ground truth remaining \\ $n_{votes}$, averaged  \end{tabular}
    &	 
        \begin{tabular}{@{}c@{}} $\mathbb{E}[n_{votes}]$,\\ Monte-Carlo \end{tabular} 

    \\\hline 
0-0    &   3.409   &   3.141    \\
1-1	   &   3.521   &   3.419    \\
2-2	   &   3.552   &   3.560    \\
3-3	   &   3.633   &   3.642    \\
4-4	   &   3.712   &   3.697    \\
5-5	   &   3.675   &   3.740    \\
\end{tabular}
\caption{Comparison of ground truth vs. Monte-Carlo expectation of key quantities, for $\delta=2$. We show the remaining time to termination vs. the Monte-Carlo estimates of $\mathbb{E}[n_{votes}]$ across all the processes that reached the state in the first column.}
\label{tab:nvotes_true_vs_EEnvotes_MC}
\end{table}

\paragraph{Scope of validation.}
The tight agreement between theoretical and empirical values in this section confirms that the closed-form derivations are correctly implemented and produce accurate predictions when applied to realistic per-item accuracy values~$p_i$ drawn from actual crowdsourcing data. Because the simulation uses sampling with replacement (restoring the i.i.d.\ assumption), the experiment is primarily a consistency check rather than a test of robustness to real-world deviations such as finite worker pools, dependent errors, or non-stationary accuracy. We investigate the sensitivity of the framework to some of these violations in Section~\ref{sec:stress_tests} below.

\subsection{Comparison with fixed-size majority vote}
\label{sec:baseline_comparison}

A natural question is whether $\delta$-margin voting offers practical advantages over the simpler \emph{fixed-size majority vote} (MV), in which a fixed committee of $n$ workers votes and the majority label is adopted (with early stopping once the majority is determined). We compare the two approaches on the \emph{Bluebirds} dataset.

For fixed-size MV, we simulate committees of $n \in \{3, 5, 7\}$ workers per item, drawing votes with replacement (as in Section~\ref{sec:simulation_setting}) and employing \emph{early stopping}: voting halts as soon as a majority is guaranteed or precluded, so the expected number of votes is less than~$n$. For $\delta$-margin voting, we use $\delta \in \{1, 2, 3, 4\}$. For each method and parameter, we run $1{,}000$ simulations per item and report quality~$Q$ (fraction of correct consensus labels) with 95\% bootstrap CIs ($B = 2{,}000$).

\begin{table}[t]
    \centering
    \def\arraystretch{1.3}
    \begin{tabular}{l|c|c|c|c}
    Method & Parameter & Quality $Q$ & 95\% CI & $\mathbb{E}[n_{\mathit{votes}}]$ \\
    \hline
    MV (fixed) & $n=3$ & 0.683 & $[0.680,\,0.686]$ & 2.40 \\
    MV (fixed) & $n=5$ & 0.708 & $[0.705,\,0.710]$ & 3.86 \\
    MV (fixed) & $n=7$ & 0.723 & $[0.720,\,0.725]$ & 5.32 \\
    \hline
    $\delta$-margin & $\delta=1$ & 0.637 & $[0.634,\,0.640]$ & 1.00 \\
    $\delta$-margin & $\delta=2$ & 0.708 & $[0.705,\,0.711]$ & 3.41 \\
    $\delta$-margin & $\delta=3$ & 0.739 & $[0.736,\,0.741]$ & 6.50 \\
    $\delta$-margin & $\delta=4$ & 0.755 & $[0.752,\,0.757]$ & 10.02 \\
    \end{tabular}
    \caption{Comparison of fixed-size majority vote (MV) with early stopping vs.\ $\delta$-margin voting on the \emph{Bluebirds} dataset (108 items, $1{,}000$ simulations per item, sampling with replacement). Quality~$Q$ is the fraction of consensus labels matching ground truth, averaged across all items and runs; 95\% bootstrap CIs are shown ($B = 2{,}000$). $\mathbb{E}[n_{\mathit{votes}}]$ is the average number of votes used per item.}
    \label{tab:mv_vs_delta}
\end{table}

Table~\ref{tab:mv_vs_delta} shows the results. At \emph{quality-matched} levels, $\delta$-margin voting uses fewer expected votes. For instance, $\delta=2$ achieves the same quality ($Q \approx 0.708$) as MV with $n=5$ while using $3.41$ votes per item on average---a $12\%$ reduction. This savings arises because $\delta$-margin voting adapts to item difficulty: easy items (high~$p$) terminate after only~$\delta$ votes, while hard items receive more votes as needed. By contrast, MV always solicits at least $\lceil n/2 \rceil + 1$ votes even for items on which all workers agree. The advantage grows with the committee size: MV with $n=7$ uses $5.32$ votes on average to achieve $Q = 0.723$, whereas $\delta=3$ achieves \emph{higher} quality ($Q = 0.739$) with $6.50$ votes, but without the hard upper bound on votes that caps MV quality.

\begin{figure}[t]
\centering
\includegraphics[width=0.7\columnwidth]{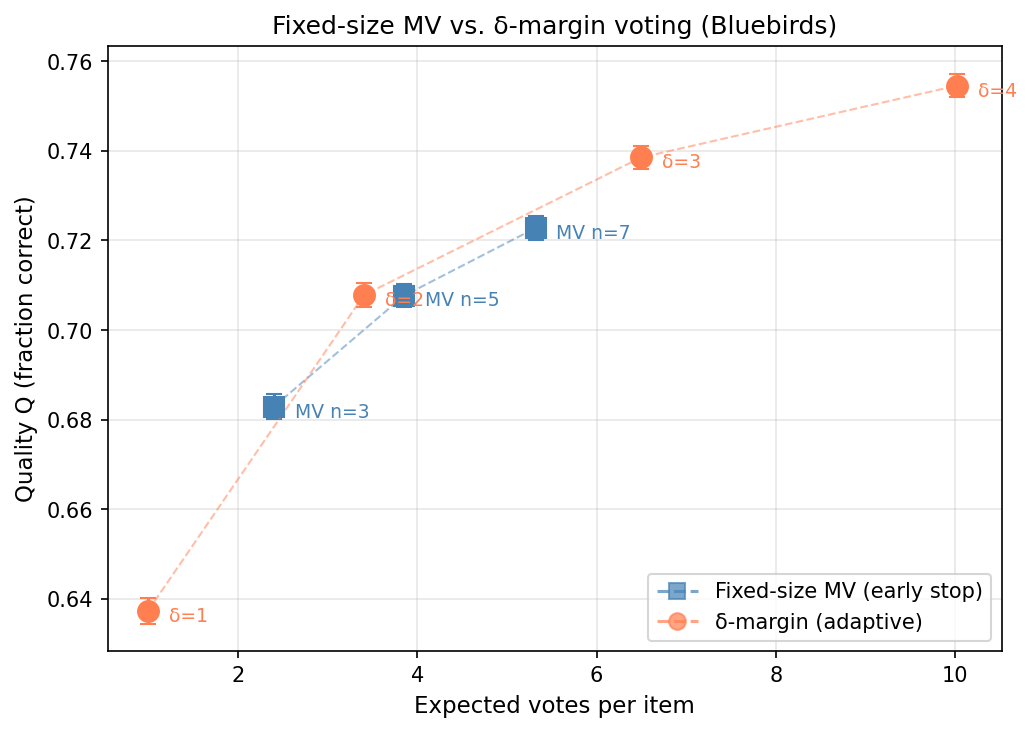}
\Description{Scatter plot comparing quality versus expected votes for fixed-size majority vote and delta-margin voting on the Bluebirds dataset.}
\caption{Quality vs.\ expected votes per item for fixed-size majority vote (MV, with early stopping) and $\delta$-margin voting on the \emph{Bluebirds} dataset. Each point represents one method/parameter setting; error bars show 95\% bootstrap CIs. At comparable quality levels, $\delta$-margin voting uses fewer expected votes.}
\label{fig:mv_vs_delta}
\end{figure}

Figure~\ref{fig:mv_vs_delta} visualizes the quality--cost trade-off. The $\delta$-margin curve lies to the left of the MV curve at comparable quality levels, confirming the efficiency advantage of adaptive stopping. Crucially, $\delta$-margin voting provides a \emph{continuous quality dial}: by choosing any~$\delta$, the practitioner can target a specific quality level, whereas MV constrains the choice to odd committee sizes with corresponding quality steps.

\subsection{Bayesian estimates with informative priors}
\label{sec:informative_priors}

Section~\ref{sec:unknown_p_experiments_MC_vs_Bluebirds} used an uninformative $Beta(1,1)$ prior on~$p$. In practice, organizations often have prior knowledge about their worker pool's accuracy (e.g., from historical data or pilot studies). We examine whether informative priors---$Beta(2,1)$ (mild belief that $p > 0.5$) and $Beta(3,1)$ (stronger belief)---improve the Bayesian quality estimates.

Using the same simulation protocol as Section~\ref{sec:unknown_p_experiments_MC_vs_Bluebirds} ($\delta=2$, and 500 simulations per item, sampling with replacement), we compute $E[Q|\delta, \alpha, \beta, n_{\max}, n_{\min}]$ from Proposition~\ref{prop:deployment_quality} for each prior at each terminal state. The \emph{Bluebirds} dataset has a mean item accuracy of $p \approx 0.636$, and approximately one-third of items have $p < 0.5$, making this a challenging test case.

All priors yield Bayesian estimates that exceed the empirical ground truth quality, which is expected: the estimates condition on the majority being correct ($H_c$), whereas for items with $p < 0.5$ the majority vote is often \emph{incorrect}. The gap between estimates and ground truth reflects the fraction of items where the model's assumption of competent workers ($p > 0.5$) is violated. For datasets where workers are reliably above chance, the estimates converge more closely to ground truth; the AML dataset in Section~\ref{sec:example} ($\bar{p}_{\text{junior}} = 0.747$, $\bar{p}_{\text{senior}} = 0.840$) demonstrates this pattern.

\subsection{Robustness: synthetic stress tests}
\label{sec:stress_tests}

We assess the robustness of the theoretical predictions to controlled violations of the model assumptions. All tests use synthetic data, allowing us to isolate individual factors.

\paragraph{Worker heterogeneity (Assumption A3).}
We draw votes from pools of 20 workers sharing the same mean accuracy $p = 0.75$ but with increasing spread: homogeneous ($p_i = 0.75$ for all), low variance ($p_i \in [0.70, 0.80]$), medium ($p_i \in [0.60, 0.90]$), and high ($p_i \in [0.51, 0.99]$). For $\delta = 3$ and $2{,}000$ items, the empirical quality remains within $0.006$ of the theoretical prediction $Q(\varphi, 3) = 0.964$ in all cases (Table~\ref{tab:stress_tests}). This validates Assumption~A3: the mean accuracy determines consensus quality under i.i.d.\ sampling, regardless of the variance across workers.

\paragraph{Inter-vote correlation.}
We introduce positive correlation by having each vote repeat the previous vote with probability~$\rho$ (and draw independently with probability $1-\rho$). This models scenarios where workers are influenced by prior votes or share common biases. Table~\ref{tab:stress_tests} shows that the theory remains accurate for mild correlation ($\rho \leq 0.05$, $|\Delta Q| \leq 0.01$), but quality degrades noticeably at higher levels: $|\Delta Q| \approx 0.02$ at $\rho = 0.10$ and $|\Delta Q| \approx 0.08$ at $\rho = 0.30$. Correlation also reduces the effective number of votes, causing the process to terminate faster than predicted.

\begin{table}[t]
    \centering
    \def\arraystretch{1.3}
    \begin{tabular}{l|c|c|c|c}
    Test condition & Parameter & Empirical $Q$ & Theory $Q$ & $|\Delta Q|$ \\
    \hline
    Heterogeneity & Homogeneous & 0.962 & 0.964 & 0.002 \\
    Heterogeneity & Low var & 0.969 & 0.964 & 0.004 \\
    Heterogeneity & Medium var & 0.971 & 0.964 & 0.006 \\
    Heterogeneity & High var & 0.962 & 0.964 & 0.003 \\
    \hline
    Correlation & $\rho = 0.00$ & 0.960 & 0.964 & 0.004 \\
    Correlation & $\rho = 0.05$ & 0.954 & 0.964 & 0.010 \\
    Correlation & $\rho = 0.10$ & 0.943 & 0.964 & 0.021 \\
    Correlation & $\rho = 0.20$ & 0.915 & 0.964 & 0.049 \\
    Correlation & $\rho = 0.30$ & 0.884 & 0.964 & 0.081 \\
    \end{tabular}
    \caption{Stress tests for $\delta$-margin voting ($\delta=3$, $p=0.75$, $2{,}000$--$3{,}000$ synthetic items per condition). ``Heterogeneity'': workers drawn from pools with the same mean $p = 0.75$ but varying spread (std shown in parentheses). ``Correlation'': moderate positive inter-vote correlation~$\rho$, where each vote repeats the previous with probability~$\rho$. The mean-accuracy prediction (Equation~\ref{eq:Q_nonrand}) is robust to heterogeneity ($|\Delta Q| \leq 0.006$) but degrades gracefully under correlation.}
    \label{tab:stress_tests}
\end{table}

\paragraph{Class imbalance.}
Under the symmetric error model (A1), the base rate (fraction of positive items) does not affect per-item quality, since accuracy $p$ is defined as the probability of a correct vote regardless of the true label. We confirm this empirically: varying the base rate from $0.1$ to $0.9$ produces empirical quality within $0.01$ of the theoretical prediction ($Q = 0.964$ for $p = 0.75$, $\delta = 3$).

\paragraph{Summary.}
The theoretical predictions are robust to worker heterogeneity and class imbalance---two common features of real-world deployments---but are sensitive to inter-vote correlation. When correlation is expected (e.g., sequential labeling with visible prior votes), the practitioner should either enforce independence (e.g., by hiding prior votes) or apply a conservative safety margin to~$\delta$.

%% file: tex_files/utils/bigtable.tex
11573 & 0.69 &  & 0.87{\tiny$\pm$0.02} & 0.93{\tiny$\pm$0.02} & 0.96{\tiny$\pm$0.01} & 0.99{\tiny$\pm$0.01} & 0.99{\tiny$\pm$0.01} & 0.84 & 0.92 & 0.96 & 0.98 & 0.99 & 3.41{\tiny$\pm$0.14} & 6.65{\tiny$\pm$0.29} & 9.44{\tiny$\pm$0.38} & 12.63{\tiny$\pm$0.47} & 15.79{\tiny$\pm$0.59} & 3.48 & 6.54 & 9.62 & 12.56 & 15.36\\ \cline{1-2}\cline{4-23}
11574 & 0.49 &  & 0.46{\tiny$\pm$0.03} & 0.47{\tiny$\pm$0.03} & 0.43{\tiny$\pm$0.03} & 0.45{\tiny$\pm$0.03} & 0.41{\tiny$\pm$0.03} & 0.47 & 0.46 & 0.45 & 0.44 & 0.42 & 3.89{\tiny$\pm$0.17} & 9.08{\tiny$\pm$0.42} & 16.25{\tiny$\pm$0.83} & 25.37{\tiny$\pm$1.23} & 35.53{\tiny$\pm$1.69} & 4.00 & 8.98 & 15.95 & 24.87 & 35.73\\ \cline{1-2}\cline{4-23}
11575 & 0.67 &  & 0.82{\tiny$\pm$0.02} & 0.88{\tiny$\pm$0.02} & 0.94{\tiny$\pm$0.01} & 0.97{\tiny$\pm$0.01} & 0.99{\tiny$\pm$0.01} & 0.80 & 0.89 & 0.94 & 0.97 & 0.98 & 3.67{\tiny$\pm$0.16} & 7.20{\tiny$\pm$0.33} & 10.60{\tiny$\pm$0.45} & 14.10{\tiny$\pm$0.57} & 17.10{\tiny$\pm$0.69} & 3.60 & 7.00 & 10.59 & 14.09 & 17.45\\ \cline{1-2}\cline{4-23}
11577 & 0.36 &  & 0.24{\tiny$\pm$0.03} & 0.15{\tiny$\pm$0.02} & 0.09{\tiny$\pm$0.02} & 0.06{\tiny$\pm$0.01} & 0.03{\tiny$\pm$0.01} & 0.24 & 0.15 & 0.09 & 0.05 & 0.03 & 3.61{\tiny$\pm$0.15} & 7.53{\tiny$\pm$0.35} & 11.49{\tiny$\pm$0.48} & 15.90{\tiny$\pm$0.73} & 19.58{\tiny$\pm$0.82} & 3.71 & 7.46 & 11.64 & 15.88 & 20.00\\ \cline{1-2}\cline{4-23}
11578 & 0.44 &  & 0.39{\tiny$\pm$0.03} & 0.35{\tiny$\pm$0.03} & 0.25{\tiny$\pm$0.03} & 0.22{\tiny$\pm$0.03} & 0.17{\tiny$\pm$0.02} & 0.37 & 0.32 & 0.26 & 0.22 & 0.18 & 4.06{\tiny$\pm$0.18} & 8.51{\tiny$\pm$0.39} & 14.53{\tiny$\pm$0.66} & 22.83{\tiny$\pm$1.04} & 30.15{\tiny$\pm$1.45} & 3.94 & 8.62 & 14.80 & 22.15 & 30.37\\ \cline{1-2}\cline{4-23}
11579 & 0.72 &  & 0.85{\tiny$\pm$0.02} & 0.94{\tiny$\pm$0.01} & 0.98{\tiny$\pm$0.01} & 0.99{\tiny$\pm$0.01} & 1.00{\tiny$\pm$0.00} & 0.87 & 0.94 & 0.98 & 0.99 & 1.00 & 3.24{\tiny$\pm$0.13} & 5.88{\tiny$\pm$0.22} & 8.74{\tiny$\pm$0.31} & 11.26{\tiny$\pm$0.40} & 13.53{\tiny$\pm$0.46} & 3.36 & 6.10 & 8.75 & 11.26 & 13.66\\ \cline{1-2}\cline{4-23}
\dots & \dots & \dots & \dots & \dots & \dots & \dots & \dots & \dots & \dots & \dots & \dots & \dots & \dots & \dots & \dots & \dots & \dots & \dots & \dots & \dots & \dots & \dots  \\
\dots & \dots & \dots & \dots & \dots & \dots & \dots & \dots & \dots & \dots & \dots & \dots & \dots & \dots & \dots & \dots & \dots & \dots & \dots & \dots & \dots & \dots & \dots  \\
\cline{1-2}\cline{4-23}
36959 & 0.74 &  & 0.91{\tiny$\pm$0.02} & 0.97{\tiny$\pm$0.01} & 0.99{\tiny$\pm$0.01} & 1.00{\tiny$\pm$0.00} & 1.00{\tiny$\pm$0.00} & 0.89 & 0.96 & 0.99 & 1.00 & 1.00 & 3.14{\tiny$\pm$0.12} & 5.91{\tiny$\pm$0.24} & 7.87{\tiny$\pm$0.27} & 9.78{\tiny$\pm$0.33} & 12.07{\tiny$\pm$0.39} & 3.23 & 5.67 & 7.98 & 10.16 & 12.27\\ \cline{1-2}\cline{4-23}
36960 & 0.79 &  & 0.93{\tiny$\pm$0.02} & 0.98{\tiny$\pm$0.01} & 1.00{\tiny$\pm$0.00} & 1.00{\tiny$\pm$0.00} & 1.00{\tiny$\pm$0.00} & 0.94 & 0.98 & 1.00 & 1.00 & 1.00 & 2.99{\tiny$\pm$0.10} & 4.90{\tiny$\pm$0.16} & 6.86{\tiny$\pm$0.21} & 8.59{\tiny$\pm$0.25} & 10.16{\tiny$\pm$0.26} & 2.97 & 4.92 & 6.72 & 8.46 & 10.17\\ \cline{1-2}\cline{4-23}
36961 & 0.77 &  & 0.92{\tiny$\pm$0.02} & 0.97{\tiny$\pm$0.01} & 0.99{\tiny$\pm$0.01} & 1.00{\tiny$\pm$0.00} & 1.00{\tiny$\pm$0.00} & 0.92 & 0.97 & 0.99 & 1.00 & 1.00 & 3.11{\tiny$\pm$0.11} & 5.31{\tiny$\pm$0.20} & 7.31{\tiny$\pm$0.25} & 9.15{\tiny$\pm$0.28} & 11.08{\tiny$\pm$0.31} & 3.10 & 5.28 & 7.31 & 9.24 & 11.13\\ \cline{1-2}\cline{4-23}
36962 & 0.77 &  & 0.93{\tiny$\pm$0.02} & 0.97{\tiny$\pm$0.01} & 0.99{\tiny$\pm$0.01} & 1.00{\tiny$\pm$0.00} & 1.00{\tiny$\pm$0.00} & 0.92 & 0.97 & 0.99 & 1.00 & 1.00 & 3.12{\tiny$\pm$0.11} & 5.13{\tiny$\pm$0.18} & 7.35{\tiny$\pm$0.26} & 9.08{\tiny$\pm$0.28} & 11.36{\tiny$\pm$0.33} & 3.10 & 5.28 & 7.31 & 9.24 & 11.13\\ \cline{1-2}\cline{4-23}
36963 & 0.90 &  & 0.99{\tiny$\pm$0.01} & 1.00{\tiny$\pm$0.00} & 1.00{\tiny$\pm$0.00} & 1.00{\tiny$\pm$0.00} & 1.00{\tiny$\pm$0.00} & 0.99 & 1.00 & 1.00 & 1.00 & 1.00 & 2.41{\tiny$\pm$0.06} & 3.80{\tiny$\pm$0.09} & 4.90{\tiny$\pm$0.10} & 6.37{\tiny$\pm$0.13} & 7.61{\tiny$\pm$0.13} & 2.45 & 3.76 & 5.03 & 6.29 & 7.55\\ \cline{1-2}\cline{4-23}
36964 & 0.85 &  & 0.97{\tiny$\pm$0.01} & 0.99{\tiny$\pm$0.01} & 1.00{\tiny$\pm$0.00} & 1.00{\tiny$\pm$0.00} & 1.00{\tiny$\pm$0.00} & 0.97 & 0.99 & 1.00 & 1.00 & 1.00 & 2.70{\tiny$\pm$0.08} & 4.22{\tiny$\pm$0.13} & 5.82{\tiny$\pm$0.16} & 7.33{\tiny$\pm$0.17} & 8.65{\tiny$\pm$0.19} & 2.70 & 4.28 & 5.77 & 7.22 & 8.67\\ \cline{1-2}\cline{4-23}

%% file: tex_files/8-example.tex
\section{Case Study: Anti-Money Laundering Alert Review}
\label{sec:example}

A bank's anti-money laundering (AML) compliance team faces a concrete design problem: an ML system flags suspicious transactions, and human investigators must review each alert and decide whether to escalate it or dismiss it as benign. The bank employs two pools of investigators---junior and senior---and wants to answer two questions: \emph{How should we configure the voting process to reach 98\% accuracy? And should we use junior investigators, senior investigators, or a mix of both?} In this section, we show how the theoretical framework developed in Sections~\ref{sec:characteristics}--\ref{sec:equiv} answers both questions directly, without pilot studies or trial-and-error experimentation.

\subsection{Dataset description}
\label{sec:aml_data}

We use a fully anonymized dataset from a real AML alert review process at a financial institution.\footnote{The dataset was provided by an anonymous collaborator who manages crowdsourcing/BPO systems for banks. The fully anonymized dataset is publicly available in the paper's supplementary materials. The investigators knew they were being evaluated on gold-standard items, but did not know which items were gold.} The dataset contains 1{,}000 alert items, each with a ground-truth label (escalate or benign), reviewed by 21 investigators at two experience levels: \emph{junior} (less experienced) and \emph{senior} (more experienced). Each investigator labeled all 1{,}000 items at each experience level, yielding $21 \times 2 \times 1{,}000 = 42{,}000$ total labels. The ground-truth base rate is 19.2\% positive (escalate), consistent with the reported ML system precision of 15--25\%.

\begin{figure}[t]
\centering
\includegraphics[width=0.85\columnwidth]{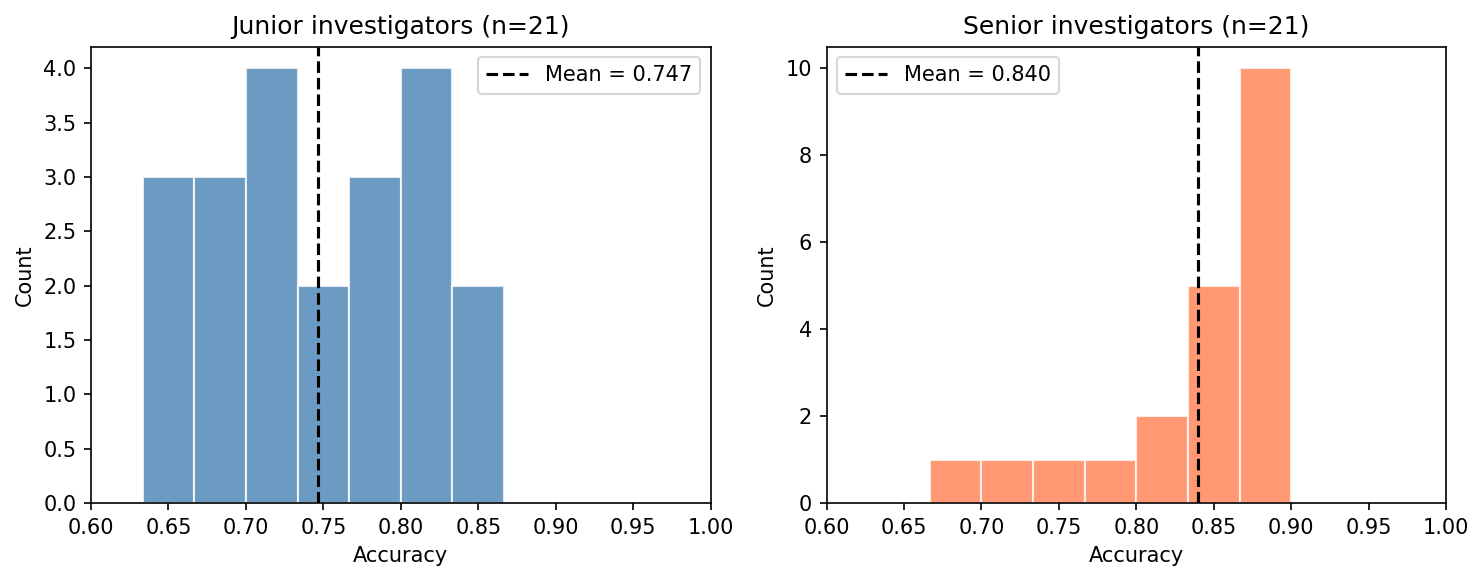}
\Description{Two histograms showing distribution of investigator accuracies in the AML dataset for junior and senior pools.}
\caption{Distribution of individual investigator accuracies in the AML dataset. Left: junior pool (mean $p = 0.747$). Right: senior pool (mean $p = 0.840$). Dashed lines indicate pool means.}
\label{fig:aml_accuracy}
\end{figure}

Table~\ref{tab:aml_pool_stats} summarizes the per-pool accuracy statistics. Figure~\ref{fig:aml_accuracy} shows the distribution of individual investigator accuracies within each pool.

\begin{table}[t]
\centering
\def\arraystretch{1.3}
\begin{tabular}{l c c c c c}
\hline
\textbf{Pool} & \textbf{$n$ investigators} & \textbf{Mean $p$} & \textbf{Mean $\varphi$} & \textbf{Range of $p$} & \textbf{Cost multiplier} \\
\hline
Junior & 21 & 0.747 & 2.95 & [0.641, 0.850] & $1\times$ \\
Senior & 21 & 0.840 & 5.25 & [0.684, 0.895] & $1.5\times$ \\
\hline
\end{tabular}
\caption{Pool-level accuracy statistics for the AML dataset. Mean accuracy $p$ is the average across investigators. The odds ratio $\varphi = p/(1-p)$. Senior investigators cost approximately 50\% more per vote.}
\label{tab:aml_pool_stats}
\end{table}

\subsection{Model validation on AML data}
\label{sec:aml_validation}

Before using the framework prescriptively, we first verify that it produces accurate predictions on this dataset, complementing the Bluebirds validation in Section~\ref{sec:theory_vs_exp}. For each of the 42 investigators (21 junior, 21 senior), we compute the oracle accuracy $p_i$ from all 1{,}000 labels and use it to derive theoretical quality $Q(\varphi_i, \delta)$ and expected votes $\mathbb{E}[n_{\mathit{votes}} | \varphi_i, \delta]$ via Theorems~\ref{th:Q_nonrand} and~\ref{th:ET}. We then simulate $\delta$-margin voting 500 times per investigator (sampling votes with replacement) and compare.

\begin{figure}[t]
\centering
\includegraphics[width=0.95\columnwidth]{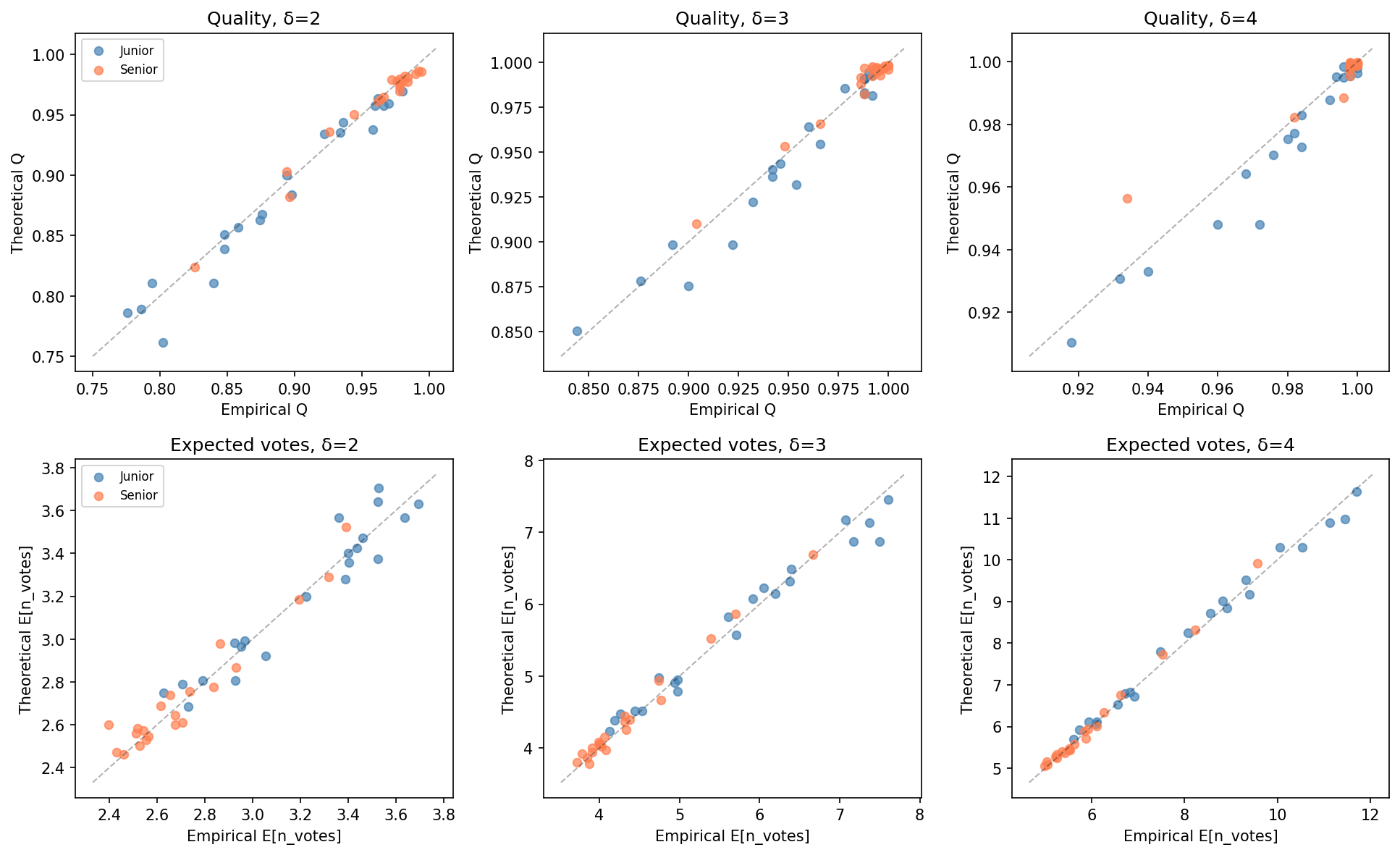}
\Description{Scatter plots of theoretical versus empirical quality and expected votes for AML investigators across multiple delta values.}
\caption{Theoretical vs.\ empirical quality $Q$ (top row) and expected votes $\mathbb{E}[n_{\mathit{votes}}]$ (bottom row) for each of the 42 investigators in the AML dataset, for $\delta \in \{2, 3, 4\}$. Blue: junior investigators; red: senior investigators. Dashed diagonal: perfect agreement.}
\label{fig:aml_theory_vs_emp}
\end{figure}

Table~\ref{tab:aml_theory_vs_emp} summarizes the results, averaged across investigators within each pool. The theoretical and empirical values agree closely: the mean absolute deviation is less than 0.005 for quality and less than 0.03 for expected votes across all configurations. Figure~\ref{fig:aml_theory_vs_emp} shows the per-investigator scatter plots.

\begin{table}[t]
\centering
\def\arraystretch{1.3}
\begin{tabular}{l c r r r r}
\hline
\textbf{Pool} & $\delta$ & $Q_{\text{th}}$ & $Q_{\text{emp}}$ & $\mathbb{E}[n]_{\text{th}}$ & $\mathbb{E}[n]_{\text{emp}}$ \\
\hline
Junior & 2 & 0.885 & 0.890 & 3.21 & 3.20 \\
Junior & 3 & 0.947 & 0.951 & 5.71 & 5.72 \\
Junior & 4 & 0.975 & 0.979 & 8.20 & 8.19 \\
Junior & 5 & 0.988 & 0.987 & 10.61 & 10.59 \\
\hline
Senior & 2 & 0.958 & 0.959 & 2.74 & 2.72 \\
Senior & 3 & 0.987 & 0.986 & 4.42 & 4.37 \\
Senior & 4 & 0.996 & 0.995 & 6.02 & 5.99 \\
Senior & 5 & 0.998 & 0.998 & 7.59 & 7.58 \\
\hline
\end{tabular}
\caption{Theory vs.\ empirical results for the AML dataset (known $p$, per-investigator). Theoretical values from Theorems~\ref{th:Q_nonrand} and~\ref{th:ET}; empirical values averaged over 500 simulation runs per investigator, then averaged across investigators within each pool.}
\label{tab:aml_theory_vs_emp}
\end{table}

\subsection{Answering the design questions}
\label{sec:aml_design}

We now apply the framework to answer the bank's two questions. Note that Section~\ref{sec:aml_validation} validated the formulas at the \emph{per-investigator} level (computing $Q(\varphi_i,\delta)$ for each investigator~$i$ and then averaging across the pool), whereas the design calculations below operate at the \emph{pool level}, treating each vote as drawn from a randomly selected investigator with the pool's mean accuracy~$\bar{p}$ (Assumption~A3). This explains why the numerical values below differ slightly from the per-investigator averages in Table~\ref{tab:aml_theory_vs_emp}.

\paragraph{Question 1: What $\delta$ is needed to reach 98\% accuracy?}
From Theorem~\ref{th:Q_nonrand}, the required $\delta$ satisfies $\varphi^{\delta} / (1 + \varphi^{\delta}) \geq Q^*$, which gives $\delta \geq \ln(Q^*/(1-Q^*)) / \ln(\varphi)$. For the two pools:

\begin{itemize}
    \item \textbf{Junior pool} ($\varphi = 2.95$): $\delta \geq \lceil 3.60 \rceil = 4$, yielding $Q(2.95, 4) = 0.987$ and $\mathbb{E}[n_{\mathit{votes}}] = 7.89$.
    \item \textbf{Senior pool} ($\varphi = 5.25$): $\delta \geq \lceil 2.35 \rceil = 3$, yielding $Q(5.25, 3) = 0.993$ and $\mathbb{E}[n_{\mathit{votes}}] = 4.35$.
\end{itemize}

\noindent Both configurations exceed the 98\% target (with 95\% bootstrap CIs of $[0.977,\, 0.993]$ for juniors and $[0.989,\, 0.996]$ for seniors), confirming the target is achievable with either pool.

\paragraph{Question 2: Junior or senior investigators?}
The choice depends on total cost per item, which combines the per-vote rate with the expected number of votes. Since senior investigators cost approximately $1.5\times$ per vote, the expected cost per item is:
\begin{align*}
\text{Junior (}\delta=4\text{):} &\quad 1.0 \times 7.89 = 7.89 \text{ units} \\
\text{Senior (}\delta=3\text{):} &\quad 1.5 \times 4.35 = 6.53 \text{ units}
\end{align*}

\noindent Despite the 50\% higher per-vote cost, the senior pool is 17\% \emph{cheaper} per item because it reaches the quality target with fewer votes ($\delta = 3$ vs.\ $\delta = 4$). This result is consistent with Theorem~\ref{theo:pay_eq}: the cost-equivalent payment ratio is $\textit{pay}(\varphi_{\text{junior}}) / \textit{pay}(\varphi_{\text{senior}}) = 0.47$, meaning that to equalize total cost at equal quality, juniors should be paid less than half what seniors are paid per vote. The bank's actual cost differential of $1.5\times$ is far less than the $2.1\times$ ratio that would equalize costs, so senior investigators offer better value.

\begin{figure}[t]
\centering
\includegraphics[width=0.7\columnwidth]{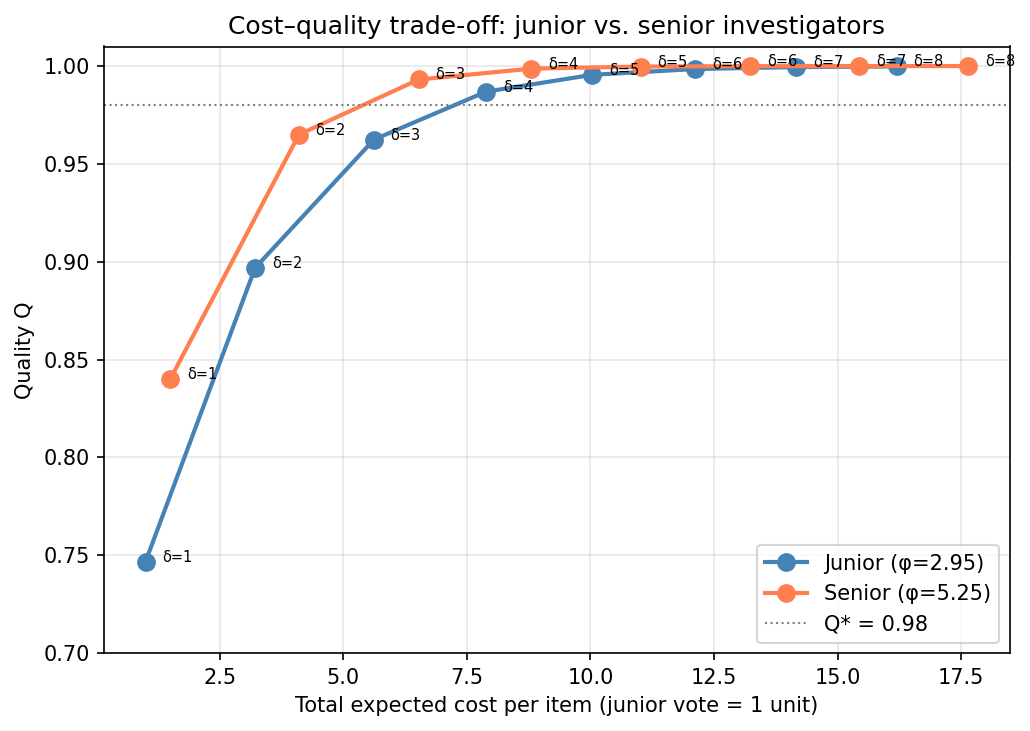}
\Description{Line plot of quality versus total cost per item for junior and senior investigator pools with a horizontal line at the 98 percent quality target.}
\caption{Cost--quality trade-off for junior ($\varphi = 2.95$) and senior ($\varphi = 5.25$) investigator pools. Each point is labeled with its $\delta$ value. Horizontal line: $Q^* = 0.98$ target. Total cost is in normalized units (junior vote = 1 unit; senior vote = 1.5 units).}
\label{fig:aml_cost_quality}
\end{figure}

Table~\ref{tab:aml_cost_quality} and Figure~\ref{fig:aml_cost_quality} summarize the cost--quality trade-off across a range of $\delta$ values for both pools. The framework also reveals an alternative: if the bank can tolerate $Q = 0.963$ (a slightly relaxed target), it can use juniors at $\delta = 3$ with $\mathbb{E}[n_{\mathit{votes}}] = 5.62$ and total cost of 5.62 units---the cheapest option of all.

\begin{table}[t]
\centering
\def\arraystretch{1.3}
\begin{tabular}{l c c c c}
\hline
\textbf{Pool} & $\delta$ & $Q$ & $\mathbb{E}[n_{\mathit{votes}}]$ & \textbf{Total cost} \\
\hline
Junior & 2 & 0.897 & 3.22 & 3.22 \\
Junior & 3 & 0.963 & 5.62 & 5.62 \\
Junior & 4 & \textbf{0.987} & \textbf{7.89} & \textbf{7.89} \\
Junior & 5 & 0.996 & 10.04 & 10.04 \\
\hline
Senior & 2 & 0.965 & 2.74 & 4.10 \\
Senior & 3 & \textbf{0.993} & \textbf{4.35} & \textbf{6.53} \\
Senior & 4 & 0.999 & 5.87 & 8.80 \\
\hline
\end{tabular}
\caption{Cost--quality trade-off for the AML dataset. Total cost is in normalized units (junior vote = 1 unit; senior vote = 1.5 units). Shaded rows indicate the minimum $\delta$ that meets the $Q^* = 0.98$ target for each pool.}
\label{tab:aml_cost_quality}
\end{table}

\subsection{Takeaway}
\label{sec:aml_takeaway}

This case study demonstrates the prescriptive value of the framework. The bank's original approach---running pilot studies with different $\delta$ values and pool configurations---would have required weeks of experimentation and thousands of labeled items. Our framework answers both design questions \emph{ex ante}, using only two inputs: an estimate of each pool's mean accuracy and the relative cost per vote. The key findings are:

\begin{enumerate}
    \item Both pools can achieve the 98\% quality target: juniors with $\delta = 4$ and seniors with $\delta = 3$ (Theorem~\ref{th:Q_nonrand}).
    \item The quality-equivalent threshold is $\delta_{\text{senior}} = \lceil \delta_{\text{junior}} \cdot \ln(\varphi_{\text{junior}}) / \ln(\varphi_{\text{senior}}) \rceil$ (Theorem~\ref{theo:delta}).
    \item Despite higher per-vote cost, the senior pool is 17\% cheaper per item at the 98\% quality target, because its higher accuracy requires fewer votes (Theorems~\ref{th:ET} and~\ref{theo:pay_eq}).
    \item The framework also enables \emph{dynamic routing}: using the Bayesian estimates from Section~\ref{sec:unknown_p}, items can be routed between pools based on real-time quality and cost estimates, assigning ambiguous items to the senior pool while routing easier items to the less expensive junior pool.
\end{enumerate}

%% file: tex_files/9-future.tex
\section{Conclusions, Practical Implications, and Future Work}
\label{sec:future}

\subsection{Summary}

Consensus voting is an essential technique in assuring quality for many AI pipelines in business and research. Due to its convenience and ease of understanding, simple majority voting is widely used to aggregate worker votes in consensus voting tasks. A variant of majority voting narrowly known in the crowdsourcing community is the $\delta$-margin voting -- a stronger, adaptive version of simple majority. However, it is commonly regarded as a heuristic approach without a comprehensive theoretical foundation. Our research aims to address this gap by proposing a model that represents the $\delta$-margin process as a Markov chain with absorbing states. In making the connection between known results in the Markov chain theory and $\delta$-majority voting, we aim to enable the missing analytical piece that allows voting processes to be designed with strong prior theoretical guarantees. AI practitioners can thus derive valuable insights into various properties, such as expected quality and time to consensus when designing pipelines that rely on consensus voting, where voting is performed either by human or LLM-based workers.

One significant advantage of our approach is its applicability to diverse settings. We accommodate heterogeneous worker qualities as long as the mean worker quality is known. Moreover, even without prior knowledge about worker quality, we demonstrate that the accumulated votes can be utilized to estimate worker quality. By leveraging our analytical results, AI pipelines can be designed to operate effectively under uncertainty.

We conducted experimental evaluations using real voting data to validate our theoretical findings. The results indicate that our estimates align well with practical outcomes, both in scenarios where worker quality is known and when it needs to be estimated dynamically. Our study contributes valuable insights into the understanding and practical implementation of $\delta$-margin majority voting.

\subsection{Practical Implications}


In critical decision-making scenarios where maintaining high levels of quality is crucial, our analysis can serve as a guiding tool during the voting process. For example, if our estimates predict a quality level below the acceptable threshold, the process owner can make informed decisions, such as increasing the delta parameter or potentially transitioning the annotation process to a different worker pool with higher expected accuracy.

Using our theoretical framework, consensus voting task designers can enhance their decision-making capabilities, optimize resource allocation, and ultimately improve the overall quality of the results. Our study empowers AI developers to take proactive, data-driven measures to ensure desired levels of quality and efficiency whenever a voting process is used.

Furthermore, in settings where individual votes come from human workers, our research includes guidelines for establishing payment schemes that align with performance. By considering the performance of individual workers, we can design a compensation structure that rewards their contributions accordingly. This approach ensures that workers who deliver high-quality results receive appropriate compensation.

When organizing workers in pools, based roughly on performance, we can offer differentiated payment schemes and even design schemes that balance required accuracy and cost. For example, easy tasks can be directed to low-accuracy and low-cost worker pools. When we detect that the item requires higher accuracy, we can escalate to a higher-quality worker pool.

\subsection{Future Work}

One limitation of our current framework is the symmetric noise assumption: we use a single accuracy parameter~$p = P(\text{vote is correct} \mid Y_i)$ regardless of the true label. In domains such as fraud detection or medical diagnostics, \emph{class-conditional error rates} (sensitivity~$\neq$ specificity) are the norm. When errors are asymmetric, the random walk transition probabilities depend on the (unknown) true label, and a single~$\delta$ threshold cannot simultaneously optimize for both error types. Extending the framework to class-conditional accuracy parameters ($p_+$ for positive items, $p_-$ for negative items) would require separate analyses per class and asymmetric absorbing boundaries, a natural direction for future work. The current symmetric model is most appropriate when class prevalence is roughly balanced or when sensitivity and specificity are approximately equal.

A second limitation is the focus on binary classification tasks. To expand the scope of our research, future work should aim to derive corresponding results applicable to non-binary classification tasks. One possible approach is a one-vs-rest decomposition, running a separate $\delta$-margin process for each class against the union of all others. This would allow us to extend our current theoretical framework to handle multi-class scenarios, though it would require class-specific accuracy assumptions (i.e., a separate $p$ for each binary sub-problem) and careful treatment of how the individual binary decisions are reconciled into a single multi-class label.

In addition, we can explore the application of our analysis to continuous tasks, such as regression problems. By providing theoretical estimates of quality and cost for such tasks, we can further enhance the understanding and optimization of voting processes in a broader range of applications.

Looking ahead, we plan to extend our analysis to other types of aggregation process, specifically investigating the interplay between quality, time to completion, and payment costs. Our goal is to develop a modular set of analyses that enable practitioners to input the structure of a task or process and obtain ex ante predictions of its behavior. This would provide valuable insights without the need to execute the process itself. For example, we can consider iterative tasks, where workers try to improve upon each other's work: Can we establish a framework that estimates the quality and the number of iterations until convergence? By investigating these aspects in future work, we aim to allow practitioners to infer cost-optimal parameters for multistage workflows in advance.

%% file: tex_files/appendices/A-MC_simulations_table.tex
\section{Monte Carlo estimates of the key targets}
\label{app:MC_sim_table}
Tables~\ref{tab:MC_alltargets_d2}, \ref{tab:MC_alltargets_d3}, \ref{tab:MC_alltargets_d4}, and \ref{tab:MC_alltargets_d5} list the results of Monte-Carlo simulations for the key quantities examined in this paper. All expectations are computed under a $\textit{Beta}(1, 1)$ prior for the item difficulty~$p$, \emph{restricted to $p > 0.5$} (i.e., conditioning on the majority class being correct; see Section~\ref{sec:pure-prior}). In particular, the expectation is estimated for the following quantities: the probability $Q$ of the correctness of the consensus vote, the expected number of votes $\mathbb{E}[n_{votes}]$ needed to reach a consensus from a given state, the variance of the latter $Var(n_{votes})$, and the payment for one worker pool. The columns $n_1$ and $n_2$ define the current state of the voting process---namely, the number of correct and incorrect votes, respectively. Note that in the fourth and fifth columns, $n_{votes}$ refers to the number of votes \emph{remaining} until the termination of the process, given the current state $(n_1, n_2).$ At terminal states (shown in \textbf{bold}), these columns are zero by definition---no further votes are needed. In contrast, $\mathbb{E}[Q]$ is the \emph{posterior probability} that the consensus label is correct, given the observed votes and the prior on~$p$. Because~$p$ is modeled as uncertain (not fixed at~1), $\mathbb{E}[Q]$ remains strictly less than~1 even at terminal states: the residual uncertainty reflects the possibility that the worker pool has moderate accuracy despite reaching consensus. The last column contains the payment rate for a single worker pool from Equation~\ref{equ:pay_p_beta}, evaluated at the posterior $\textit{Beta}(\alpha,\beta) = \textit{Beta}(1{+}n_1,\, 1{+}n_2)$ using the full formula $\textit{pay}(\alpha,\beta) \propto \tfrac{\alpha-\beta}{\alpha+\beta}\bigl(\psi(\alpha)-\psi(\beta)\bigr) + \tfrac{2}{\alpha+\beta}$. This quantity is independent of $\delta$. We compute the payment value for the terminal states only; values are meaningful only as ratios (see Section~\ref{sec:pay_worker_unknown_acc}).

\paragraph{Indexing convention.}
The tables list only states with $n_1 \geq n_2$, that is, states where the majority class has at least as many votes as the minority class. This is a natural indexing choice because the observable quantity during voting is the ordered pair $(n_{\max}, n_{\min})$, not which class is ``correct.'' During deployment, the practitioner maps $n_{\max} \to n_1$ and $n_{\min} \to n_2$ (under $H_c$) to look up the expected quality, cost, and variance in the table. Since the tables condition on $H_c$ (i.e., restrict the prior to $p > 0.5$), the mirror state $(n_2, n_1)$ with $n_2 > n_1$ would correspond to a different conditioning ($H_i$) and would generally yield different expected cost and variance, so one cannot infer mirror-state values by simply swapping indices. Under the \emph{full (unrestricted)} symmetric prior $\textit{Beta}(1,1)$ on $[0,1]$, the swap $(n_1, n_2) \leftrightarrow (n_2, n_1)$ is equivalent to replacing $p$ with $1-p$, which leaves cost and variance unchanged and gives complementary quality $\mathbb{E}[Q \mid n_2, n_1] = 1 - \mathbb{E}[Q \mid n_1, n_2]$. Under the truncated prior ($p > 0.5$), this full symmetry no longer holds. The reported $\mathbb{E}[Q]$ values at diagonal states (e.g., $\mathbb{E}[Q] = 0.847$ at $(0,0)$ for $\delta=2$) exceed $0.5$ precisely because of the truncation; under the full prior, diagonal states would give $\mathbb{E}[Q] = 0.5$ exactly.

\begin{table}[t]
    \centering
    \def\arraystretch{1.5}
    \begin{tabular}{c|c|c|c|c|c} 
    \hline
      \multicolumn{6}{c}{$\bm{\delta = 2}$}\\  
      \hline
      $n_1$  &
      $n_2$  &
      $\mathbf{\mathbb{E}[Q]}$  &  
      $\mathbf{\mathbb{E}[\mathbb{E}[n_{votes}]]}$  & 
      $\mathbf{\mathbb{E}[Var(n_{votes})]}$  &
      $\mathbf{pay^*}$ \\\hline\hline
      0  &  0  &  0.847  &  3.138  &  3.998  &  \\ 
      1  &  0  &  0.874  &  1.758  &  2.862  &  \\ 
      1  &  1  &  0.790  &  3.423  &  5.128  &  \\ 
      \textbf{2}  &  \textbf{0}  &  0.898  &  0.000  &  0.000  &  1.250  \\
      2  &  1  &  0.815  &  2.036  &  4.013  &  \\ 
      2  &  2  &  0.756  &  3.561  &  5.754  &  \\ 
      \textbf{3}  &  \textbf{1}  &  0.840  &  0.000  &  0.000  &  0.611  \\
      3  &  2  &  0.778  &  2.183  &  4.695  &  \\ 
      3  &  3  &  0.732  &  3.640  &  6.129  &  \\ 
      \textbf{4}  &  \textbf{2}  &  0.801  &  0.000  &  0.000  &  0.396  \\
      4  &  3  &  0.752  &  2.282  &  5.140  &  \\ 
      4  &  4  &  0.714  &  3.697  &  6.401  &  \\ 
      \textbf{5}  &  \textbf{3}  &  0.772  &  0.000  &  0.000  &  0.290  \\
      5  &  4  &  0.732  &  2.365  &  5.491  &  \\ 
      5  &  5  &  0.700  &  3.737  &  6.589  &  \\ 
      \hline
\end{tabular}
    \caption{Results of Monte Carlo estimation of the key quantities, when starting with a prior $\textit{Beta}(1, 1)$ restricted to $p > 0.5$ for the item difficulty $p$. For $\delta=2$. 
    }
    \label{tab:MC_alltargets_d2}
\end{table}
%
\begin{table}[ht!]
    \centering
    \def\arraystretch{1.3}
    \begin{tabular}{c|c|c|c|c|c} 
      \hline
      \multicolumn{6}{c}{$\bm{\delta = 3}$}\\  
      \hline
      $n_1$  &
      $n_2$  &
      $\mathbf{\mathbb{E}[Q]}$  &  
      $\mathbf{\mathbb{E}[\mathbb{E}[n_{votes}]]}$  & 
      $\mathbf{\mathbb{E}[Var(n_{votes})]}$  &
      $\mathbf{pay^*}$ \\\hline\hline
0  &  0  &  0.891  &  5.836  &  17.582  &  \\ 
1  &  0  &  0.915  &  3.978  &  12.988  &  \\ 
1  &  1  &  0.845  &  6.663  &  23.886  &  \\ 
2  &  0  &  0.935  &  1.935  &  6.917  &  \\ 
2  &  1  &  0.869  &  4.755  &  19.062  &  \\ 
2  &  2  &  0.815  &  7.119  &  27.623  &  \\ 
\textbf{3}  &  \textbf{0}  &  0.951  &  0.000  &  0.000  &  1.500  \\
3  &  1  &  0.892  &  2.409  &  11.096  &  \\ 
3  &  2  &  0.838  &  5.236  &  22.972  &  \\ 
3  &  3  &  0.793  &  7.408  &  30.275  &  \\ 
\textbf{4}  &  \textbf{1}  &  0.912  &  0.000  &  0.000  &  0.750  \\
4  &  2  &  0.860  &  2.719  &  14.168  &  \\ 
4  &  3  &  0.814  &  5.559  &  25.806  &  \\ 
4  &  4  &  0.775  &  7.606  &  32.138  &  \\ 
\textbf{5}  &  \textbf{2}  &  0.880  &  0.000  &  0.000  &  0.483  \\
5  &  3  &  0.834  &  2.933  &  16.482  &  \\ 
5  &  4  &  0.794  &  5.799  &  27.956  &  \\ 
5  &  5  &  0.760  &  7.771  &  33.694  &  \\ 
\textbf{6}  &  \textbf{3}  &  0.854  &  0.000  &  0.000  &  0.350  \\
6  &  4  &  0.813  &  3.106  &  18.239  &  \\ 
6  &  5  &  0.778  &  5.989  &  29.810  &  \\ 
6  &  6  &  0.748  &  7.897  &  34.921  &  \\ 
\textbf{7}  &  \textbf{4}  &  0.833  &  0.000  &  0.000  &  0.271  \\
7  &  5  &  0.797  &  3.241  &  19.700  &  \\ 
7  &  6  &  0.765  &  6.137  &  31.006  &  \\ 
7  &  7  &  0.737  &  7.988  &  35.976  &  \\ 
      \hline
\end{tabular}
    \caption{Results of Monte Carlo estimation of the key quantities, when starting with a prior $\textit{Beta}(1, 1)$ restricted to $p > 0.5$ for the item difficulty $p$. For $\delta=3$. 
    }
    \label{tab:MC_alltargets_d3}
\end{table}

%
\begin{table}[t]
    \def\arraystretch{1.6}
    \begin{tabular}{cc} 
    \footnotesize
    \hspace{-2cm}
    \raisebox{-\height}{\bgroup
    \def\arraystretch{1.2}
    \begin{tabular}{c|c|c|c|c|c}
      \hline
      \multicolumn{6}{c}{$\bm{\delta = 4}$}\\  
      \hline
      $n_1$  &
      $n_2$  &
      $\mathbf{\mathbb{E}[Q]}$  &  
      $\mathbf{\mathbb{E}[\mathbb{E}[n_{votes}]]}$  & 
      $\mathbf{\mathbb{E}[Var(n_{votes})]}$  &
      $\mathbf{pay^*}$ \\\hline\hline
0  &  0  &  0.917  &  8.915  &  45.960  &  \\ 
1  &  0  &  0.937  &  6.501  &  33.656  &  \\ 
1  &  1  &  0.879  &  10.455  &  64.657  &  \\ 
2  &  0  &  0.953  &  4.078  &  21.512  & \\ 
2  &  1  &  0.900  &  7.979  &  51.128  &  \\ 
2  &  2  &  0.853  &  11.383  &  76.509  &  \\ 
3  &  0  &  0.966  &  1.903  &  9.857  &  \\ 
3  &  1  &  0.920  &  5.199  &  35.158  &  \\ 
3  &  2  &  0.874  &  8.910  &  62.978  &  \\ 
3  &  3  &  0.833  &  11.948  &  84.664  &  \\ 
\textbf{4}  &  \textbf{0}  &  0.976  &  0.000  &  0.000  &  1.722  \\
4  &  1  &  0.937  &  2.469  &  17.644  &  \\ 
4  &  2  &  0.894  &  5.932  &  45.913  &  \\ 
4  &  3  &  0.853  &  9.555  &  71.839  &  \\ 
4  &  4  &  0.817  &  12.395  &  90.737  &  \\ 
\textbf{5}  &  \textbf{1}  &  0.951  &  0.000  &  0.000  &  0.892  \\
5  &  2  &  0.912  &  2.871  &  24.364  &  \\ 
5  &  3  &  0.873  &  6.474  &  53.891  &  \\ 
5  &  4  &  0.836  &  10.012  &  78.803  &  \\ 
5  &  5  &  0.803  &  12.727  &  96.691  &  \\ 
\hline
\end{tabular}\egroup}
      &
      %
    \raisebox{-\height}{\bgroup
    \def\arraystretch{1.3}
    \begin{tabular}{c|c|c|c|c|c}
    \hline
      \multicolumn{6}{c}{$\bm{\delta = 4}$}\\  
      \hline
      $n_1$  &
      $n_2$  &
      $\mathbf{\mathbb{E}[Q]}$  &  
      $\mathbf{\mathbb{E}[\mathbb{E}[n_{votes}]]}$  & 
      $\mathbf{\mathbb{E}[Var(n_{votes})]}$  &
      $\mathbf{pay^*}$ \\\hline\hline
\textbf{6}  &  \textbf{2}  &  0.928  &  0.000  &  0.000  &  0.580  \\
6  &  3  &  0.891  &  3.198  &  29.625  &  \\ 
6  &  4  &  0.855  &  6.885  &  60.942  &  \\ 
6  &  5  &  0.821  &  10.382  &  84.901  &  \\ 
6  &  6  &  0.791  &  13.031  &  100.967  &  \\ 
\textbf{7}  &  \textbf{3}  &  0.908  &  0.000  &  0.000  &  0.420  \\
7  &  4  &  0.873  &  3.442  &  34.271  &  \\ 
7  &  5  &  0.839  &  7.253  &  66.604  &  \\ 
7  &  6  &  0.809  &  10.736  &  89.723  &  \\ 
7  &  7  &  0.781  &  13.261  &  104.608  &  \\ 
\textbf{8}  &  \textbf{4}  &  0.890  &  0.000  &  0.000  &  0.324  \\
8  &  5  &  0.867  &  3.656  &  37.907  &  \\ 
8  &  6  &  0.826  &  7.518  &  71.345  &  \\ 
8  &  7  &  0.797  &  10.998  &  94.118  &  \\ 
8  &  8  &  0.772  &  13.463  &  107.939  &  \\ 
\textbf{9}  &  \textbf{5}  &  0.874  &  0.000  &  0.000  &  0.261  \\
9  &  6  &  0.843  &  3.831  &  41.596  &  \\ 
9  &  7  &  0.814  &  7.765  &  75.524  &  \\ 
9  &  8  &  0.787  &  11.237  &  97.239  &  \\ 
9  &  9  &  0.764  &  13.602  &  110.709  &  \\ 
 \hline
      \end{tabular}\egroup}
\end{tabular}
    \caption{Results of Monte Carlo estimation of the key quantities, when starting with a prior $\textit{Beta}(1, 1)$ restricted to $p > 0.5$ for the item difficulty $p$. For $\delta=4$. 
    }
    \label{tab:MC_alltargets_d4}
\end{table}
%
\begin{table}[t]
    \centering
    \def\arraystretch{1.7}
    \begin{tabular}{cc}
    \footnotesize
    \hspace{-2cm}
    \raisebox{-\height}{\bgroup
    \def\arraystretch{1.3}
    \begin{tabular}{c|c|c|c|c|c}
      \hline
      \multicolumn{6}{c}{$\bm{\delta = 5}$}\\  
      \hline
      $n_1$  &
      $n_2$  &
      $\mathbf{\mathbb{E}[Q]}$  &  
      $\mathbf{\mathbb{E}[\mathbb{E}[n_{votes}]]}$  & 
      $\mathbf{\mathbb{E}[Var(n_{votes})]}$  &
      $\mathbf{pay^*}$ \\\hline\hline
0  &  0  &  0.933  &  12.145  &  94.191  &  \\ 
1  &  0  &  0.950  &  9.187  &  68.850  &  \\ 
1  &  1  &  0.901  &  14.659  &  135.330  &  \\ 
2  &  0  &  0.964  &  6.393  &  45.822  &  \\ 
2  &  1  &  0.920  &  11.570  &  105.911  &  \\ 
2  &  2  &  0.879  &  16.055  &  161.141  &  \\ 
3  &  0  &  0.975  &  3.909  &  26.198  &  \\ 
3  &  1  &  0.937  &  8.267  &  77.027  &  \\ 
3  &  2  &  0.898  &  12.966  &  132.362  &  \\ 
3  &  3  &  0.861  &  17.097  &  181.086  &  \\ 
4  &  0  &  0.982  &  1.783  &  10.863  &  \\ 
4  &  1  &  0.952  &  5.146  &  48.012  &  \\ 
4  &  2  &  0.916  &  9.500  &  101.607  &  \\ 
4  &  3  &  0.880  &  14.044  &  152.959  &  \\ 
4  &  4  &  0.847  &  17.821  &  197.148  &  \\ 
\textbf{5}  &  \textbf{0}  &  0.988  &  0.000  &  0.000  &  1.917  \\
5  &  1  &  0.963  &  2.381  &  21.910  &  \\ 
5  &  2  &  0.932  &  6.080  &  66.724  &  \\ 
5  &  3  &  0.898  &  10.485  &  120.725  &  \\ 
5  &  4  &  0.865  &  14.823  &  169.516  &  \\ 
5  &  5  &  0.834  &  18.422  &  209.422  &  \\ 
\textbf{6}  &  \textbf{1}  &  0.973  &  0.000  &  0.000  &  1.028  \\
6  &  2  &  0.945  &  2.842  &  31.572  &  \\ 
6  &  3  &  0.914  &  6.760  &  82.731  &  \\ 
6  &  4  &  0.882  &  11.203  &  136.651  & \\ 
6  &  5  &  0.851  &  15.509  &  182.890  &  \\ 
6  &  6  &  0.823  &  18.978  &  221.973  &  \\ 
\textbf{7}  &  \textbf{2}  &  0.957  &  0.000  &  0.000  &  0.679  \\
7  &  3  &  0.929  &  3.238  &  40.730  &  \\ 
7  &  4  &  0.899  &  7.350  &  95.839  &  \\ 
7  &  5  &  0.869  &  11.855  &  151.603  & \\ 
7  &  6  &  0.840  &  16.051  &  195.441  & \\ 
7  &  7  &  0.814  &  19.336  &  230.199  & \\ 
\hline
      \end{tabular}\egroup}
      &
      %
    \raisebox{-\height}{\bgroup
    \def\arraystretch{1.3}
    \begin{tabular}{c|c|c|c|c|c}
      \hline
      \multicolumn{6}{c}{$\bm{\delta = 5}$}\\  
      \hline
      $n_1$  &
      $n_2$  &
      $\mathbf{\mathbb{E}[Q]}$  &  
      $\mathbf{\mathbb{E}[\mathbb{E}[n_{votes}]]}$  & 
      $\mathbf{\mathbb{E}[Var(n_{votes})]}$  &
      $\mathbf{pay^*}$ \\\hline\hline  
\textbf{8}  &  \textbf{3}  &  0.942  &  0.000  &  0.000  &  0.494  \\
8  &  4  &  0.914  &  3.556  &  49.011  &  \\ 
8  &  5  &  0.885  &  7.842  &  106.722  & \\ 
8  &  6  &  0.856  &  12.360  &  162.937  &  \\ 
8  &  7  &  0.830  &  16.518  &  205.669  &  \\ 
8  &  8  &  0.805  &  19.708  &  239.370  &  \\ 
\textbf{9}  &  \textbf{4}  &  0.928  &  0.000  &  0.000  &  0.382  \\
9  &  5  &  0.901  &  3.789  &  55.714  &  \\ 
9  &  6  &  0.873  &  8.251  &  118.670  &  \\ 
9  &  7  &  0.846  &  12.771  &  172.994  & \\ 
9  &  8  &  0.820  &  16.854  &  215.285  & \\ 
9  &  9  &  0.797  &  20.008  &  247.119  & \\ 
\textbf{10}  &  \textbf{5}  &  0.915  &  0.000  &  0.000  &  0.308  \\
10  &  6  &  0.888  &  4.022  &  62.404  &  \\ 
10  &  7  &  0.861  &  8.588  &  126.756  &  \\ 
10  &  8  &  0.836  &  13.141  &  181.694  &  \\ 
10  &  9  &  0.812  &  17.251  &  223.283  &  \\ 
10  &  10  &  0.790  &  20.295  &  253.224  &  \\ 
\textbf{11}  &  \textbf{6}  &  0.903  &  0.000  &  0.000  &  0.255  \\
11  &  7  &  0.877  &  4.239  &  68.460  &  \\ 
11  &  8  &  0.851  &  8.894  &  134.953  &  \\ 
11  &  9  &  0.827  &  13.505  &  190.358  &  \\ 
11  &  10  &  0.804  &  17.536  &  231.317  &  \\ 
11  &  11  &  0.783  &  20.540  &  258.980  &  \\ 
     \hline
      \end{tabular}\egroup}
\end{tabular}
    \caption{Results of Monte Carlo estimation of the key quantities, when starting with a prior $\textit{Beta}(1, 1)$ restricted to $p > 0.5$ for the item difficulty $p$. For $\delta=5$. 
    }
    \label{tab:MC_alltargets_d5}
\end{table}
\clearpage


\subsection{Decision tables for informative priors}
\label{app:informative_prior_tables}

The preceding tables used the uninformative $\textit{Beta}(1,1)$ prior. In practice, a requester often has prior knowledge that the worker pool performs above chance---for instance, from historical data, qualification tests, or domain experience. This subsection provides analogous decision tables for two informative priors:
\begin{itemize}
    \item $\textit{Beta}(2,1)$: a mild prior belief that workers are above chance ($\mathbb{E}[p] = 2/3$, moderate concentration);
    \item $\textit{Beta}(3,1)$: a stronger prior belief ($\mathbb{E}[p] = 3/4$, tighter concentration around higher accuracy).
\end{itemize}
\noindent All expectations are again restricted to $p > 0.5$ (conditioning on the majority class being correct; see Section~\ref{sec:pure-prior}), and computed via numerical integration over the posterior $\textit{Beta}(\alpha+n_1, \beta+n_2)$ truncated to $[0.5, 1]$. For these asymmetric priors, the symmetry argument from the $\textit{Beta}(1,1)$ case no longer applies exactly: swapping $n_1 \leftrightarrow n_2$ does not produce complementary quality because the prior is not symmetric in $p \mapsto 1-p$. Nevertheless, we continue to list only states with $n_1 \geq n_2$, since these correspond to the practically relevant regime where the majority class has received at least as many votes.

\paragraph{Effect of the informative prior.} Comparing across the three prior settings reveals the practical value of prior knowledge. At the start state $(0,0)$ with $\delta = 2$: the expected quality is $\mathbb{E}[Q] = 0.847$ under $\textit{Beta}(1,1)$, $0.874$ under $\textit{Beta}(2,1)$, and $0.898$ under $\textit{Beta}(3,1)$. The expected number of remaining votes decreases correspondingly ($3.14 \to 3.02 \to 2.90$), reflecting the fact that a stronger prior concentrating mass on higher $p$ yields higher quality estimates and shorter expected voting time. This effect is even more pronounced at larger $\delta$: for $\delta = 5$, the start-state quality improves from $0.933$ to $0.950$ to $0.964$, and the expected votes decrease from $12.1$ to $11.0$ to $10.0$.

\paragraph{Payment values.} All payment values (here and in the $\textit{Beta}(1,1)$ tables above) use the full formula from Equation~\ref{equ:pay_p_beta}, including the $+2/(\alpha{+}\beta)$ covariance correction term. Because the informative priors have $\alpha \neq \beta$ even before observing any votes, the payment values at the first terminal state differ from the $\textit{Beta}(1,1)$ tables (e.g., for posterior $(\alpha,\beta)=(4,1)$: $\textit{pay}^* = 1.50$ under both $\textit{Beta}(1,1)$ with $(n_1,n_2)=(3,0)$ and $\textit{Beta}(2,1)$ with $(n_1,n_2)=(2,0)$, since the posterior is the same). As always, these values are meaningful only as \emph{ratios}---see Section~\ref{sec:pay_worker_unknown_acc}.

\vspace{0.25in}
\noindent\textbf{Tables for $\textit{Beta}(2,1)$ prior} \nopagebreak

\input{tex_files/appendices/A-MC_tables_beta21}

\vspace{0.25in}
\noindent\textbf{Tables for $\textit{Beta}(3,1)$ prior} \nopagebreak

\input{tex_files/appendices/A-MC_tables_beta31}

%% file: tex_files/appendices/A-MC_tables_beta21.tex

\begin{table}[t]
    \centering
    \def\arraystretch{1.5}
    \begin{tabular}{c|c|c|c|c|c} 
    \hline
      \multicolumn{6}{c}{$\bm{\delta = 2}$}\\  
      \hline
      $n_1$  &
      $n_2$  &
      $\mathbf{\mathbb{E}[Q]}$  &  
      $\mathbf{\mathbb{E}[\mathbb{E}[n_{votes}]]}$  & 
      $\mathbf{\mathbb{E}[Var(n_{votes})]}$  &
      $\mathbf{pay^*}$ \\\hline\hline
0  &  0  &  0.874  &  3.019  &  3.485  &   \\ 
1  &  0  &  0.898  &  1.655  &  2.435  &   \\ 
1  &  1  &  0.815  &  3.334  &  4.738  &   \\ 
\textbf{2}  &  \textbf{0}  &  0.919  &  0.000  &  0.000  &    1.500 \\ 
2  &  1  &  0.840  &  1.940  &  3.621  &   \\ 
2  &  2  &  0.778  &  3.493  &  5.424  &   \\ 
\textbf{3}  &  \textbf{1}  &  0.863  &  0.000  &  0.000  &    0.750 \\ 
3  &  2  &  0.801  &  2.108  &  4.351  &   \\ 
3  &  3  &  0.752  &  3.590  &  5.865  &   \\ 
\textbf{4}  &  \textbf{2}  &  0.823  &  0.000  &  0.000  &    0.483 \\ 
4  &  3  &  0.772  &  2.221  &  4.852  &   \\ 
4  &  4  &  0.732  &  3.655  &  6.174  &   \\ 
\textbf{5}  &  \textbf{3}  &  0.792  &  0.000  &  0.000  &    0.350 \\ 
5  &  4  &  0.750  &  2.303  &  5.220  &   \\ 
5  &  5  &  0.716  &  3.703  &  6.404  &   \\ 
      \hline
\end{tabular}
    \vspace{0.5cm}
    \caption{Decision table for a $\textit{Beta}(2, 1)$ prior restricted to $p > 0.5$. For $\delta=2$. 
    }
    \label{tab:MC_alltargets_d2_beta21}
\end{table}
\newpage
\begin{table}[t]
    \centering
    \def\arraystretch{1.5}
    \begin{tabular}{c|c|c|c|c|c} 
    \hline
      \multicolumn{6}{c}{$\bm{\delta = 3}$}\\  
      \hline
      $n_1$  &
      $n_2$  &
      $\mathbf{\mathbb{E}[Q]}$  &  
      $\mathbf{\mathbb{E}[\mathbb{E}[n_{votes}]]}$  & 
      $\mathbf{\mathbb{E}[Var(n_{votes})]}$  &
      $\mathbf{pay^*}$ \\\hline\hline
0  &  0  &  0.915  &  5.457  &  14.651  &   \\ 
1  &  0  &  0.935  &  3.677  &  10.459  &   \\ 
1  &  1  &  0.869  &  6.360  &  21.246  &   \\ 
2  &  0  &  0.951  &  1.777  &  5.430  &   \\ 
2  &  1  &  0.892  &  4.484  &  16.506  &   \\ 
2  &  2  &  0.838  &  6.867  &  25.361  &   \\ 
\textbf{3}  &  \textbf{0}  &  0.964  &  0.000  &  0.000  &    1.722 \\ 
3  &  1  &  0.912  &  2.237  &  9.396  &   \\ 
3  &  2  &  0.859  &  4.987  &  20.633  &   \\ 
3  &  3  &  0.814  &  7.200  &  28.261  &   \\ 
\textbf{4}  &  \textbf{1}  &  0.929  &  0.000  &  0.000  &    0.892 \\ 
4  &  2  &  0.880  &  2.552  &  12.402  &   \\ 
4  &  3  &  0.834  &  5.340  &  23.684  &   \\ 
4  &  4  &  0.794  &  7.438  &  30.445  &   \\ 
\textbf{5}  &  \textbf{2}  &  0.899  &  0.000  &  0.000  &    0.580 \\ 
5  &  3  &  0.854  &  2.785  &  14.769  &   \\ 
5  &  4  &  0.814  &  5.604  &  26.055  &   \\ 
5  &  5  &  0.778  &  7.619  &  32.163  &   \\ 
\textbf{6}  &  \textbf{3}  &  0.873  &  0.000  &  0.000  &    0.420 \\ 
6  &  4  &  0.833  &  2.967  &  16.692  &   \\ 
6  &  5  &  0.797  &  5.811  &  27.961  &   \\ 
6  &  6  &  0.765  &  7.760  &  33.556  &   \\ 
\textbf{7}  &  \textbf{4}  &  0.852  &  0.000  &  0.000  &    0.324 \\ 
7  &  5  &  0.815  &  3.113  &  18.290  &   \\ 
7  &  6  &  0.782  &  5.978  &  29.534  &   \\ 
7  &  7  &  0.753  &  7.875  &  34.712  &   \\ 
      \hline
\end{tabular}
    \vspace{0.5cm}
    \caption{Decision table for a $\textit{Beta}(2, 1)$ prior restricted to $p > 0.5$. For $\delta=3$. 
    }
    \label{tab:MC_alltargets_d3_beta21}
\end{table}
\newpage
\begin{table}[t]
    \def\arraystretch{1.6}
    \begin{tabular}{cc} 
    \footnotesize
    \hspace{-2cm}
    \raisebox{-\height}{\bgroup
    \def\arraystretch{1.2}
    \begin{tabular}{c|c|c|c|c|c}
      \hline
      \multicolumn{6}{c}{$\bm{\delta = 4}$}\\  
      \hline
      $n_1$  &
      $n_2$  &
      $\mathbf{\mathbb{E}[Q]}$  &  
      $\mathbf{\mathbb{E}[\mathbb{E}[n_{votes}]]}$  & 
      $\mathbf{\mathbb{E}[Var(n_{votes})]}$  &
      $\mathbf{pay^*}$ \\\hline\hline
0  &  0  &  0.937  &  8.145  &  37.116  &   \\ 
1  &  0  &  0.953  &  5.912  &  26.449  &   \\ 
1  &  1  &  0.900  &  9.815  &  55.744  &   \\ 
2  &  0  &  0.966  &  3.714  &  16.362  &   \\ 
2  &  1  &  0.920  &  7.400  &  43.218  &   \\ 
2  &  2  &  0.874  &  10.809  &  68.211  &   \\ 
3  &  0  &  0.976  &  1.732  &  7.309  &   \\ 
3  &  1  &  0.937  &  4.775  &  29.203  &   \\ 
3  &  2  &  0.894  &  8.366  &  55.376  &   \\ 
3  &  3  &  0.853  &  11.492  &  77.468  &   \\ 
\textbf{4}  &  \textbf{0}  &  0.983  &  0.000  &  0.000  &    1.917 \\ 
4  &  1  &  0.951  &  2.260  &  14.239  &   \\ 
4  &  2  &  0.912  &  5.524  &  39.402  &   \\ 
4  &  3  &  0.873  &  9.064  &  64.793  &   \\ 
4  &  4  &  0.836  &  11.999  &  84.739  &   \\ 
\textbf{5}  &  \textbf{1}  &  0.962  &  0.000  &  0.000  &    1.028 \\ 
5  &  2  &  0.928  &  2.662  &  20.290  &   \\ 
5  &  3  &  0.891  &  6.091  &  47.729  &   \\ 
5  &  4  &  0.855  &  9.600  &  72.390  &   \\ 
5  &  5  &  0.821  &  12.394  &  90.662  &   \\ 
 \hline
      \end{tabular}
\egroup}
      &
      %
    \raisebox{-\height}{\bgroup
    \def\arraystretch{1.3}
    \begin{tabular}{c|c|c|c|c|c}
      \hline
      \multicolumn{6}{c}{$\bm{\delta = 4}$}\\  
      \hline
      $n_1$  &
      $n_2$  &
      $\mathbf{\mathbb{E}[Q]}$  &  
      $\mathbf{\mathbb{E}[\mathbb{E}[n_{votes}]]}$  & 
      $\mathbf{\mathbb{E}[Var(n_{votes})]}$  &
      $\mathbf{pay^*}$ \\\hline\hline
\textbf{6}  &  \textbf{2}  &  0.942  &  0.000  &  0.000  &    0.679 \\ 
6  &  3  &  0.908  &  2.982  &  25.528  &   \\ 
6  &  4  &  0.873  &  6.540  &  54.689  &   \\ 
6  &  5  &  0.839  &  10.028  &  78.695  &   \\ 
6  &  6  &  0.809  &  12.712  &  95.613  &   \\ 
\textbf{7}  &  \textbf{3}  &  0.923  &  0.000  &  0.000  &    0.494 \\ 
7  &  4  &  0.890  &  3.244  &  30.088  &   \\ 
7  &  5  &  0.857  &  6.908  &  60.617  &   \\ 
7  &  6  &  0.826  &  10.380  &  84.037  &   \\ 
7  &  7  &  0.797  &  12.974  &  99.830  &   \\ 
\textbf{8}  &  \textbf{4}  &  0.906  &  0.000  &  0.000  &    0.382 \\ 
8  &  5  &  0.874  &  3.463  &  34.091  &   \\ 
8  &  6  &  0.843  &  7.217  &  65.742  &   \\ 
8  &  7  &  0.814  &  10.677  &  88.638  &   \\ 
8  &  8  &  0.787  &  13.196  &  103.479  &   \\ 
\textbf{9}  &  \textbf{5}  &  0.890  &  0.000  &  0.000  &    0.308 \\ 
9  &  6  &  0.860  &  3.650  &  37.636  &   \\ 
9  &  7  &  0.830  &  7.480  &  70.229  &   \\ 
9  &  8  &  0.803  &  10.930  &  92.652  &   \\ 
9  &  9  &  0.778  &  13.385  &  106.674  &   \\ 
 \hline
      \end{tabular}
\egroup}
\end{tabular}
    \vspace{0.5cm}
    \caption{Decision table for a $\textit{Beta}(2, 1)$ prior restricted to $p > 0.5$. For $\delta=4$. 
    }
    \label{tab:MC_alltargets_d4_beta21}
\end{table}
\newpage
\begin{table}[t]
    \def\arraystretch{1.6}
    \begin{tabular}{cc} 
    \footnotesize
    \hspace{-2cm}
    \raisebox{-\height}{\bgroup
    \def\arraystretch{1.2}
    \begin{tabular}{c|c|c|c|c|c}
      \hline
      \multicolumn{6}{c}{$\bm{\delta = 5}$}\\  
      \hline
      $n_1$  &
      $n_2$  &
      $\mathbf{\mathbb{E}[Q]}$  &  
      $\mathbf{\mathbb{E}[\mathbb{E}[n_{votes}]]}$  & 
      $\mathbf{\mathbb{E}[Var(n_{votes})]}$  &
      $\mathbf{pay^*}$ \\\hline\hline
0  &  0  &  0.950  &  11.012  &  74.371  &   \\ 
1  &  0  &  0.964  &  8.293  &  52.654  &   \\ 
1  &  1  &  0.920  &  13.573  &  114.103  &   \\ 
2  &  0  &  0.975  &  5.757  &  34.176  &   \\ 
2  &  1  &  0.937  &  10.575  &  87.949  &   \\ 
2  &  2  &  0.898  &  15.158  &  141.843  &   \\ 
3  &  0  &  0.982  &  3.538  &  19.157  &   \\ 
3  &  1  &  0.952  &  7.519  &  62.305  &   \\ 
3  &  2  &  0.916  &  12.101  &  114.540  &   \\ 
3  &  3  &  0.880  &  16.279  &  163.130  &   \\ 
4  &  0  &  0.988  &  1.636  &  7.821  &   \\ 
4  &  1  &  0.963  &  4.676  &  38.077  &   \\ 
4  &  2  &  0.932  &  8.789  &  85.386  &   \\ 
4  &  3  &  0.898  &  13.228  &  135.767  &   \\ 
4  &  4  &  0.865  &  17.131  &  180.313  &   \\ 
\textbf{5}  &  \textbf{0}  &  0.992  &  0.000  &  0.000  &    2.087 \\ 
5  &  1  &  0.973  &  2.165  &  16.876  &   \\ 
5  &  2  &  0.945  &  5.560  &  55.052  &   \\ 
5  &  3  &  0.914  &  9.768  &  104.721  &   \\ 
5  &  4  &  0.882  &  14.108  &  153.328  &   \\ 
5  &  5  &  0.851  &  17.809  &  194.641  &   \\ 
\textbf{6}  &  \textbf{1}  &  0.980  &  0.000  &  0.000  &    1.156 \\ 
6  &  2  &  0.957  &  2.604  &  25.758  &   \\ 
6  &  3  &  0.929  &  6.271  &  70.063  &   \\ 
6  &  4  &  0.899  &  10.556  &  121.232  &   \\ 
6  &  5  &  0.868  &  14.823  &  168.220  &   \\ 
6  &  6  &  0.840  &  18.364  &  206.861  &   \\ 
\textbf{7}  &  \textbf{2}  &  0.967  &  0.000  &  0.000  &    0.776 \\ 
 \hline
      \end{tabular}
\egroup}
      &
      %
    \raisebox{-\height}{\bgroup
    \def\arraystretch{1.3}
    \begin{tabular}{c|c|c|c|c|c}
      \hline
      \multicolumn{6}{c}{$\bm{\delta = 5}$}\\  
      \hline
      $n_1$  &
      $n_2$  &
      $\mathbf{\mathbb{E}[Q]}$  &  
      $\mathbf{\mathbb{E}[\mathbb{E}[n_{votes}]]}$  & 
      $\mathbf{\mathbb{E}[Var(n_{votes})]}$  &
      $\mathbf{pay^*}$ \\\hline\hline
7  &  3  &  0.942  &  2.973  &  34.074  &   \\ 
7  &  4  &  0.914  &  6.861  &  83.362  &   \\ 
7  &  5  &  0.885  &  11.209  &  135.555  &   \\ 
7  &  6  &  0.856  &  15.418  &  181.082  &   \\ 
7  &  7  &  0.830  &  18.830  &  217.464  &   \\ 
\textbf{8}  &  \textbf{3}  &  0.954  &  0.000  &  0.000  &    0.570 \\ 
8  &  4  &  0.928  &  3.289  &  41.744  &   \\ 
8  &  5  &  0.901  &  7.360  &  95.213  &   \\ 
8  &  6  &  0.873  &  11.762  &  148.139  &   \\ 
8  &  7  &  0.846  &  15.925  &  192.348  &   \\ 
8  &  8  &  0.820  &  19.229  &  226.785  &   \\ 
\textbf{9}  &  \textbf{4}  &  0.941  &  0.000  &  0.000  &    0.442 \\ 
9  &  5  &  0.915  &  3.562  &  48.787  &   \\ 
9  &  6  &  0.888  &  7.791  &  105.845  &   \\ 
9  &  7  &  0.861  &  12.239  &  159.313  &   \\ 
9  &  8  &  0.836  &  16.363  &  202.329  &   \\ 
9  &  9  &  0.812  &  19.574  &  235.069  &   \\ 
\textbf{10}  &  \textbf{5}  &  0.928  &  0.000  &  0.000  &    0.357 \\ 
10  &  6  &  0.903  &  3.801  &  55.257  &   \\ 
10  &  7  &  0.877  &  8.167  &  115.446  &   \\ 
10  &  8  &  0.851  &  12.655  &  169.323  &   \\ 
10  &  9  &  0.827  &  16.746  &  211.254  &   \\ 
10  &  10  &  0.804  &  19.877  &  242.496  &   \\ 
\textbf{11}  &  \textbf{6}  &  0.916  &  0.000  &  0.000  &    0.296 \\ 
11  &  7  &  0.891  &  4.014  &  61.210  &   \\ 
11  &  8  &  0.866  &  8.499  &  124.167  &   \\ 
11  &  9  &  0.842  &  13.023  &  178.359  &   \\ 
11  &  10  &  0.819  &  17.086  &  219.297  &   \\ 
11  &  11  &  0.797  &  20.145  &  249.204  &   \\ 
 \hline
      \end{tabular}
\egroup}
\end{tabular}
    \vspace{0.5cm}
    \caption{Decision table for a $\textit{Beta}(2, 1)$ prior restricted to $p > 0.5$. For $\delta=5$. 
    }
    \label{tab:MC_alltargets_d5_beta21}
\end{table}
\newpage
%

%% file: tex_files/appendices/A-MC_tables_beta31.tex

\begin{table}[t]
    \centering
    \def\arraystretch{1.5}
    \begin{tabular}{c|c|c|c|c|c} 
    \hline
      \multicolumn{6}{c}{$\bm{\delta = 2}$}\\  
      \hline
      $n_1$  &
      $n_2$  &
      $\mathbf{\mathbb{E}[Q]}$  &  
      $\mathbf{\mathbb{E}[\mathbb{E}[n_{votes}]]}$  & 
      $\mathbf{\mathbb{E}[Var(n_{votes})]}$  &
      $\mathbf{pay^*}$ \\\hline\hline
0  &  0  &  0.898  &  2.903  &  3.009  &   \\ 
1  &  0  &  0.919  &  1.562  &  2.042  &   \\ 
1  &  1  &  0.840  &  3.240  &  4.322  &   \\ 
\textbf{2}  &  \textbf{0}  &  0.935  &  0.000  &  0.000  &    1.722 \\ 
2  &  1  &  0.863  &  1.848  &  3.219  &   \\ 
2  &  2  &  0.801  &  3.419  &  5.080  &   \\ 
\textbf{3}  &  \textbf{1}  &  0.883  &  0.000  &  0.000  &    0.892 \\ 
3  &  2  &  0.823  &  2.026  &  3.987  &   \\ 
3  &  3  &  0.772  &  3.531  &  5.578  &   \\ 
\textbf{4}  &  \textbf{2}  &  0.843  &  0.000  &  0.000  &    0.580 \\ 
4  &  3  &  0.792  &  2.149  &  4.530  &   \\ 
4  &  4  &  0.750  &  3.607  &  5.932  &   \\ 
\textbf{5}  &  \textbf{3}  &  0.812  &  0.000  &  0.000  &    0.420 \\ 
5  &  4  &  0.769  &  2.240  &  4.935  &   \\ 
5  &  5  &  0.733  &  3.662  &  6.196  &   \\ 
      \hline
\end{tabular}
    \vspace{0.5cm}
    \caption{Decision table for a $\textit{Beta}(3, 1)$ prior restricted to $p > 0.5$. For $\delta=2$. 
    }
    \label{tab:MC_alltargets_d2_beta31}
\end{table}
\newpage
\begin{table}[t]
    \centering
    \def\arraystretch{1.5}
    \begin{tabular}{c|c|c|c|c|c} 
    \hline
      \multicolumn{6}{c}{$\bm{\delta = 3}$}\\  
      \hline
      $n_1$  &
      $n_2$  &
      $\mathbf{\mathbb{E}[Q]}$  &  
      $\mathbf{\mathbb{E}[\mathbb{E}[n_{votes}]]}$  & 
      $\mathbf{\mathbb{E}[Var(n_{votes})]}$  &
      $\mathbf{pay^*}$ \\\hline\hline
0  &  0  &  0.935  &  5.109  &  11.980  &   \\ 
1  &  0  &  0.951  &  3.407  &  8.345  &   \\ 
1  &  1  &  0.892  &  6.048  &  18.575  &   \\ 
2  &  0  &  0.964  &  1.644  &  4.220  &   \\ 
2  &  1  &  0.912  &  4.201  &  14.100  &   \\ 
2  &  2  &  0.859  &  6.602  &  22.944  &   \\ 
\textbf{3}  &  \textbf{0}  &  0.973  &  0.000  &  0.000  &    1.917 \\ 
3  &  1  &  0.929  &  2.078  &  7.808  &   \\ 
3  &  2  &  0.880  &  4.725  &  18.297  &   \\ 
3  &  3  &  0.834  &  6.975  &  26.105  &   \\ 
\textbf{4}  &  \textbf{1}  &  0.943  &  0.000  &  0.000  &    1.028 \\ 
4  &  2  &  0.899  &  2.393  &  10.721  &   \\ 
4  &  3  &  0.854  &  5.103  &  21.505  &   \\ 
4  &  4  &  0.814  &  7.245  &  28.520  &   \\ 
\textbf{5}  &  \textbf{2}  &  0.915  &  0.000  &  0.000  &    0.679 \\ 
5  &  3  &  0.873  &  2.633  &  13.107  &   \\ 
5  &  4  &  0.833  &  5.390  &  24.046  &   \\ 
5  &  5  &  0.797  &  7.450  &  30.435  &   \\ 
\textbf{6}  &  \textbf{3}  &  0.891  &  0.000  &  0.000  &    0.494 \\ 
6  &  4  &  0.852  &  2.824  &  15.091  &   \\ 
6  &  5  &  0.815  &  5.617  &  26.115  &   \\ 
6  &  6  &  0.782  &  7.612  &  31.995  &   \\ 
\textbf{7}  &  \textbf{4}  &  0.869  &  0.000  &  0.000  &    0.382 \\ 
7  &  5  &  0.833  &  2.980  &  16.767  &   \\ 
7  &  6  &  0.800  &  5.802  &  27.835  &   \\ 
7  &  7  &  0.769  &  7.742  &  33.295  &   \\ 
      \hline
\end{tabular}
    \vspace{0.5cm}
    \caption{Decision table for a $\textit{Beta}(3, 1)$ prior restricted to $p > 0.5$. For $\delta=3$. 
    }
    \label{tab:MC_alltargets_d3_beta31}
\end{table}
\newpage
\begin{table}[t]
    \def\arraystretch{1.6}
    \begin{tabular}{cc} 
    \footnotesize
    \hspace{-2cm}
    \raisebox{-\height}{\bgroup
    \def\arraystretch{1.2}
    \begin{tabular}{c|c|c|c|c|c}
      \hline
      \multicolumn{6}{c}{$\bm{\delta = 4}$}\\  
      \hline
      $n_1$  &
      $n_2$  &
      $\mathbf{\mathbb{E}[Q]}$  &  
      $\mathbf{\mathbb{E}[\mathbb{E}[n_{votes}]]}$  & 
      $\mathbf{\mathbb{E}[Var(n_{votes})]}$  &
      $\mathbf{pay^*}$ \\\hline\hline
0  &  0  &  0.953  &  7.482  &  29.215  &   \\ 
1  &  0  &  0.966  &  5.402  &  20.341  &   \\ 
1  &  1  &  0.920  &  9.179  &  47.242  &   \\ 
2  &  0  &  0.976  &  3.401  &  12.311  &   \\ 
2  &  1  &  0.937  &  6.842  &  35.809  &   \\ 
2  &  2  &  0.894  &  10.246  &  60.090  &   \\ 
3  &  0  &  0.983  &  1.597  &  5.401  &   \\ 
3  &  1  &  0.951  &  4.389  &  23.623  &   \\ 
3  &  2  &  0.912  &  7.834  &  47.834  &   \\ 
3  &  3  &  0.873  &  10.996  &  69.913  &   \\ 
\textbf{4}  &  \textbf{0}  &  0.988  &  0.000  &  0.000  &    2.087 \\ 
4  &  1  &  0.962  &  2.075  &  11.258  &   \\ 
4  &  2  &  0.928  &  5.128  &  33.286  &   \\ 
4  &  3  &  0.891  &  8.571  &  57.489  &   \\ 
4  &  4  &  0.855  &  11.560  &  77.756  &   \\ 
\textbf{5}  &  \textbf{1}  &  0.972  &  0.000  &  0.000  &    1.156 \\ 
5  &  2  &  0.942  &  2.460  &  16.758  &   \\ 
5  &  3  &  0.908  &  5.707  &  41.507  &   \\ 
5  &  4  &  0.873  &  9.147  &  65.447  &   \\ 
5  &  5  &  0.839  &  12.001  &  84.208  &   \\ 
 \hline
      \end{tabular}
\egroup}
      &
      %
    \raisebox{-\height}{\bgroup
    \def\arraystretch{1.3}
    \begin{tabular}{c|c|c|c|c|c}
      \hline
      \multicolumn{6}{c}{$\bm{\delta = 4}$}\\  
      \hline
      $n_1$  &
      $n_2$  &
      $\mathbf{\mathbb{E}[Q]}$  &  
      $\mathbf{\mathbb{E}[\mathbb{E}[n_{votes}]]}$  & 
      $\mathbf{\mathbb{E}[Var(n_{votes})]}$  &
      $\mathbf{pay^*}$ \\\hline\hline
\textbf{6}  &  \textbf{2}  &  0.954  &  0.000  &  0.000  &    0.776 \\ 
6  &  3  &  0.923  &  2.776  &  21.734  &   \\ 
6  &  4  &  0.890  &  6.175  &  48.558  &   \\ 
6  &  5  &  0.857  &  9.611  &  72.141  &   \\ 
6  &  6  &  0.826  &  12.358  &  89.634  &   \\ 
\textbf{7}  &  \textbf{3}  &  0.937  &  0.000  &  0.000  &    0.570 \\ 
7  &  4  &  0.906  &  3.041  &  26.195  &   \\ 
7  &  5  &  0.874  &  6.563  &  54.667  &   \\ 
7  &  6  &  0.843  &  9.996  &  77.867  &   \\ 
7  &  7  &  0.814  &  12.654  &  94.277  &   \\ 
\textbf{8}  &  \textbf{4}  &  0.920  &  0.000  &  0.000  &    0.442 \\ 
8  &  5  &  0.890  &  3.266  &  30.192  &   \\ 
8  &  6  &  0.860  &  6.891  &  60.015  &   \\ 
8  &  7  &  0.830  &  10.321  &  82.831  &   \\ 
8  &  8  &  0.803  &  12.903  &  98.305  &   \\ 
\textbf{9}  &  \textbf{5}  &  0.905  &  0.000  &  0.000  &    0.357 \\ 
9  &  6  &  0.876  &  3.461  &  33.784  &   \\ 
9  &  7  &  0.847  &  7.173  &  64.737  &   \\ 
9  &  8  &  0.819  &  10.600  &  87.182  &   \\ 
9  &  9  &  0.794  &  13.117  &  101.838  &   \\ 
 \hline
      \end{tabular}
\egroup}
\end{tabular}
    \vspace{0.5cm}
    \caption{Decision table for a $\textit{Beta}(3, 1)$ prior restricted to $p > 0.5$. For $\delta=4$. 
    }
    \label{tab:MC_alltargets_d4_beta31}
\end{table}
\newpage
\begin{table}[t]
    \def\arraystretch{1.6}
    \begin{tabular}{cc} 
    \footnotesize
    \hspace{-2cm}
    \raisebox{-\height}{\bgroup
    \def\arraystretch{1.2}
    \begin{tabular}{c|c|c|c|c|c}
      \hline
      \multicolumn{6}{c}{$\bm{\delta = 5}$}\\  
      \hline
      $n_1$  &
      $n_2$  &
      $\mathbf{\mathbb{E}[Q]}$  &  
      $\mathbf{\mathbb{E}[\mathbb{E}[n_{votes}]]}$  & 
      $\mathbf{\mathbb{E}[Var(n_{votes})]}$  &
      $\mathbf{pay^*}$ \\\hline\hline
0  &  0  &  0.964  &  9.971  &  56.927  &   \\ 
1  &  0  &  0.975  &  7.496  &  39.383  &   \\ 
1  &  1  &  0.937  &  12.532  &  94.450  &   \\ 
2  &  0  &  0.982  &  5.230  &  25.054  &   \\ 
2  &  1  &  0.952  &  9.677  &  71.172  &   \\ 
2  &  2  &  0.916  &  14.206  &  122.407  &   \\ 
3  &  0  &  0.988  &  3.244  &  13.824  &   \\ 
3  &  1  &  0.963  &  6.852  &  49.298  &   \\ 
3  &  2  &  0.932  &  11.224  &  96.911  &   \\ 
3  &  3  &  0.898  &  15.420  &  144.534  &   \\ 
4  &  0  &  0.992  &  1.518  &  5.584  &   \\ 
4  &  1  &  0.973  &  4.265  &  29.507  &   \\ 
4  &  2  &  0.945  &  8.094  &  70.745  &   \\ 
4  &  3  &  0.914  &  12.400  &  118.245  &   \\ 
4  &  4  &  0.882  &  16.354  &  162.714  &   \\ 
\textbf{5}  &  \textbf{0}  &  0.995  &  0.000  &  0.000  &    2.239 \\ 
5  &  1  &  0.980  &  1.984  &  12.847  &   \\ 
5  &  2  &  0.957  &  5.102  &  44.675  &   \\ 
5  &  3  &  0.929  &  9.084  &  89.496  &   \\ 
5  &  4  &  0.899  &  13.335  &  136.300  &   \\ 
5  &  5  &  0.868  &  17.102  &  178.039  &   \\ 
\textbf{6}  &  \textbf{1}  &  0.986  &  0.000  &  0.000  &    1.275 \\ 
6  &  2  &  0.967  &  2.390  &  20.504  &   \\ 
6  &  3  &  0.942  &  5.800  &  58.718  &   \\ 
6  &  4  &  0.914  &  9.896  &  105.948  &   \\ 
6  &  5  &  0.885  &  14.102  &  151.840  &   \\ 
6  &  6  &  0.856  &  17.718  &  191.206  &   \\ 
\textbf{7}  &  \textbf{2}  &  0.975  &  0.000  &  0.000  &    0.869 \\ 
 \hline
      \end{tabular}
\egroup}
      &
      %
    \raisebox{-\height}{\bgroup
    \def\arraystretch{1.3}
    \begin{tabular}{c|c|c|c|c|c}
      \hline
      \multicolumn{6}{c}{$\bm{\delta = 5}$}\\  
      \hline
      $n_1$  &
      $n_2$  &
      $\mathbf{\mathbb{E}[Q]}$  &  
      $\mathbf{\mathbb{E}[\mathbb{E}[n_{votes}]]}$  & 
      $\mathbf{\mathbb{E}[Var(n_{votes})]}$  &
      $\mathbf{pay^*}$ \\\hline\hline
7  &  3  &  0.954  &  2.743  &  28.021  &   \\ 
7  &  4  &  0.928  &  6.392  &  71.546  &   \\ 
7  &  5  &  0.901  &  10.579  &  120.484  &   \\ 
7  &  6  &  0.873  &  14.746  &  165.399  &   \\ 
7  &  7  &  0.846  &  18.236  &  202.688  &   \\ 
\textbf{8}  &  \textbf{3}  &  0.963  &  0.000  &  0.000  &    0.645 \\ 
8  &  4  &  0.941  &  3.052  &  35.186  &   \\ 
8  &  5  &  0.915  &  6.902  &  83.227  &   \\ 
8  &  6  &  0.888  &  11.163  &  133.425  &   \\ 
8  &  7  &  0.861  &  15.297  &  177.365  &   \\ 
8  &  8  &  0.836  &  18.680  &  212.818  &   \\ 
\textbf{9}  &  \textbf{4}  &  0.951  &  0.000  &  0.000  &    0.503 \\ 
9  &  5  &  0.928  &  3.324  &  41.927  &   \\ 
9  &  6  &  0.903  &  7.346  &  93.873  &   \\ 
9  &  7  &  0.877  &  11.669  &  145.029  &   \\ 
9  &  8  &  0.851  &  15.774  &  188.024  &   \\ 
9  &  9  &  0.827  &  19.065  &  221.843  &   \\ 
\textbf{10}  &  \textbf{5}  &  0.940  &  0.000  &  0.000  &    0.407 \\ 
10  &  6  &  0.916  &  3.566  &  48.230  &   \\ 
10  &  7  &  0.891  &  7.737  &  103.602  &   \\ 
10  &  8  &  0.866  &  12.114  &  155.502  &   \\ 
10  &  9  &  0.842  &  16.193  &  197.594  &   \\ 
10  &  10  &  0.819  &  19.402  &  229.950  &   \\ 
\textbf{11}  &  \textbf{6}  &  0.929  &  0.000  &  0.000  &    0.339 \\ 
11  &  7  &  0.905  &  3.782  &  54.113  &   \\ 
11  &  8  &  0.881  &  8.085  &  112.522  &   \\ 
11  &  9  &  0.857  &  12.509  &  165.010  &   \\ 
11  &  10  &  0.833  &  16.564  &  206.246  &   \\ 
11  &  11  &  0.811  &  19.701  &  237.281  &   \\ 
 \hline
      \end{tabular}
\egroup}
\end{tabular}
    \vspace{0.5cm}
    \caption{Decision table for a $\textit{Beta}(3, 1)$ prior restricted to $p > 0.5$. For $\delta=5$. 
    }
    \label{tab:MC_alltargets_d5_beta31}
\end{table}
\newpage

%% file: tex_files/appendices/A-payment-proof.tex
\section{Proof of the expected payment under a Beta prior}
\label{app:payment_proof}

This appendix provides a formal derivation of Equation~\ref{equ:pay_p_beta}, restated below as a proposition.

\begin{propo}
\label{prop:pay_beta}
Let $p \sim \textit{Beta}(\alpha,\beta)$ with $\alpha, \beta > 0$, and let $\varphi = p/(1-p)$ denote the odds. Then the expected cost-equivalent payment rate from Theorem~\ref{theo:pay_eq},
\begin{displaymath}
\textit{pay}(\alpha,\beta) \;=\; \mathbb{E}_{\textit{Beta}(\alpha,\beta)}\!\left[\ln\varphi \cdot \frac{\varphi - 1}{\varphi + 1}\right] \;=\; \mathbb{E}_{\textit{Beta}(\alpha,\beta)}\!\left[(2p-1)\,\ln\!\frac{p}{1-p}\right],
\end{displaymath}
satisfies
\begin{displaymath}
\textit{pay}(\alpha,\beta) \;=\; \frac{\alpha - \beta}{\alpha + \beta}\,\bigl(\psi(\alpha) - \psi(\beta)\bigr) \;+\; \frac{2}{\alpha + \beta},
\end{displaymath}
where $\psi(\cdot)$ is the digamma function.
\end{propo}

\begin{proof}
Write $f(p) = \ln\frac{p}{1-p}$ and expand the integrand:
\begin{displaymath}
(2p-1)\,f(p) \;=\; 2p\,f(p) \;-\; f(p).
\end{displaymath}

\paragraph{Step 1: Shift identity for Beta expectations.}
For any integrable function $f$ and $p \sim \textit{Beta}(\alpha,\beta)$, the identity
\begin{equation}
\mathbb{E}_{\textit{Beta}(\alpha,\beta)}[p \cdot f(p)] \;=\; \frac{\alpha}{\alpha+\beta}\;\mathbb{E}_{\textit{Beta}(\alpha+1,\beta)}[f(p)]
\label{equ:beta_shift}
\end{equation}
follows directly from the Beta density: $p \cdot \frac{p^{\alpha-1}(1-p)^{\beta-1}}{B(\alpha,\beta)} = \frac{B(\alpha+1,\beta)}{B(\alpha,\beta)} \cdot \frac{p^{\alpha}(1-p)^{\beta-1}}{B(\alpha+1,\beta)} = \frac{\alpha}{\alpha+\beta} \cdot \textit{Beta}(p;\alpha+1,\beta)$, using the identity $B(\alpha+1,\beta)/B(\alpha,\beta)=\alpha/(\alpha+\beta)$.

\paragraph{Step 2: Digamma expectation.}
The log-odds under a Beta distribution has a well-known expectation~\citep{archer14a}:
\begin{equation}
\mathbb{E}_{\textit{Beta}(\alpha,\beta)}\!\left[\ln\frac{p}{1-p}\right] \;=\; \psi(\alpha) - \psi(\beta).
\label{equ:logodds_expect}
\end{equation}

\paragraph{Step 3: Assembling the result.}
Using Equations~\ref{equ:beta_shift} and~\ref{equ:logodds_expect}:
\begin{align*}
\mathbb{E}[(2p-1)\,f(p)]
&= 2\,\mathbb{E}[p\,f(p)] - \mathbb{E}[f(p)] \\[4pt]
&= \frac{2\alpha}{\alpha+\beta}\,\mathbb{E}_{\textit{Beta}(\alpha+1,\beta)}[f(p)] \;-\; \bigl(\psi(\alpha)-\psi(\beta)\bigr) \\[4pt]
&= \frac{2\alpha}{\alpha+\beta}\,\bigl(\psi(\alpha+1)-\psi(\beta)\bigr) \;-\; \bigl(\psi(\alpha)-\psi(\beta)\bigr).
\end{align*}
Applying the digamma recurrence $\psi(\alpha+1) = \psi(\alpha) + 1/\alpha$:
\begin{align*}
&= \frac{2\alpha}{\alpha+\beta}\left(\psi(\alpha) + \frac{1}{\alpha} - \psi(\beta)\right) - \bigl(\psi(\alpha)-\psi(\beta)\bigr) \\[4pt]
&= \frac{2\alpha}{\alpha+\beta}\,\bigl(\psi(\alpha)-\psi(\beta)\bigr) + \frac{2}{\alpha+\beta} - \bigl(\psi(\alpha)-\psi(\beta)\bigr) \\[4pt]
&= \left(\frac{2\alpha}{\alpha+\beta} - 1\right)\bigl(\psi(\alpha)-\psi(\beta)\bigr) + \frac{2}{\alpha+\beta} \\[4pt]
&= \frac{\alpha-\beta}{\alpha+\beta}\,\bigl(\psi(\alpha)-\psi(\beta)\bigr) + \frac{2}{\alpha+\beta}.
\end{align*}
The last equality uses $\frac{2\alpha}{\alpha+\beta} - 1 = \frac{2\alpha - \alpha - \beta}{\alpha+\beta} = \frac{\alpha-\beta}{\alpha+\beta}$.
\end{proof}

\paragraph{Interpretation.} The formula decomposes into two terms. The first term, $\frac{\alpha-\beta}{\alpha+\beta}(\psi(\alpha)-\psi(\beta))$, equals $\mathbb{E}[2p-1] \cdot \mathbb{E}[\ln\varphi]$ and captures the payment that would arise if accuracy $(2p-1)$ and informativeness $(\ln\varphi)$ were independent. The second term, $\frac{2}{\alpha+\beta}$, is a covariance correction: under the Beta distribution, higher-accuracy workers also provide more informative votes, so the expected payment exceeds the product of the individual expectations. As the prior concentrates (i.e., $\alpha+\beta \to \infty$), this correction vanishes and the formula reduces to the deterministic result from Theorem~\ref{theo:pay_eq}. Note also that when $\alpha = \beta$ (symmetric prior), $\mathbb{E}[2p-1]=0$, and the payment reduces to the pure covariance term $\frac{2}{\alpha+\beta}$, which is always positive---reflecting the fact that even a symmetric pool has positive expected informativeness.